\documentclass[aps,pre,twocolumn,superscriptaddress]{revtex4-1}
\usepackage{graphicx}
\usepackage{dcolumn}
\usepackage{bm}
\usepackage{amssymb}
\usepackage{amsmath}
\usepackage{color}
\usepackage{hyperref}
\usepackage{subfigure}
\usepackage{textcomp}
\usepackage{multirow}
\extrafloats{100}

\usepackage[usenames,dvipsnames]{xcolor}
\usepackage[normalem]{ulem}

\hypersetup{
     colorlinks=true,       
     linkcolor=blue,          
     citecolor=blue,        
     filecolor=magenta,      
     urlcolor=black         
}

\begin{document}

\title{The dynamics of scattering in undulatory active collisions}

\author{Jennifer M. Rieser}
\email{jennifer.rieser@physics.gatech.edu}
\affiliation{School of Physics, Georgia Institute of Technology, Atlanta, GA, 30332, USA}

\author{Perrin E. Schiebel}
\affiliation{School of Physics, Georgia Institute of Technology, Atlanta, GA, 30332, USA}

\author{Arman Pazouki}
\affiliation{Department of Mechanical Engineering, California State University, Los Angeles, CA, 90032, USA}

\author{Feifei Qian}
\altaffiliation[Now at ]{Department of Electrical and Systems Engineering, University of Pennsylvania, Philadelphia, PA 19104, USA}
\affiliation{School of Electrical and Computer Engineering, Georgia Institute of Technology, Atlanta, GA, 30332, USA}

\author{Zachary Goddard}
\affiliation{School of Mechanical Engineering, Georgia Institute of Technology, Atlanta, GA, 30332, USA}

\author{Andrew Zangwill}
\affiliation{School of Physics, Georgia Institute of Technology, Atlanta, GA, 30332, USA}

\author{Dan Negrut}
\affiliation{Department of Mechanical Engineering, University of Wisconsin, Madison, WI, 53706, USA}

\author{Daniel I. Goldman}
\email{daniel.goldman@physics.gatech.edu}
\affiliation{School of Physics, Georgia Institute of Technology, Atlanta, GA, 30332, USA}

\date{\today}


\begin{abstract}
Natural and artificial self-propelled systems must manage environmental interactions during movement. Such interactions, which we refer to as active collisions, are fundamentally different from momentum-conserving interactions studied in classical physics, largely because the internal driving of the locomotor can lead to persistent contact with heterogeneities. Here, we experimentally and numerically study the effects of active collisions on a laterally-undulating sensory-deprived robophysical model, whose dynamics are applicable to self-propelled systems across length scales and environments. The robot moves via spatial undulation of body segments, with a nearly-linear center-of-geometry trajectory. Interactions with a single rigid post scatter the robot, and these deflections are proportional to the head-post contact duration. The distribution of scattering angles is smooth and strongly-peaked directly behind the post. Interactions with a single row of evenly-spaced posts (with inter-post spacing $d$) produce distributions reminiscent of far-field diffraction patterns: as $d$ decreases, distinct secondary peaks emerge as large deflections become more likely. Surprisingly, we find that the presence of multiple posts does not change the nature of individual collisions; instead, multi-modal scattering patterns arise from multiple posts altering the likelihood of individual collisions to occur. As $d$ decreases, collisions near the leading edges of the posts become more probable, and we find that these interactions are associated with larger deflections. Our results, which highlight the surprising dynamics that can occur during active collisions of self-propelled systems, can inform control principles for locomotors in complex terrain and facilitate design of task-capable active matter.
\end{abstract}

\maketitle

\section{Introduction}

Biological and artificial systems must manage mechanical interactions with the environment to generate and sustain movement. These interactions come in a myriad of forms, from repeated collisions with rigid ground~\cite{Holmes:2006ku} to managing and manipulating flowable substrates like granular media~\cite{Hosoi:2015gy} and fluids~\cite{childress2012natural}. We refer to the interactions between self-propelled systems and heterogeneities in the surrounding environment as \emph{active collisions}, and, as noted in~\cite{marchetti2013}, the damped and driven nature of active systems, conservation of momentum does not apply to to collisions alone.  As a result, the intuitive framework of introductory classical mechanics is unable to capture the diverse and rich behavior arising from active collisions. 

Whether the interactions are amongst like individuals or between an individual and a heterogeneity, many share a common feature: the driving allows for \emph{persistent} interactions. These interactions are an important factor in many systems spanning a wide range of length scales, from the aggregation of bacteria near surfaces to form biofilms~\cite{drescher2011fluid}, the self-assembly and disassembly of colloidal clusters~\cite{palacci2013living}, and the scattering of spermatozoa and \emph{Chlamydomonas} from surfaces~\cite{Kantsler:2013ge} to locomotion of animals and robots.

Such interactions are often detrimental to individual's ability to move, for example, \emph{E. coli} experience a speed reduction near walls~\cite{frymier1995three}, self-propelled rods are geometrically captured by cylindrical obstacles~\cite{takagi2014hydrodynamic}, and collections of self propelled particles can become jammed in disordered landscapes~\cite{reichhardt2014active}. However, if properly utilized, these interactions can benefit locomotion.  For example, rapidly running cockroaches use exoskeletal interactions to maneuver through grass~\cite{Li:2015jn} and clutter~\cite{Spagna:2007kf}, snakes use body parts to propel from bark and rubble~\cite{kelley1997effects}, and \emph{C. elegans} use structure in their environments to propel themselves faster~\cite{park2008enhanced}. In robotics, properly tuned dynamical systems can take advantage of periodic mechanical interactions to produce sustained movement~\cite{Spagna:2007kf,McGeer:1990uk,Coleman:1998uk,Saranli:2001vk} and properly-timed tail-ground interactions improve performance on yielding substrates~\cite{mcinroe2016tail} and can reduce the effects of collisions~\cite{Qian:2015gz}.

In active systems, interactions and collisions with the environment can persist for long durations, and only when the velocity is directed away from the obstacle or boundary is detachment possible (provided the individual can overcome any other pinning forces and torques). In the microscopic realm, the direction of driving is typically modeled stochastically (arising from Brownian motion) and can include a rotational diffusion term~\cite{bechinger2016active}. The strength of the driving and the size of the orientational variations dictate the duration of the interaction as well as the outcome. While the mechanisms by which the orientation can change is different in macroscopic systems, typically either induced by environmental interactions or inherent in the self-propulsion, the ability to reorient remains important for breaking contact with and maneuvering through obstacles. A recent study of environmentally-induced passive reorientation found that a robophysical cockroach was more successful in traversing narrow openings when biologically-inspired body vibrations were added~\cite{Li2017mm}.

Here, we study a system in which the orientation is inherent in the self-propulsion of a long slender locomotor that uses \emph{undulatory propulsion}, in which body bends are generated and subsequently propagated down the body to produce movement~\cite{gray1953undulatory}, to move through heterogeneous terrain. This mode of locomotion is observed over a broad range of length scales and produces effective movement in a wide range of environments, from swimming in fluids (e.g., spermatozoa~\cite{gray1955propulsion}, nematodes~\cite{gray1964locomotion}, and aquatic vertebrates~\cite{sfakiotakis1999review,gillis1996undulatory}) to slithering on and within granular materials (e.g., nematodes~\cite{juarez2010motility}, lizards~\cite{maladen2009undulatory}, and snakes~\cite{sharpe2015locomotor}) to traversing complex environments (e.g., nematodes~\cite{lockery2008artificial,park2008enhanced,majmudar2012experiments} and snakes~\cite{gray1950kinetics,kelley1997effects}). In this paper, we will focus on \emph{lateral undulation}, in which body bends only occur in the horizontal plane. Despite this restriction, this form of propulsion is still quite general, being the only mode of locomotion shared by all limbless terrestrial vertebrates~\cite{gans1975tetrapod}).

The versatility and maneuverability of limbless animals has inspired an ongoing effort to develop control schemes which will enable robotic snakes to move with similar ability in complex terrain and within confined environments, including through disaster sites and rubble~\cite{Murphy2008}. Recent studies have made progress creating control schemes to enhance movement in some terrain~\cite{transeth2008snake,Liljeback:2010eb,Liljeback:2010tg,travers2016shape}.  Despite their successes,  these controllers can be complicated and are often specific to the environment.  We hypothesize that these controllers may be simpler and more general if developed from a fundamental understanding of the physical interactions mediating locomotion in complex environments, and we begin to make progress by studying the interactions between a locomotor and heterogeneities.

In this paper, we take a robophysics~\cite{Aguilar:2016bq} approach, building upon our previous work~\cite{Qian:2015ua} to explore the nature of the interactions underlying active collisions occurring undulatory self-propulsion in dissipative environments. In Section~\ref{sec:expAndSim}, we describe the details of our experiment and simulations, and we show that the nature of the dissipation in our system is similar to other highly-damped environments, such as viscous fluids and granular materials, in which undulatory propulsion is an effective mode of locomotion. In Section~\ref{sec:interaction}, we begin our investigation of active collisions by characterizing the interactions between our robot and a single obstacle, and we find that these interactions rotate the robot's trajectory. When initial conditions are sampled evenly, we find that, counter-intuitively, the probability distribution of angles is highest in the shadow of the obstacle. This motivates the simple model in Section~\ref{sec:model} which captures key features of these non-momentum-conserving collisions: collisions near the leading edge of the obstacle can result in larger scattering events. In Section~\ref{sec:interactions_multi}, we explore interactions that arise when multiple obstacles are present, and we find that scattering patterns produced are reminiscent of far-field diffraction:  as obstacle density increases, scattering patterns broaden and preferred reorientations emerge. We demonstrate in Section~\ref{sec:times} that, despite the complexity of robot-post interactions, the contact duration of head predicts scattering angle and that a single robot-obstacle collision dominates the resulting reorientation. In Section~\ref{sec:remap}, we show that individual collision states are unaltered by the presence of multiple obstacles, and we reveal how the scattering patterns generated by multiple obstacles arise from enhancing the likelihood of collision states near the leading edge of the obstacles.

\section{Methods}\label{sec:expAndSim}

To gain physical insight into active collisions during undulatory self-propulsion, we adopted a robophysical approach and create a laterally-undulating robotic snake whose simple control scheme enabled repeatable experiments. Experiments of the robot moving through an environment with a few simple obstacles allowed for a thorough characterization of interactions and provided a statistical picture of how active collisions altered the robot's trajectory. Experimentally-validated multi-body physics simulations~\cite{pazouki2017compliant} are in good agreement with experiments and enabled a broader range of parameter variation. An overview of the experiment and simulation are provided in this section. For further details, see~\cite{supplemental}.

\subsection{Experiments}

Our robophysical snake is shown in Fig.~\ref{fig:robot}a. $13$ segments were connected together by $12$ servo motors, each of which was oriented so that actuation controlled the angular position within the horizontal plane. Body bends were produced by commanding the angular position, $\zeta_i$, of each motor, $i$, to vary sinusoidally in time: $\zeta_i(t) = \zeta_{max} \sin(2\pi i/N -2\pi f t)$, creating a \emph{serpenoid} curve~\cite{Hirose:1993} (see Fig.~\ref{fig:robot}b). Here, $N = 12$ is the number of motors along the body, $f = 0.15$~Hz is the frequency of undulation, and $\zeta_{max} = 40^{\circ}$ is the angular amplitude. Each bend originated at the head, and as time progressed, was passed sequentially along the body from head to tail.

Translational motion of the robot was achieved from the motor-angle actuation through a frictional anisotropy, created by affixing a pair of passive wheels (connected by an axle) to the bottom of each robot segment~\cite{hirose2009snake}, see Fig.~\ref{fig:robot}a. To characterize this direction-dependent friction, we performed separate experiments in which a wheel assembly (two wheels connected by an axle) was attached to a force transducer that was mounted to a robot arm. The wheel-pair was translated by the robot arm across a rubber substrate at a constant speed as forces were recorded. Steady-state forces as a function of $\psi$, the angle between the velocity and the wheel rolling direction, are shown in Fig.~\ref{fig:robot}c. For comparison, previously-determined drag forces for a submerged rod translated through granular material as well as through a viscous fluid are shown.

Markers atop each motor allow for tracking of individual robot segments throughout an experiment. A typical low-slip trajectory of the robot resulting from the serpenoid motion and the wheel-ground interaction is shown in Fig.~\ref{fig:robot}d. Experimentally-measured local joint angles, $\zeta_i$, throughout this trajectory (determined from the segment positions) are shown in the space-time plot in  Fig.~\ref{fig:robot}e. The head-to-tail wave progression is confirmed by the diagonal stripes, and the consistency of these stripes throughout four undulations shows that the robot motors reliably followed the prescribed motion.

Heterogeneous environments were created by anchoring rigid, vertical, force-sensitive cylindrical posts to an otherwise homogeneous substrate. Experiments were performed with both a single post (shown in Fig.~\ref{fig:setups}a) as well as five posts (shown in Fig.~\ref{fig:setups}b). In the five-post setup, posts were uniformly-spaced and placed in a single row which was oriented transverse to the initial heading of the robot. In each experiment, four cameras recorded the positions of markers atop each robot segment as the robot traversed the terrain, initially heading toward, interacting with, and subsequently moving beyond the post(s). To characterize the interactions between the robot and the post(s), the robot was initialized to always start in the same configuration: the ``S" shape shown in Fig.~\ref{fig:robot}a. The robot is then placed so that its head is within a box of dimension $L_x \times L_z$, where $L_x$ is set by either the amplitude of the robot (single post) or the center-to-center distance between posts (multi-post) and $L_z$ is set by the distance traveled by the robot in a single undulation cycle. Outside of this region, interactions would either be repeated or the snake would always entirely miss the post.

\subsection{Simulations}

\begin{figure}[ht!]
	\includegraphics[width=\columnwidth]{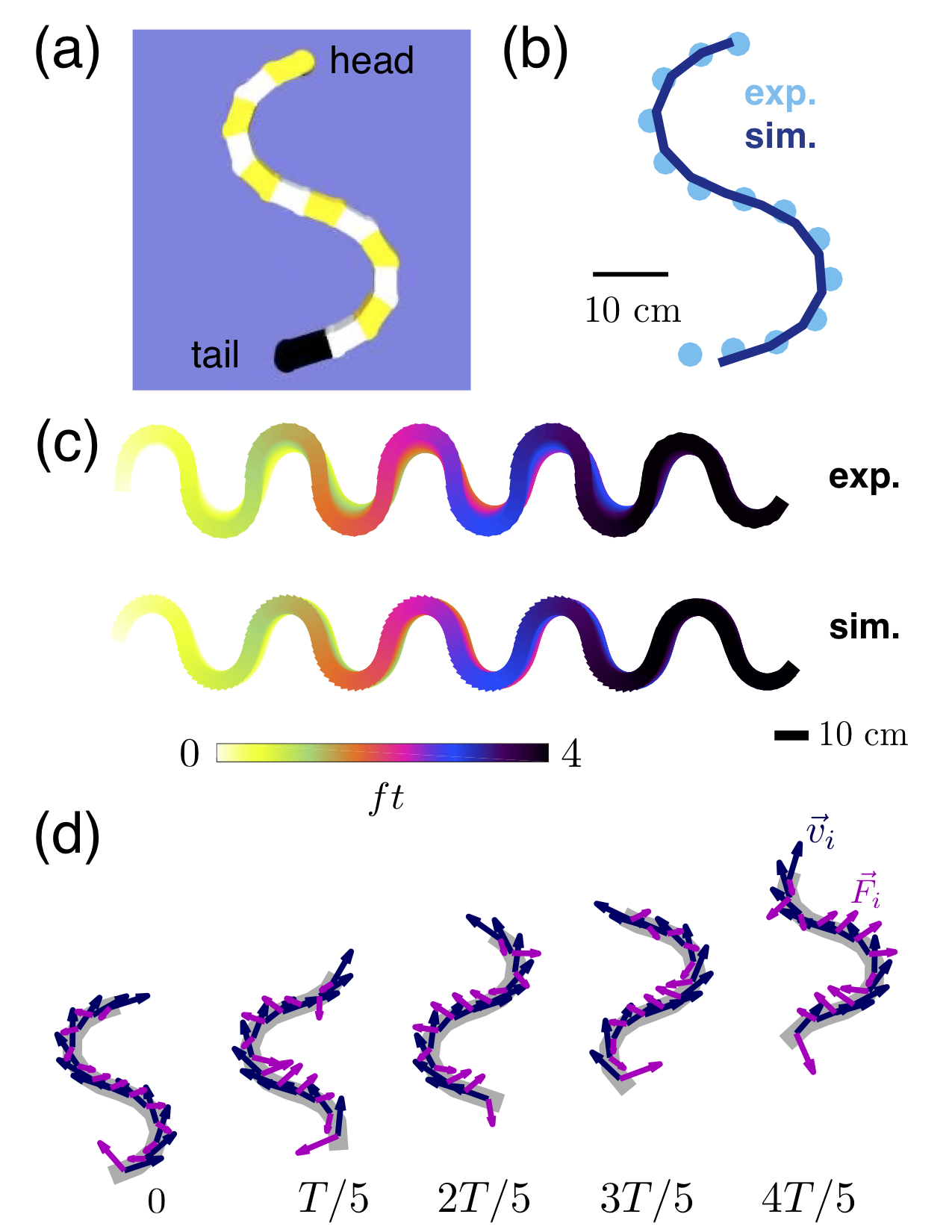}
	\caption{\textbf{Shape and trajectory matching between experiment and simulation.} (a) Snapshot of simulation snake. (b) Experimental and simulated robot shapes. Dots represent the center of each segment on the snake. The solid line connects joints of simulated robot. (c) Position and configuration snapshots of experimental (top) and simulated (bottom) robots in a post-free environment, each colored by elapsed time. (d) Snapshots of robot shapes (equally spaced in time) throughout one undulation cycle. Dark blue vectors show instantaneous velocities of each segment and dark magenta vectors show instantaneous ground-friction forces acting on each segment.}
	\label{fig:sim}
\end{figure}

In addition to the robophysical experiments, we developed and experimentally-validated simulations using Chrono~\cite{Chrono2016}, an open-source multi-body physics simulation environment. Physical parameters from the experiment were used to match the geometry of the experimental and simulated robots, and all snake segments were modeled as boxes linked together with revolute joints. The simulated snake was given a spherical nose cap (of diameter equal to the snake width) which extended from front edge of the head segment; similarly, a vertical cylinder cap extended from the back face of the tail segment. To prevent any unphysical behavior arising from the interactions of inner-segment box corners with other objects, all joints were enclosed by spheres. A snapshot of the simulated snake is shown in Fig.~\ref{fig:sim}a, and a comparison between experimental and simulated robot shapes is shown Fig.~\ref{fig:sim}b. To match the kinematics of the experiment and the simulation in a homogeneous environment, we used fits to the measured wheel-friction force relations~\cite{supplemental} in Fig.~\ref{fig:robot}d for each robot segment~\cite{wheelnote}. The experimental and resulting simulated trajectories through a homogeneous environment are shown in Fig.~\ref{fig:sim}c. The six snapshots in Fig.~\ref{fig:sim}d show the simulated body shapes, segment friction forces, and resulting segment velocities throughout a single undulation cycle.

To create heterogeneous environments in the simulations analogous to those in the experiments, vertical posts were modeled as immovable upright cylinders with high contact stiffness modulus, and robot-post contacts were treated as locally elastic deformations parameterized by the geometric overlaps between robot segments and posts~\cite{johnson1987contact}. The contact stiffness modulus of the simulated posts was tuned to improve agreement between the experimental and simulated trajectories and robot-post forces (to see force comparisons as well as a table of simulation parameters and values, see~\cite{supplemental}). A representative example of the resulting agreement is shown Fig.~\ref{fig:scatter} for a single-post and a multi-post interaction. In each case, the experimental and simulated collisions are nearly identical (occurring at the same location on the peg and in the same part of the undulation phase), and the final trajectories align to within $4\%$.

\section{Results and Discussion}

While most experimental and simulation details were covered in and will be confined to the previous section, we note here that the nature of the environmental interactions in our system is similar to other highly-dissipative systems in which undulatory locomotion is an effective mode of locmotion. Fig.~\ref{fig:robot}c shows dissipation forces for our system, as well as for movement within sand and viscous fluids, making the study of highly-damped wheeled robotic systems potentially relevant to movement within other environments and over a broad range of length scales.

In the work presented here, we investigated the collisional dynamics of an extended, self-propelled, undulating locomotor moving through and interacting with obstacles within a highly-dissipatve environment. We explored both a simple and more complex terrain, and we find that the former can provide insight into the latter.

\subsection{Scattering from a single post}\label{sec:interaction}

We begin with a simple heterogeneous terrain: a single vertical post firmly anchored to an otherwise homogeneous substrate (a schematic is shown in Fig.~\ref{fig:setups}a). With this setup, we investigated how a robotic snake interacted with the post, and how this interaction depended upon the details of the collisions. For all experiments presented here, lateral undulation was achieved by programming the robotic snake to repeatedly propagate a traveling wave from head to tail down its body. The snake had no sensing capabilities, and could therefore only passively react to collisions with environmental heterogeneities. 

Experimental and computational trajectories for a single, similar interaction are shown in the left and right panels of Fig.~\ref{fig:scatter}a. This interaction is representative of the dynamics observed traversing the terrain: the robot initially traveled toward, collided with, and subsequently emerged from the interaction traveling in a new direction. As shown in the right panel of Fig.~\ref{fig:scatter}a, the simulated trajectory matches the experiment, demonstrating that simulations were able to reproduce the interaction and resulting dynamics observed in the experiments. 

These trajectories also show the low-slip movement of the robot through the environment: segments along the body (shown as colors) typically follow the path traced by the head (shown in gray). We find that the head trajectory more faithfully describes the robot's motion than the more commonly-used center-of-geometry (CoG, shown as the dashed line in Fig.~\ref{fig:robot}e). Upon closer inspection, this is not surprising: given the elongated and wavy body, body-segment trajectories rarely intersect the CoG trajectory, and the CoG spends most of its time visiting regions of space that are never occupied by any segment. The CoG can be particularly problematic in describing interactions with heterogeneous environments because it can travel through regions of space that are completely inaccessible (e.g., blocked by a rigid obstacle), even when there is no actual interaction. We therefore use the head trajectories rather than the more standard center of geometry as a simplifying description and representation of the kinematics throughout interaction. 

\begin{figure}[ht!]
	\includegraphics[width=\columnwidth]{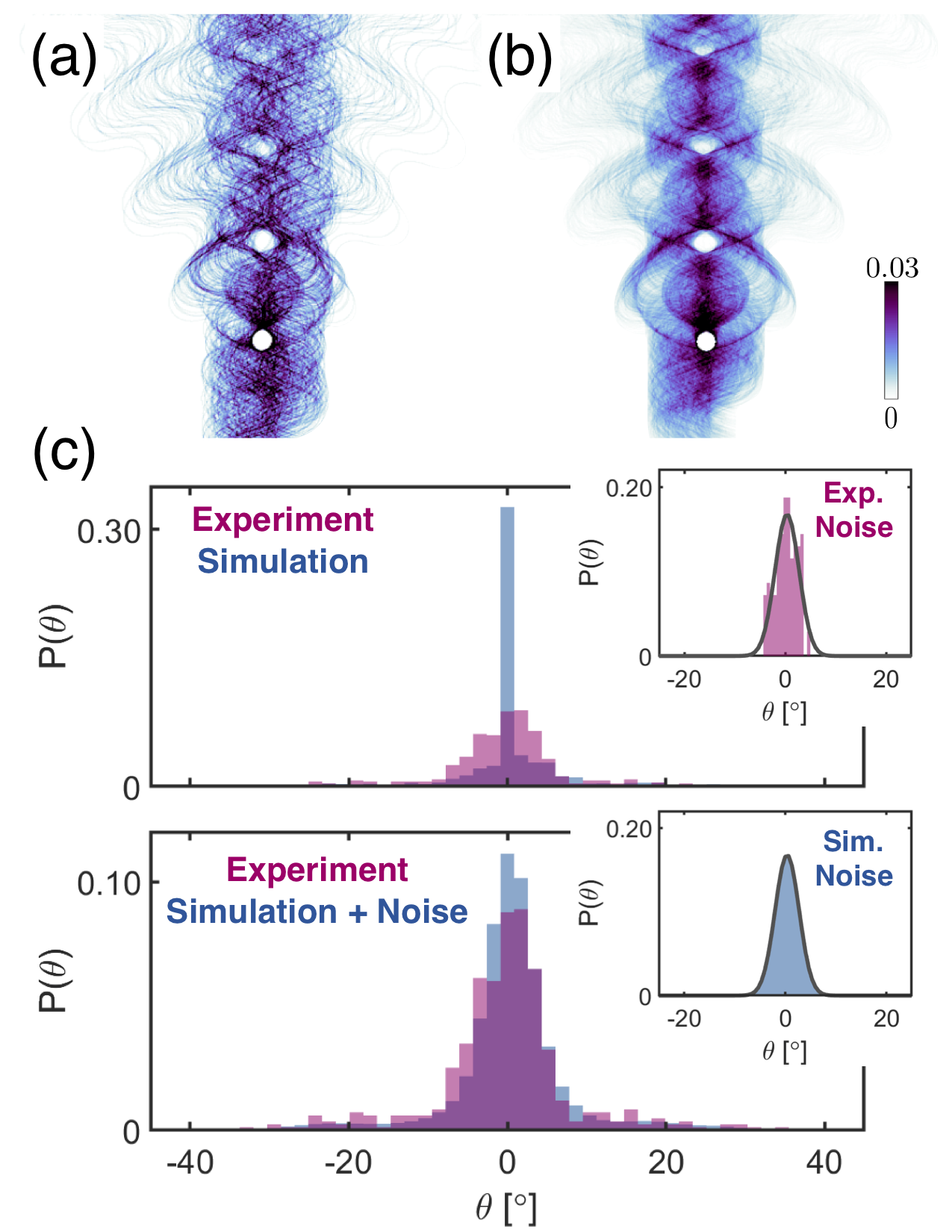}
	\caption{\textbf{Single-post scattering patterns and distributions.} Probability maps of head trajectories for (a) $481$~experimental and (b) $2,998$~simulated snake-post interactions. Here, the color scale indicates fraction of trials passing through each pixel. (c) Top: Scattering angle distributions for both experimental (maroon) and simulated (blue) snake-post interactions with a single post. Inset: Trajectory angles for $104$ experiments in which the robot and the post did not interact (no contact forces were recorded by the post). The curve shows a normalized Gaussian fit to the data, with mean $\theta_0 = 0.4^\circ \pm 0.1^\circ$ and standard deviation $\sigma_\theta = 2.4^\circ \pm 0.1^\circ$. Uncertainty in fit parameters indicate $95\%$ confidence intervals. Bottom: To estimate the effect of the experimental error in robot placement would have on the simulation distribution, a noise-value was drawn from the Gaussian fit and added to each simulation scattering angle. This process was repeated $10,000$ times, and the resulting simulation distribution is shown in blue. The experimental distribution is shown again in maroon for comparison. Inset: The distribution of noise values added to simulation angles is shown in blue, and the gray curve shows the Gaussian fit to the experimental noise distribution.}
	\label{fig:singlepegheatmaps}
\end{figure}

To visualize how collision-induced deflections affect the spatial density of trajectories after the post, we create a binary image of the head trajectory for each experiment. Any region in space that was visited by the robot is assigned a value of unity, and all other pixels are set to zero. Summing these images over many trials produces a \emph{probability map}: regions of space that are more frequently visited by the robot have higher values, and less-frequently-visited regions have lower values. Fig.~\ref{fig:singlepeg} shows how this probability map evolves as more experimental trajectories are added. When all initial conditions are uniformly- and densely-sampled, a structured pattern emerges. Fig.~\ref{fig:singlepegheatmaps}a-b shows these probability maps for experiment and simulation.

We note that there are features in this probability map which are quite different from what we would expect from the momentum-conserving particle collisions of introductory classical mechanics. First, there are periodic excluded regions (``images" of the post) beyond and directly behind the post. These are caused by the combination of (1) the physical constraint that the robot cannot penetrate or move the post guarantees that, at the post, there is a forbidden region (the robot is instead rotated) and (2) the low-slip trajectory enforced by the frictional anisotropy ensures that these forbidden regions will reappear during subsequent undulations and occur at integer multiples of  $v_0T$ (the distance traveled in a single undulation cycle, see Fig.~\ref{fig:scatter}a). 

While such forbidden regions could result from collisions in introductory mechanics, their structure would be quite different. For example, in the predicted scattering pattern of a ball initially traveling along a straight path toward a fixed obstacle, momentum-conserving final trajectories would either miss the obstacle completely or bounce off and scatter backward. In contrast, we find that no single robot trial results in back-scattering or reflection; in all cases, the robot makes it past the post and continues to travel forward.  

Finally, after the robot-post collision, the most likely places to find the robot are directly behind the post. We quantify this observation by measuring the distribution of scattering angles, $\theta$ (defined in Fig.~\ref{fig:scatter}). In both experiments and simulations, the most probable $\theta$ corresponds to zero deflection (see Fig.~\ref{fig:singlepegheatmaps}c). As shown in the top distribution, the central peak is much sharper in the simulation than in the experiment, however, we find that this discrepancy can be explained by the experimental error introduced in the robot's initial heading arising from manual placement of the robot at the beginning of each experiment (see inset in the top distribution of Fig.~\ref{fig:singlepegheatmaps}c).  When we account for this error by artificially adding noise that is representative of experimental error to the simulation scattering angles, we recover a simulation distribution which is much closer to that of the experiment (see bottom plot in Fig.~\ref{fig:singlepegheatmaps}c for resulting distribution and inset for distribution of added noise).

\subsection{Single post collision durations}\label{sec:singleForces}

\begin{figure}[ht!]
	\includegraphics[width=\columnwidth]{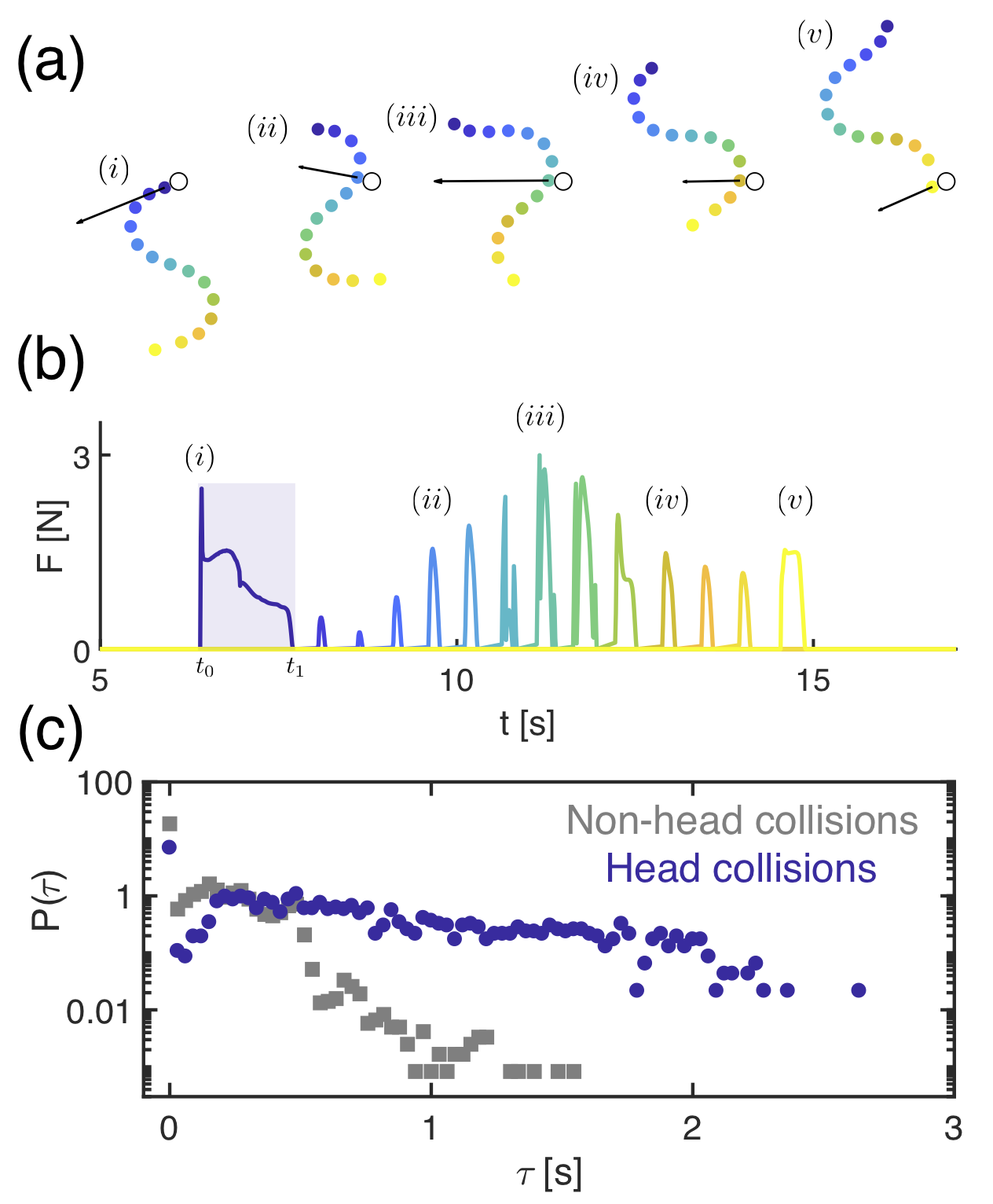}
	\caption{\textbf{Single-post forces in simulation.} (a) Five snapshots throughout an interaction between the simulated robot and a single post for a trajectory that scattered $13.9^\circ$ to the left. The maximum force exerted by the post on the segment in contact at each time is indicated by the arrow. (b) Forces experienced by each segment as the robot interacts with the post. Colors correspond to segment colors is (a). All segments can have comparable maximum forces, but the duration of the head contact (indicated by the shaded region) tends to be longer than the other segments.(c) Distributions of contact durations for head-post collisions (blue) and all other segment-post contact durations (gray). }
	\label{fig:singlepegforces}
\end{figure}

While the central peak in the scattering angles is surprising, approximately half of the simulated trials involved no interaction between the robot and the post. If we restrict our distribution to include only simulations in which a collision occurred, how does this change the distribution? To address this question, we use forces from the simulations to distinguish trials with at least one collision from those without any collisions. Fig.~\ref{fig:singlepegforces}a-b shows snapshots of the simulated robot as it interacted with the post as well as time traces of the forces experienced by each segment as the robot moved past the post.

Using the contact forces, we distinguish these trials according to whether the simulated robot interacted with the post, and we find that $50\%$ have no interaction, $40\%$ of trajectories had head-post collisions, and $10\%$ had only non-head interactions. Surprisingly, we find that when only considering interactions that involved the head of the robot, the distribution remains strongly peaked around no deflection. This result is unlike the familiar momentum-conserving collisions of introductory mechanics, which would have scattered particles away from the post.  

\begin{figure}[ht!]
	\includegraphics[width=\columnwidth]{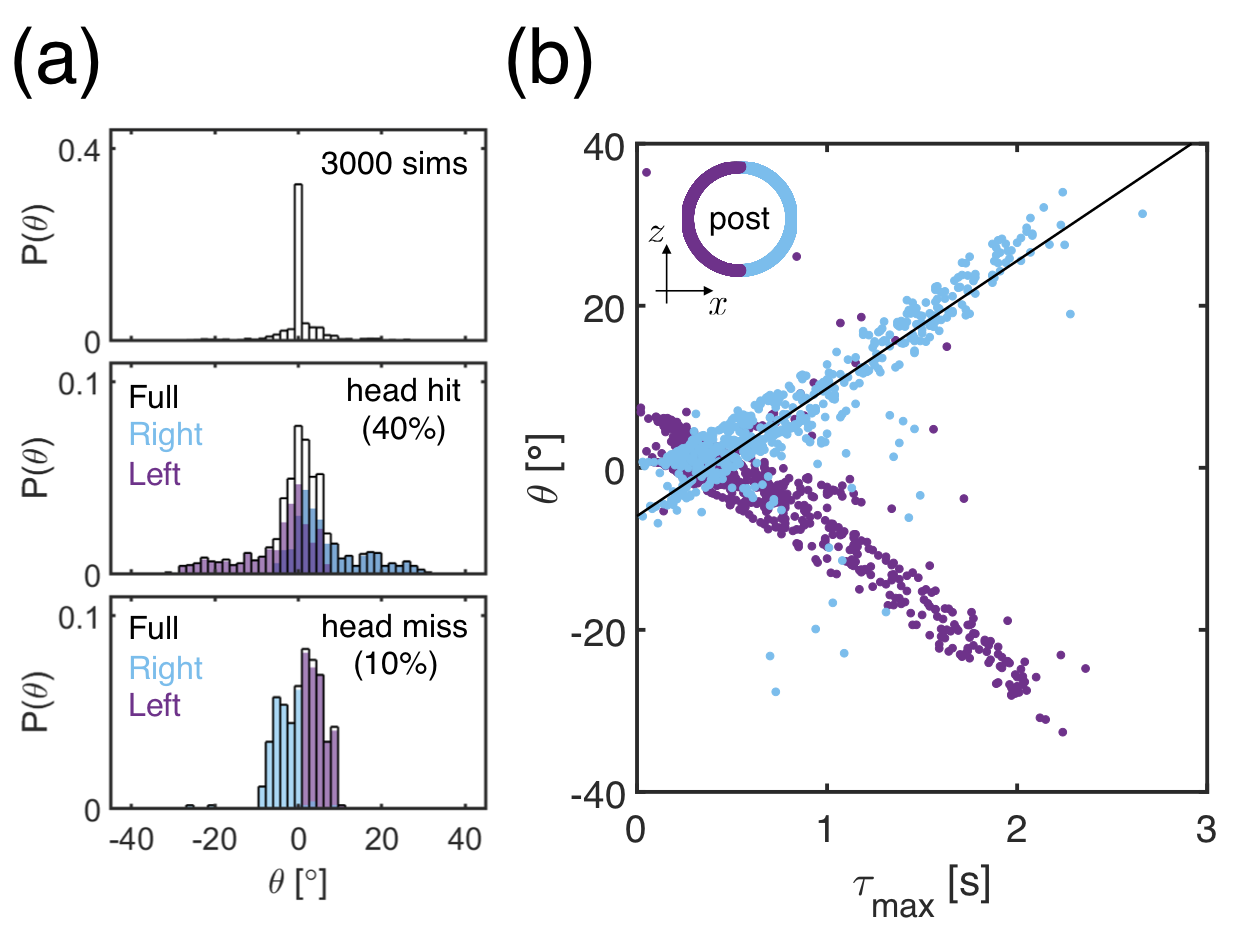}
	\caption{\textbf{Single-post scattering correlates with head-post contact duration.} (a) Distribution of scattering angles for single-post collisions. The top distribution contains $3,000$ simulations.  In $40\%$ of these simulations, the head of the simulated robot collided with the post; another $10\%$ involved only non-head interactions; and the remaining $50\%$ did not interact with the post at all.  The middle plot shows the distribution for simulations in which the head interacted with the post.  The black outline shows the entire distribution, and the colored distributions indicate how the robot scattered based on which side of the post was hit.  In nearly all cases, the robot is rotated away from the post as a result of the interaction (e.g., colliding with the left side of the post (purple) mostly resulted in negative scattering angles). The bottom distribution shows the resulting deflections for interactions which did not involve the head of the simulated robot. Here, the simulated robot is nearly always rotated inward toward the post (e.g., colliding with the left side of the post resulted in positive scattering angles.) (b) For interactions which involved the head of the robot, the resulting scattering angle varies linearly with the duration of the head-post contact. The points are colored by which side of the post was hit, and a linear fit to the blue points is shown in black.}
	\label{fig:singlepegpdfs}
\end{figure}

Most trajectories that did not have a head-post collision did not scatter. Again, due to the low-slip motion of the robot, all body segments tended to reliably follow the path traced by the head. Therefore, if the head did not hit, likely no segment did. However, the slip was not zero, therefore, some trajectories did interact even if the head missed the post. These interactions explain the non-zero scattering events shown in the bottom distribution in Fig.~\ref{fig:singlepegpdfs}. Interestingly, and also different from the collisions of introductory mechanics, nearly all of these trajectories are \emph{attractive}, i.e., the robot was rotated inward toward the post as a result of the interaction. This is also in contrast to the scattering that resulted from head-post collisions, in which most interactions rotated the robot away from the post and caused primarily \emph{repulsive} trajectories.

Given that (1) the robot trajectory reliably follows the path traced by the head (shown in Fig.~\ref{fig:scatter}a), (2) most robot-post interactions involved a head collision (shown in Fig.~\ref{fig:singlepegpdfs}a), and (3) these head collisions persist for longer than other segment-post interactions (shown in Fig.~\ref{fig:singlepegforces}b-c),  we begin to investigate the nature of the robot-post interactions by focusing our attention on the head-post collision.  To quantify the persistence of this interaction, we define the contact duration, $\tau_{max} = t_1 - t_0$. Surprisingly, we find that the scattering angle, $\theta$, varies linearly with $\tau_{max}$, see Fig.~\ref{fig:singlepegpdfs}b.  

\begin{figure}[ht!]
	\includegraphics[width=\columnwidth]{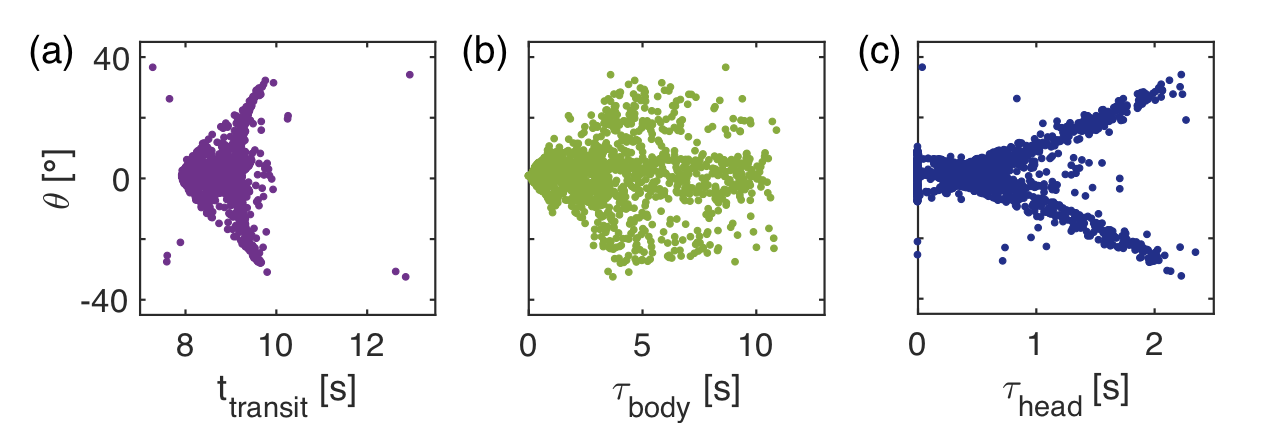}
	\caption{\textbf{Dependence of scattering angle on robot-post interaction times.} (a) Scattering angle dependence on total time required for the robot to pass by the post, determined from positions of the head and tail. The correlation coefficient for $|\theta|$ and $t_{transit}$  is $0.52$. (b) Scattering angle dependence on total robot-post contact duration, determined by summing all individual segment contact durations. The correlation coefficient for $|\theta|$ and $\tau_{body}$  is $0.36$. (c) Scattering angle dependence on head-post contact duration, determined from window of non-zero forces on the head segment. The correlation coefficient for $|\theta|$ and $\tau_{max}$  is $0.82$.
	 }
	\label{fig:times}
\end{figure}

To test the importance of the head-post contact duration relative to other interaction durations, we also calculate two other times associated with the robot-post interaction. First, the transit time, $t_{transit}$, is defined as by the temporal window which starts when the tip of the nose crosses the plane set by the leading edge of the post (the part of the post that is initially closest to the robot) and ends when the tip of the tail crosses the plane set by the trailing edge of the post (the part of the post that is initially farthest to the robot). Second, the total contact time, $\tau_{body}$, is determined by summing all temporal windows for which the forces on any robot segment were non-zero. Fig.~\ref{fig:times} shows how $\theta$ depends on $t_{transit}$, $\tau_{body}$, and $\tau_{max}$.  

While all times show some correlation with $|\theta|$, the head-post duration is by far the best predictor of the resulting scattering angle. This result is surprising: not only is the duration of the head collision an excellent predictor of the scattering angle, adding durations of other segments greatly reduces the correlation and therefore the predictability of the final outcome. Given this result and that the head trajectory provides a simplifying description of the robot's trajectory throughout the interaction, we next focus on understanding the interactions between the head of the robot and the post.

\subsection{Single post active collision model}\label{sec:model}

\begin{figure}[ht!]
	\includegraphics[width=\columnwidth]{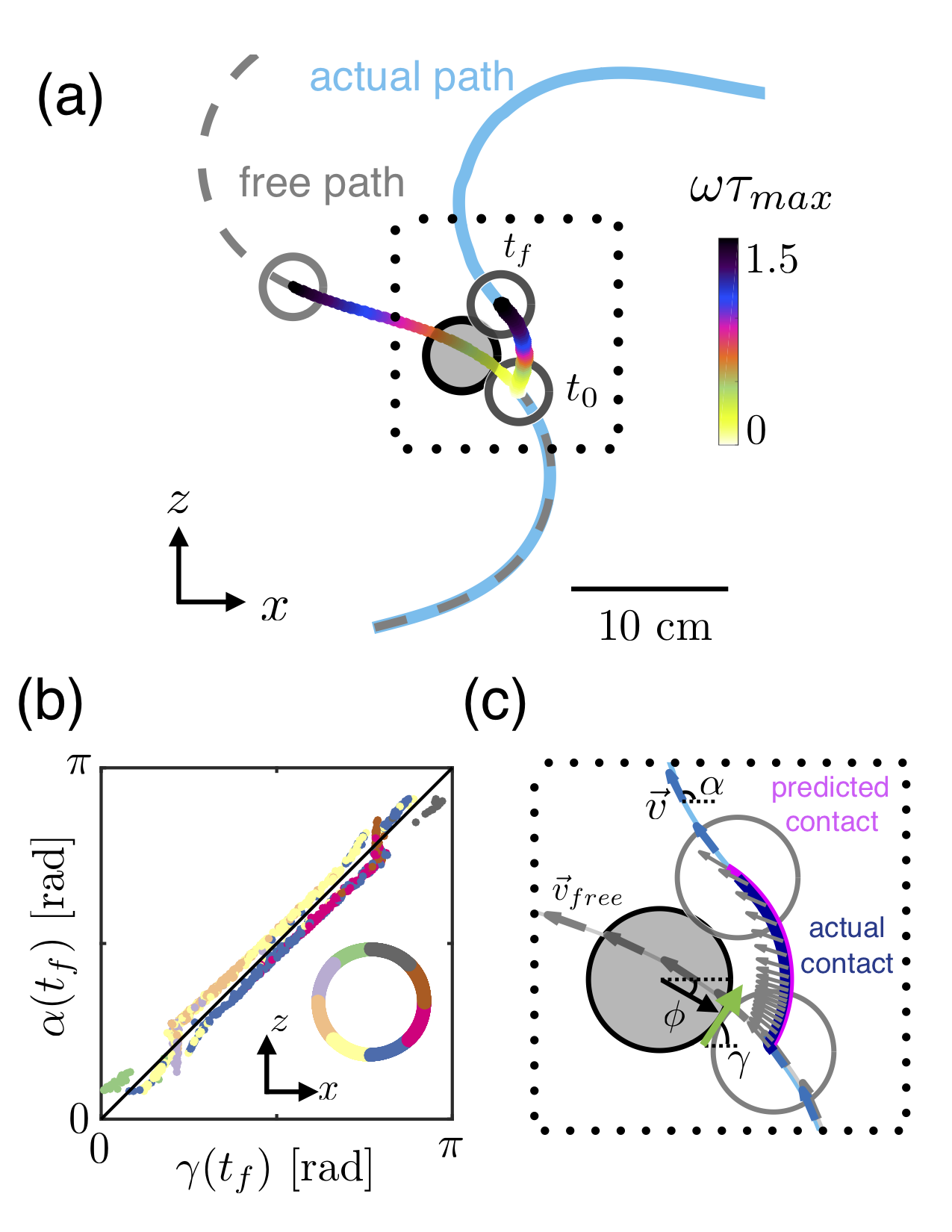}
	\caption{\textbf{Single-particle single-post collisions.} (a) Schematic for the interaction of the robot head (gray circles) with the post (black circle). The actual path of the robot (blue, $\theta = 18.5^{\circ}$) and the corresponding trajectory for the unobstructed robot (gray). The path while in contact is colored by duration, and a comparison of the unobstructed head path over the same duration is also shown to pass through the post and follow the gray dashed line. (b) From single-post simulations, velocity orientation, $\alpha$, vs post tangent angle, $\gamma$, at the final contact time, $t_f$. At $t_f$, the velocity vector has a small component that points away from the center of the post. For collisions occurring on the left side of the post, this means $\alpha(t_f) \gtrsim \gamma(t_f)$ (lighter colored points) and for collisions on the right side of the post, $\alpha(t_f) \lesssim \gamma(t_f)$ (darker colored points). The solid line shows $\alpha(t_f) = \gamma(t_f)$. (c) Zoom-in of the outlined region in (a). Velocity vectors are shown along each trajectory. Actual and model-predicted head contact trajectories shown in blue and magenta, respectively. The velocity vectors of the unobstructed particle (resulting from commanded positions) overlaid onto the actual obstructed path.  For this interaction, $\omega \tau_{sim} = 1.54$ is slightly shorter than the predicted duration of $\omega \tau_{pred} = 1.65$. }
	\label{fig:singlepost}
\end{figure}

Motivated by the observation that scattering events are typically dominated by the head-post collision, we next study single post interactions and develop a self-propelled particle model. Fig.~\ref{fig:singlepost}a shows head trajectories for colliding and unobstructed events originating from the same initial condition. The presence of the rigid, impenetrable post prevents the unobstructed path, forcing the robot to temporarily follow the post surface. The actual and unobstructed paths originating from the same initial condition are identical until contact is established. The initial contact time, $t_0$, and impact location on the post, $\phi$, are well predicted from geometry (the first time and location for which the post and particle overlap). The final contact time, $t_f$, coincides with a non-zero velocity component pointing away from the center of the post, see Fig.~\ref{fig:singlepost}b.

These observations motivate a simple model in which we treat the head as a self-propelled circular particle, illustrated in Fig.~\ref{fig:singlepost}c. We assume that the velocity, $\vec{v}$, determined by differentiating the unobstructed trajectory, is driven as a function of time. During post contact, the particle moves with velocity $\vec{v}_{tan}$, the component of $\vec{v}$ that is locally tangent to the post surface. We further assume that there is zero friction along the post surface, allowing the particle to achieve the full $\vec{v}_{tan}$, and we also require $\vec{v}_{tan} \cdot \hat{z} \geq 0$ (i.e., the particle can not travel backward). The particle maintains contact with the post until the driving velocity aligns with the post tangent. Therefore, the predicted duration is set by the amount of time required for the velocity vector to sweep through enough of a cycle to reorient and align with or barely exceed the local post tangent.  

To describe the head's position in the undulation cycle at $t_0$, we choose to define the wave phase, $\eta$, which, unlike the velocity vector orientation, uniquely specifies location in the undulation cycle. Given that the primary oscillation direction is transverse to the average heading, which is initially in the $+z$-direction, we define the phase as $\eta = \tan^{-1}(\dot{x}/\omega x)$. 

\begin{figure*}[h!]
	\includegraphics[width=2\columnwidth]{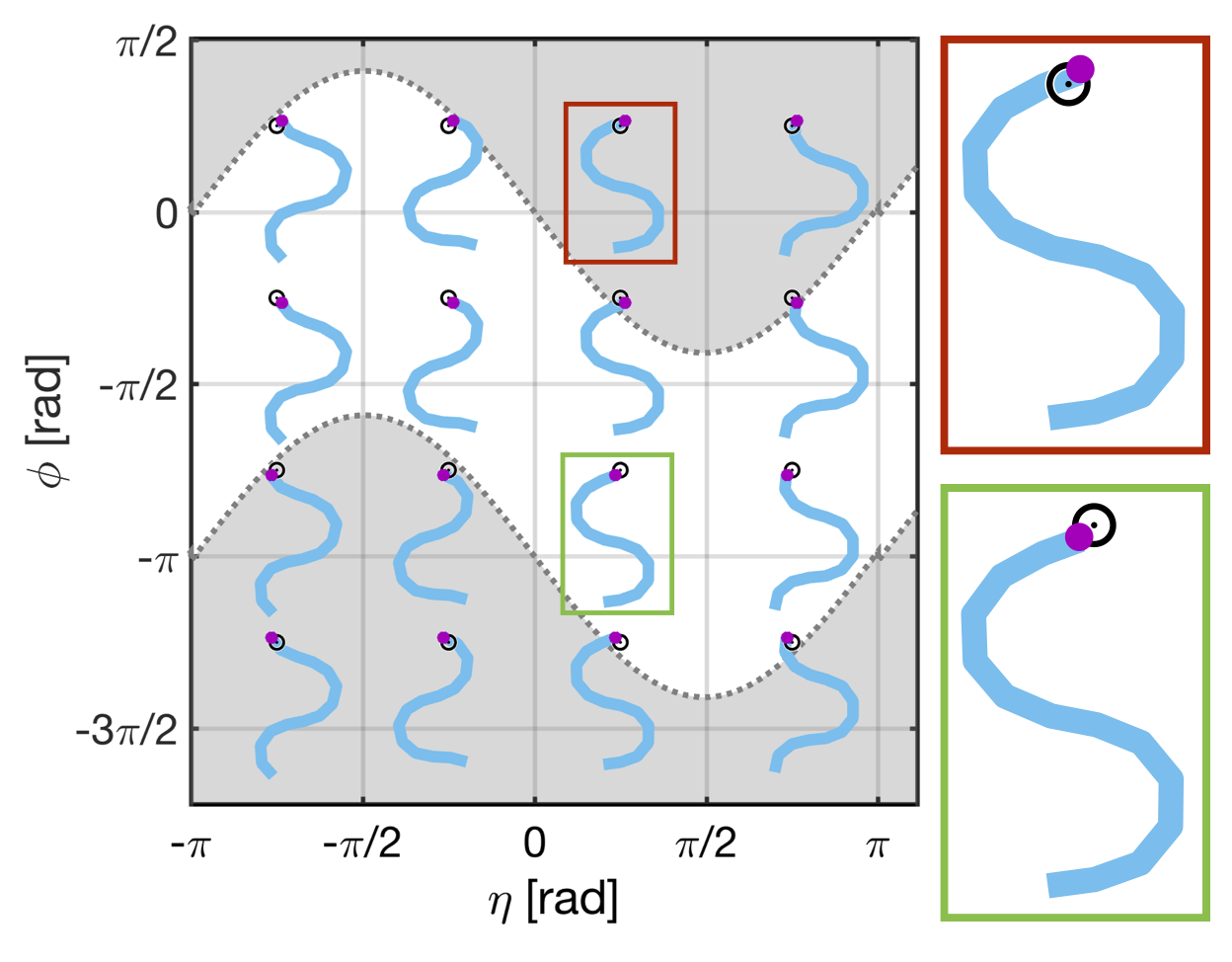}
	\caption{\textbf{Single-post collision state space.} Sketches depicting snake configuration and impact location for single-post collision states. The shaded gray region indicates states that are not allowed because they require the robot to travel through the post to reach the correct configuration. On the right, magnified versions of two configurations are shown: one allowed (outlined in light green) and one forbidden (outlined in dark red). Boxes of corresponding colors in the main plot identify these states within this space.}
	\label{fig:sketches}
\end{figure*}

Using $\eta$ and $\phi$ to characterize collision states, we have reduced the snake-post interaction to a single head-post collision. Fig.~\ref{fig:sketches} depicts the physical configuration of the snake and the post for several of these collision states. Accessing the states in the shaded gray regions would require the robot to travel through the post, therefore these states are forbidden. States within the white band are allowed, and the dashed line between the two regions indicates the boundary between allowed and disallowed states.

We use the model to predict $\omega \tau$ for all possible collision states, and we compare with results from simulation in Fig.~\ref{fig:singlepoststates}a. The structure of $\omega \tau$ as a function of $\eta$ and $\phi$ is qualitatively similar: both are contained within the same region, whose boundaries are identical to those of Fig.~\ref{fig:sketches}, and while there are quantitative differences between the simulation and the prediction~\cite{supplemental}, the dependence of the duration on the collision state is qualitatively similar. 

As shown in Fig.~\ref{fig:singlepoststates}b, the model and simulation follow the same trend for most $\phi$: the maximum duration increases monotonically as impact locations approach the leading edge of the post (except near $\phi \sim -\pi/2$, where the condition $\vec{v}_{tan} \cdot \hat{z} \geq 0$ results in a much longer predicted contact than the actual, grazing contact in the simulation~\cite{supplemental}). This trend can be understood by considering the range of velocity angles, $\alpha$, swept out in a single undulation, compared to the local post tangent, $\gamma$, see Fig.~\ref{fig:singlepoststates}c. As $\gamma$ decreases, a larger fraction of velocity vectors have a component pointing inward toward the center of the post, resulting in more states that can be pinned. Given that (1) any state with an infinitesimal velocity component pointing into the center of the post will be pinned and (2) that the pinning persists until the local tangent vector and velocity orientation align, Fig.~\ref{fig:singlepoststates}c demonstrates that (on the right side of the post) the longest duration pinning will result from a collision occurring with velocity angle pointing just inside the post tangent and increasing into the post (i.e., $\alpha \gtrsim \gamma$ and $\dot{\alpha} > 0$). In this case, the velocity vector must continue to increase until $\alpha = \alpha_{max}$, at which point $\dot{\alpha}$ changes sign and the velocity vector continues to point into the post until $\alpha$ aligns with $\gamma$. As $\phi \to -\pi/2$, the increasingly shallow post tangent results in larger maximum pinning times.    

The model provides a framework for describing active collisions in damped-driven systems. Heterogeneities in the environment impose geometric constraints which can prevent active particles from fully utilizing their internal driving to produce movement. The degree to which obstacles hinder locomotion depends on the details of the driving and the shape of the obstacle: for an undulating locomotor interacting with round posts, we find that the duration of the interactions is set by the undulation phase and post impact location at the initial time of contact. The locomotor is ``stuck" to and can only move along the surface of the obstacle until the velocity vector reorients and has a component pointing away from the obstacle. We note that, in this picture, the contact duration is qualitatively equivalent to the reorientation time of many other active matter systems (see e.g.,~\cite{bechinger2016active}).  However, unlike active Brownian systems and those which experience a purely passive reorientation~\cite{Li2017mm}, the reorientation is largely inherent in the driving of the locomotor. For a periodically-driven locomotor interacting with a single post, we find that collision durations (and corresponding locomotor reorientations) increase as initial impact locations approach the leading edge of the post.   

\subsection{Scattering through multiple posts}\label{sec:interactions_multi}

Since many environments are more complex than a single heterogeneity, we next explore how the of multiple obstacles alters the observed dynamics. Here, five evenly-spaced vertical posts were firmly anchored to an otherwise homogeneous substrate (a schematic is shown in Fig.~\ref{fig:setups}b). Experimental and computational trajectories for a single, similar interaction are shown in the left and right panels of Fig.~\ref{fig:scatter}b. Similar to the single-post interactions, here, the robot is rotated by the collisions with the posts. A probability map of multi-post trajectories shows the likelihood of the robot to occupy points in space after the collision. Fig.~\ref{fig:multiBuildup} shows the evolution of a probability map as more experimental trajectories are added.  When the initial conditions were evenly sampled (shown in the rightmost panel), a structured pattern appeared and the presence of preferred trajectories emerged. 

\begin{figure}[ht!]
	\includegraphics[width=\columnwidth]{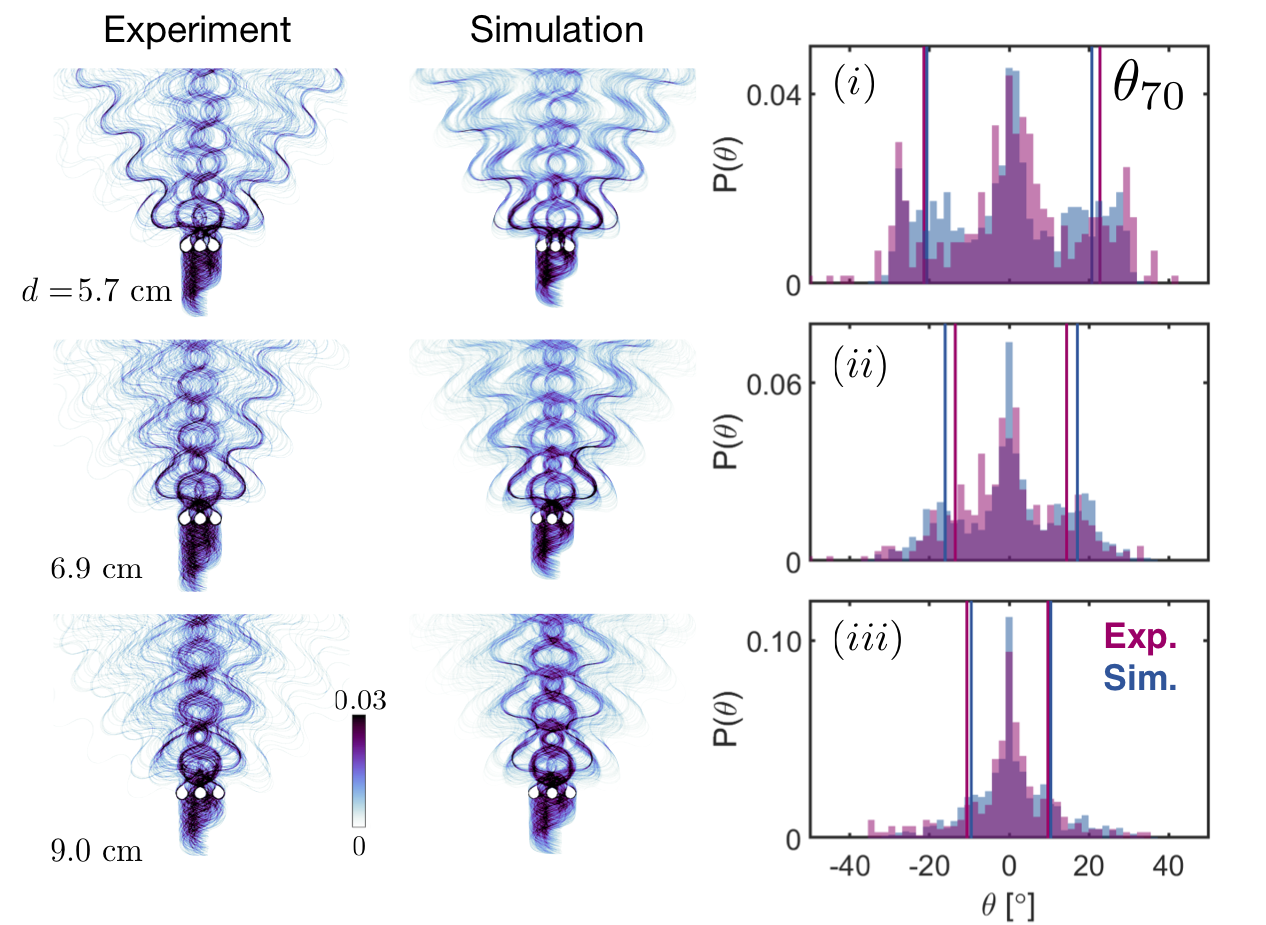}
	\caption{\textbf{Scattering patterns and distributions varying post spacing.} Probability maps of robot head trajectories for three post configurations in  experiment (left column) and simulation (middle column). Here, the color scale indicates fraction of trials passing through each pixel. $d$ is constant across each row and increases down each column. Right column: Experiment (maroon) and simulation (blue) scattering angle distributions for four post spacings, $d = 5.7$~cm, $d = 6.9$~cm, $d = 9.0$~cm, each of which contains at least $300$~trials. Vertical lines show the angles associated with the outer $\pm15\%$ of each distribution (i.e., the $15^{\mathrm{th}}$ and $85^{\mathrm{th}}$ quantiles).}
	\label{fig:preferredPaths}
\end{figure}

Probability maps and their corresponding scattering angle distributions (Fig.~\ref{fig:preferredPaths}) reveal that all $\theta$-distributions have a central peak around zero and are symmetric.  We observe distinct secondary peaks for small $d$; as $d$ increases, these off-center peaks become less prominent and eventually vanish, leaving only the central peak.  Given this qualitative change in the structure of these distributions, we measure the overall spread of the distribution using the quantile value, $\theta_q$ (the $\theta$ value for which $q\%$ of the distribution is below $\theta_{q}$). Since the $\theta$-distribution is nearly symmetric about $0$, we compute $\theta_{70}$ for the $|\theta|$-distribution.

Fig.~\ref{fig:diffraction}a shows that $\theta_{70}$ values decrease with increasing $d$, confirming that the weight of the distributions shifts inward as spacing increases. The dependence of $\theta_{70}$ on the post spacing is well-described by the function $\theta_{70} = (180 / \pi)(D/d)$, where $D$ is a fit parameter~\cite{fitnote}. Variation of $\zeta_{max}$ (defined in Fig.~\ref{fig:robot}b) in simulation reveals that this functional form is valid over an intermediate range of $\zeta_{max}$ (see Fig.~\ref{fig:diffraction}b), with $D$ set by $2\ell\sin\zeta_{max}$, the distance (along the post-plane direction) swept out by each segment during a single period (see Fig.~\ref{fig:diffraction}b-c). Outside of this intermediate $\zeta_{max}$-range, the spacing-dependence is qualitatively different~\cite{supplemental}.  

Features of these scattering distributions are a consequence of persistent, non-momentum-conserving collisions that arise in driven systems: first, even when interacting with multiple posts, there is a strong central peak; second, large reorientations are more frequent for small d and tend to occur at preferred directions. This produces secondary peaks in the scattering distributions which become more prominent as spacing decreases.

\subsection{Multi-post collision durations}\label{sec:times} 

Given the importance of the head-post contact duration for the single-post environment, we again explore the relationship between the contact duration of the head with the posts.  In the multi-post geometry, the head can have multiple collisions which can involve more than one post.  However, we find that there is typically one head collision that dominates, therefore, we start by examining the single collision with the maximal head-post contact duration, and we restrict our analysis to simulations which had at least one head-post collision~\cite{supplemental}. 

\begin{figure}[h!]
	\includegraphics[width=\columnwidth]{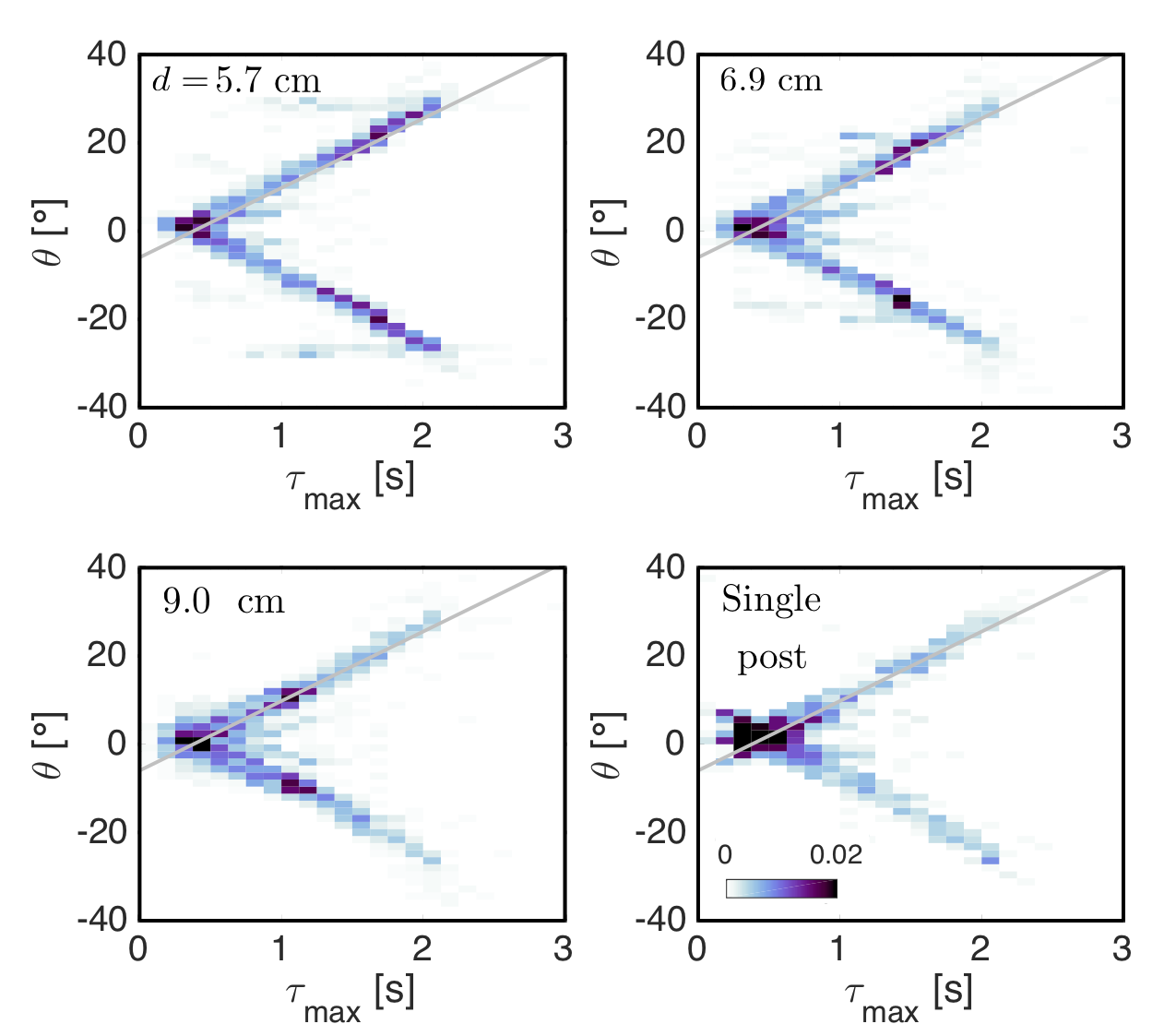}
	\caption{\textbf{Scattering angle dependence on head-post contact duration.}  Scattering angle, $\theta$, depends linearly on $\tau_{max}$, the longest head-post contact duration, even in the presence of multiple posts.  The gray line, determined by fitting the single-peg data in the bottom right, is the same in all plots and shows that this trend is independent of post-spacing, $d$. The underlying color scale represents the two-dimensional probability map and shows that the density of points shifts inward along the $\theta$ vs $\tau$ line as spacing increases. The plot in the bottom right shows the probability map version of the single post data shown in Fig.~\ref{fig:singlepegpdfs}b.} 
	\label{fig:contactDuration}
\end{figure}

Fig.~\ref{fig:contactDuration} shows that, even in the multi-post configuration, $\theta$ depends linearly on $\tau_{max}$, and that this relationship is independent of $d$. When each plot is viewed as a probability map, the dependence on $d$ is clear: the density of points along this line shifts toward lower $\theta$ and $\tau_{max}$ as $d$ increases. Given this linear relationship, we expect that the spread of both the $\tau_{max}$ and $\theta$ distributions should exhibit a similar dependence on spacing.  

We explore this potential similarity by comparing the qualitative dependence $\tau_{max}$- and $\theta$-quantiles on the spacing. Fig.~\ref{fig:collapse}a shows distributions of $\tau_{max}$ for three $d$. We again choose the $70^{\mathrm{th}}$ quantile to characterize the spread of the distributions. Fig.~\ref{fig:collapse}b shows the qualitatively similar spacing-dependence of $\tau_{\mathrm{max},70}$ and $\theta_{70}$. This correspondence is robust, holding over a range of undulation frequencies, $f$, and angular amplitudes, $\zeta_{max}$. When $\theta_{70}$- and $\tau_{\mathrm{max},70}$-distributions are scaled by $\zeta_{max}$ and $f$, respectively, all data collapse to a single line, see Fig.~\ref{fig:collapse}c. 

It is surprising that a single curve describes the dynamics over such a range of wave parameters and post-spacings. We have neglected many details of the interactions that occur along the robot body as it traverses the post array and have shown that we can reduce the system to a single interaction: the longest-duration collision. Not only does this indicate that the resulting dynamics are dominated by the longest head-peg interaction, but it also suggests that only one post is important in a given single- or multi-post scattering event.

\subsection{The origin of broadening distributions}\label{sec:remap}

\begin{figure}[ht!]
	\includegraphics[width=1\columnwidth]{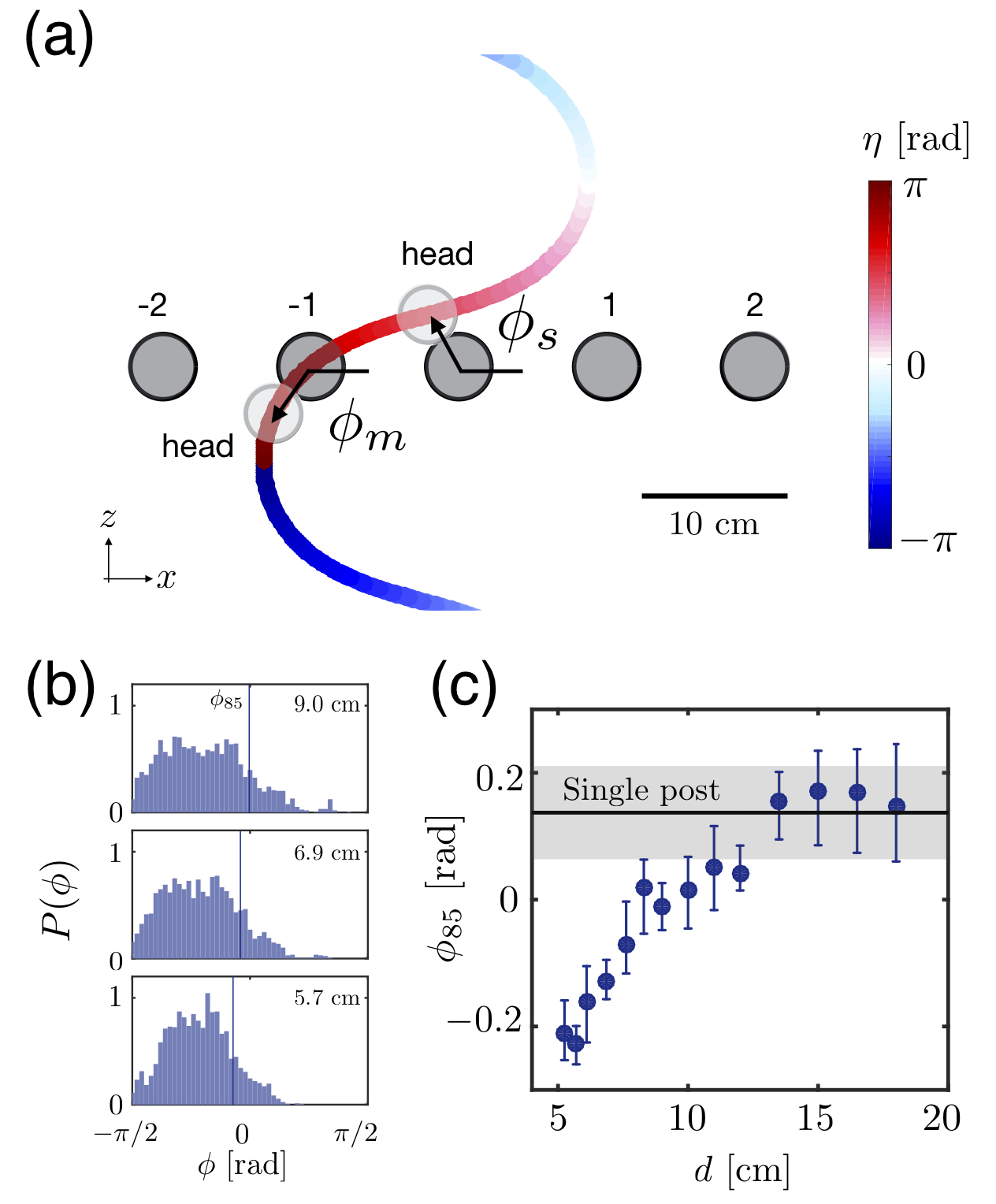}
	\caption{\textbf{Multi-post scattering results from shifted likelihood of single-post collision states.} (a) Post impact location and wave phase at collision are shifted when multiple posts are present. (b) $\phi$ distributions (reflected about $-\pi/2$) for three $d$. As $d$ decreases (down the column), less of peg surface is accessible to the robot, shifting the tails of the distributions toward the leading edge of the post. The vertical lines show $\phi_{85}$, the $85^{\mathrm{th}}$ quantile of each $\phi$-distribution. (c) $\phi_{85}$ as a function of $d$. Errorbars show the bootstrapping-estimated $95\%$ confidence interval for each $\phi_{85}$ value. }
	\label{fig:phiShifts}
\end{figure}

To understand how active collisions in the presence of multiple posts can generate the observed scattering patterns, we examine the unobstructed path of the robot. This path is shifted to coincide with an initial condition that results in a collision for both $d = 5.7$~cm as well as for the single post, see Fig.~\ref{fig:phiShifts}a. From this picture, we see that the single-post collision, which occurs opposite the leading surface of the central post, becomes inaccessible in the multi-post scenario. Instead, a collision with the post immediately to the left precedes the single-post interaction. This new collision with an adjacent post occurs closer to the leading surface of the post, which, at least in the single post case, can result in a longer-duration collision. 

We test the extent to which this observation holds by exploring the dependence of the impact location on spacing.  If our hypothesis is true, we expect that as spacing decreases, states opposite the leading edge of the post will become inaccessible and the tail of the $\phi$ distribution will shift toward the leading edge of the post.  To quantify the tail of the $\phi$ distribution, we choose $\phi_{85}$, the $85^\mathrm{th}$ quantile of the $\phi$ distribution.  This distribution is symmetric about the center line of the post, therefore, we reflect collisions that occurred on the left side of the post about the center line. The resulting distributions for three $d$ are shown in Fig.~\ref{fig:phiShifts}b, and the dependence of $\phi_{85}$ on $d$ is shown in Fig.~\ref{fig:phiShifts}c.  As predicted, the tails of these distributions shift toward the leading edge of the post as $d$ decreases.

\begin{figure}[ht!]
	\includegraphics[width=1\columnwidth]{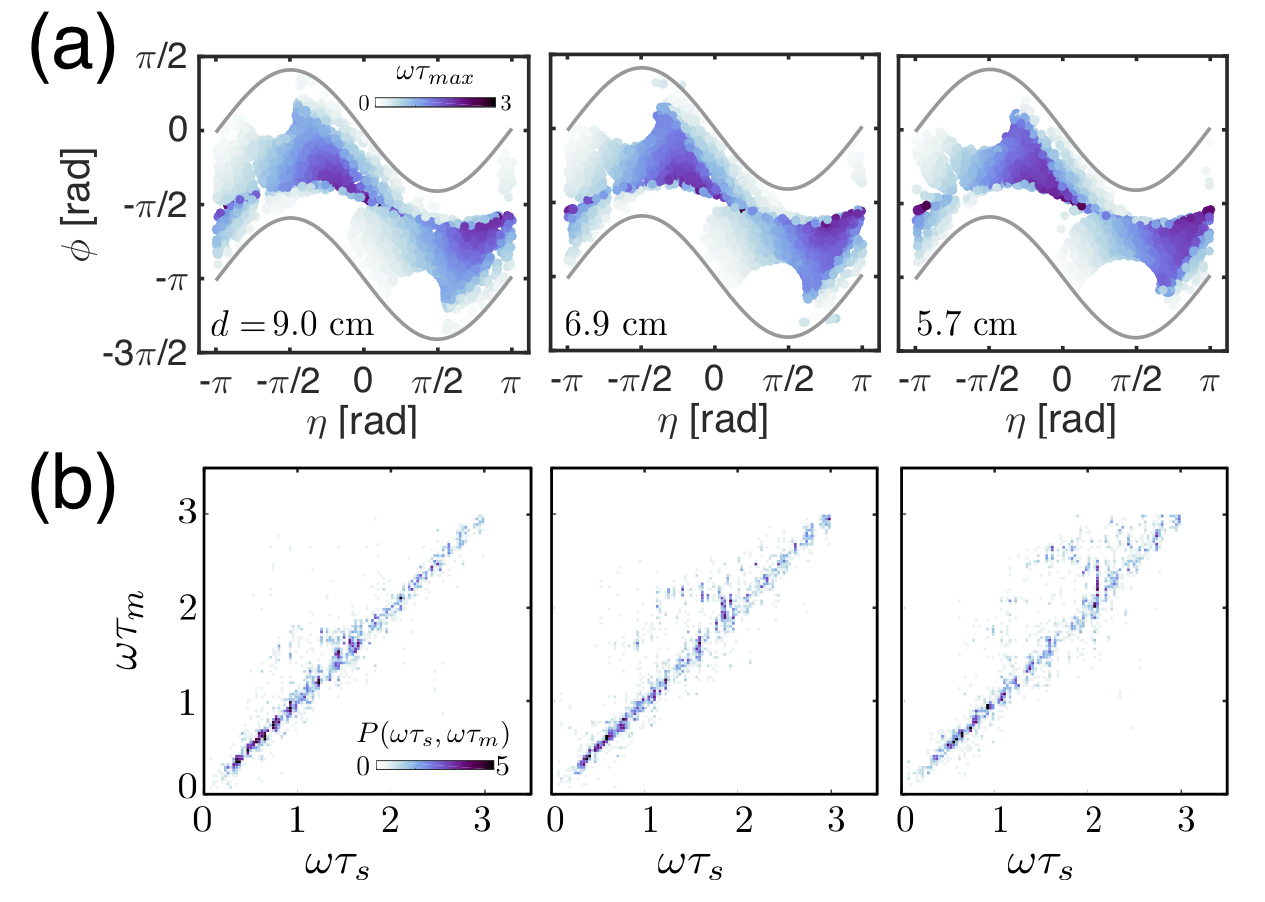}
	\caption{\textbf{Accessible collision states for different post spacing.} (a) Possible collision states in $(\eta,\phi)$-space, colored by contact duration. As $d$ decreases (left to right), fewer states are accessible.  Gray lines indicate boundaries of possible states in the single-post scenario. (b) Probability map of multi-post contact duration, $\omega \tau_m$ as a function of $\omega \tau_s$ the duration nearest single-post collision state in $(\eta,\phi)$-space.}
	\label{fig:remapping}
\end{figure}

Not only is the impact location altered by the presence of multiple posts, but it is clear from Fig.~\ref{fig:phiShifts}a that the phase of the undulation cycle upon impact is also changed. Scatter plots in Fig.~\ref{fig:remapping}a show how these collision states in $(\eta,\phi)$-space depend on spacing. As $d$ decreases, fewer states are accessible to the robot, and the states that become inaccessible are those away from the leading edge of the post. Aside from this restriction on allowed states, the dependence of $\tau_{max}$ on $\eta$ and $\phi$ is nearly the same.  This suggests that collision states are largely independent of $d$. 

To test the similarity of collision states for different post configurations, we compare the single post collision state closest to (i.e., smallest Euclidean distance in the $(\eta,\phi)$-space from) each multi-post state in $(\eta,\phi)$-space~\cite{supplemental}. If the states are indeed the same, we expect the contact durations associated with the single and multi-post state should be identical. Fig.~\ref{fig:remapping}c shows the probability maps of three multi-post durations as a function of the nearest single-post state. For all three $d$, the preponderance of the data falls along the $\omega \tau_m = \omega \tau_s$ line, confirming that adjacent posts act primarily to shift the probabilities of single-post collision states. As the spacing decreases, single-post states near the top of the post occur with greatly-reduced probability (and some are even eliminated completely) as trajectories are ``remapped" to a different single-post collision state occurring at an adjacent post. These shifted collisions tend to occur closer to the leading surface of the post than the original collision, often resulting in longer durations than the single-post state that was replaced. Given the linear relationship between duration and scattering angle, the remapping from shorter to longer durations shifts power from the central peak of the $\theta$-distributions out to the tails, creating and bolstering secondary off-center peaks. 

\begin{figure*}[ht!]
	\includegraphics[width=\textwidth]{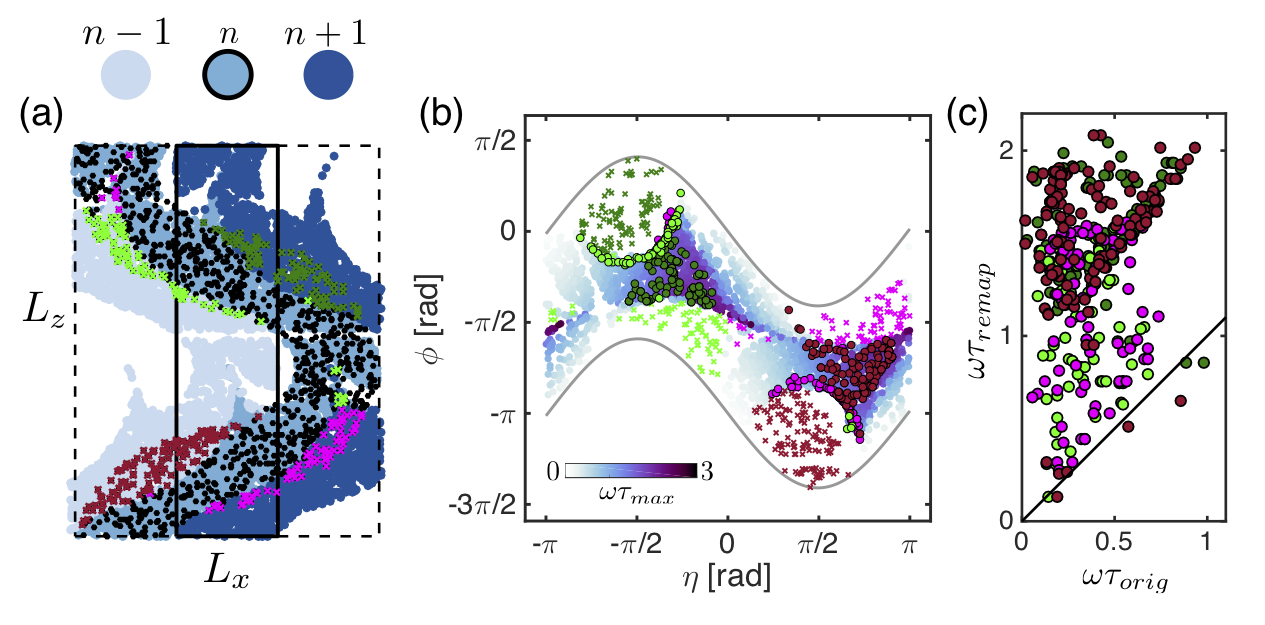}
	\caption{\textbf{Single-post states are shifted by the presence of multiple posts} (a) Tiled initial conditions for $d = 5.7$~cm are shown in blue, and single-post initial conditions are overlaid in black. All single-post points hit the central (and only) post present in that configuration, outlined in black. The color of the multi-post initial conditions indicates which post was hit. The `x' points indicate four regions which no longer hit the central post when multiple posts were present. Instead, they collided with the adjacent post. (b) Multi-post collision states in $(\eta,\phi)$-space, colored by duration of contact. The colored `x' regions here show the same four regions highlighted in (a), and the circles of the corresponding color show the collision state that occurs on the adjacent post. (c) In nearly all cases, the new collisions that occur on an adjacent post had significantly longer durations, $\tau_{remap}$, than the original single-post collision, $\tau_{orig}$. The line shows $\omega\tau_{remap} = \omega \tau_{orig}$.}
	\label{fig:remappedRegions}
\end{figure*}

To explore how single-post states are shifted by the presence of multiple posts, we identify the multi-post point closest to each single post point in $(x,z)$ space. To do this, we tiled the multi-post initial conditions box (e.g., for $d = 5.7$~cm, the solid box in Fig.~\ref{fig:remappedRegions}a) by shifting all points within this region over by $\pm m L_x$, where $m$ is an integer and $L_x$ is the transverse dimension of the initial conditions box. Outlines for shifts of $m = \pm 1$ are shown as the dashed boxes in Fig.~\ref{fig:remappedRegions}a. The points within each box show the starting point for the head of the robot, and the colors indicate which post was involved in the longest-duration collision with the head of the robot. When initial conditions were shifted, a different post was centered in front of the box, and given that all initial conditions boxes are identical, the post number associated with a collision in a box shifted by $m$ post must also be shifted by $m$.  

In Fig.~\ref{fig:remappedRegions}a, the multi-post points for $d = 5.7$~cm are shown in varying shades of blue, and the single-post points (all of which hit the central post, outlined in black) are overlaid in black. To identify how the single post points were shifted around in $(\eta,\phi)$-space, we determined the $xz$-distance between each single-post point and the nearest multi-post point, $\delta_{xz} = \sqrt{(x_s-x_m)^2+(z_s-z_m)^2}$, which was rarely larger than $0.5$~cm. The colored `x' markers in Fig.~\ref{fig:remappedRegions}a identify four regions which hit post $n$ in the single-post case but were involved in more significant collisions with adjacent posts in the multi-post case. How these regions were shifted around in $(\eta,\phi)$-space is shown in Fig.~\ref{fig:remappedRegions}. `x' points were shifted to the circular points of the same color. Fig.~\ref{fig:remappedRegions}c shows nearly all of the remapped points had significantly longer durations, $\tau_{remap}$, than the duration of the original state in the single-post case, $\tau_{orig}$.

These results confirm that single-post collision states are largely unaltered by the presence of multiple posts, even when $d$ is small. Instead, multiple posts serve to restrict the collision states accessible to the robot. As $d$ decreases, low-duration states occurring near the top of the posts become inaccessible and are replaced by longer-duration collisions near the leading edge of an adjacent peg. Stated another way, scattering events with small reorientations are preferentially remapped to larger-angle scattering events.

\section{Conclusions}

The results presented here provide a striking example of the dynamics that can arise in self-propelled systems when environmental heterogeneities are present. To explore the nature of the interactions that can occur during undulatory self-propulsion, we created a robophysical snake-like robot which self-deforms by propagating a wave of joint-angle variations from head to tail. Passive wheels affixed to the bottom of the robot enable the robot to translate by creating a highly-dissipative coupling between these self-deformations to the surrounding environment. We find that the nature of this dissipation is similar to that of both viscous fluids (relevant for swimmers in low Reynolds number fluids) and granular materials (relevant for movement on and within sand). This suggests that our results may be relevant to systems spanning a broad range of length scales and environments.

Interactions with a single obstacle (a rigid vertical post) scatter the robot, and, unlike momentum-conserving collisions in non-active systems, the distribution of scattering angles produced by interactions with a single post is strongly-peaked directly behind the post.  When multiple posts are present, secondary peaks emerge in the distribution due to an increase in the number of large scattering events, especially as post density increases.  
Surprisingly, we find that the collisions are not altered by the presence of multiple posts; instead, the likelihood of collisions shifts so that there are more interactions which produce large-scattering events.  In all cases, the resulting scattering angle is linearly proportional to the duration of the collision. A simple model reveals that this collision duration is qualitatively equivalent to the reorientation times discussed in many other active matter systems (see e.g.,~\cite{bechinger2016active}) and sets the outcome of the interaction. This understanding provides a starting point for manipulating either locomotor behavior or the surrounding environment to produce a desired outcome.

Simulations allowed for broader parameter variation and revealed that, like movement through similarly highly-dissipative environments, our results are independent of the frequency of undulation.  That is, the linear relationship between the head-post contact duration,  $\omega \tau_{max}$, and the resulting scattering angle remains the same for a broad range of frequencies.  Variation of the angular amplitude, $\zeta_{max}$, also did not significantly alter the linear dependence between the duration and the scattering angle, suggesting that our results are valid for a range of waveforms and undulation frequencies. More broadly, there are other periodically-driven macroscopic systems which produce similar scattering patterns, including the bouncing fluid droplets~\cite{Couder:2006ix,Bush:2015ev} as well as biological snakes~\cite{schiebel2018}, and it would be interesting to explore the potential connections between these systems as well as to test the extent to which the observed behavior may be a robust and generic feature of periodically-driven active systems.

There are several other parameters in our system, relevant to both biological and robotic locomotion, that would be interesting to explore in future studies.  For instance, the number of waves on the body of  an undulatory locomotor as well as the overall length can vary across individuals. While the head-obstacle collision may still dominate, we suspect that these details as well as the details of the dissipation may be important for determining the outcome.  Additionally, obstacle size and shape would be interesting to explore.  Given our model-predicted tangency-condition for breaking free from an obstacle, we expect that shape should affect the resulting distributions, and, at some point, obstacle size should also matter (for a very large post, for instance, there may be very few small deflections).  Finally, it would also be interesting to explore how our results change for more complex arrangement of obstacles as well  for deformable and/or moveable obstacles.  We would expect that scattering patterns should change (likely with fewer large scattering events) as obstacles become less rigid.

We close by noting that robophysics provides a useful approach for exploring the nature of active collisions across scales and environments because it enables controlled experiments and systematic parameter variation while avoiding the complexities and unknowns of numerical collision-modeling, and the variability and controllability difficulties found in living systems. Robophysics is widely-applicable and amenable to other modes of locomotion, body morphologies, and obstacle configurations and geometries. With an understanding of active collisions, these interactions could be used to mitigate or even utilize interactions with heterogeneities for different classes and environments for natural and artifical locomotors, e.g. in legged~\cite{li2013, Li:2015jn}, undulatory~\cite{jayne1986kinematics,majmudar2012experiments}, sidewinding~\cite{marvi2014sidewinding}, wheeled and tracked vehicles~\cite{iagnemma2003experimental,wong2009terramechanics} and even aerial systems~\cite{floreano2010flying,turpin2012trajectory}. Alternatively, environments could be designed to direct the motion of self-propelled systems, for instance, to correct for (e.g.,~\cite{Kantsler:2013ge}) or selectively enhance scattering effects. Finally, structured environments could also be used to modify the duration of these interactions, which, given the importance of the interaction duration on the dynamics of active systems, could have broad implications for collective behavior in biological and artificial systems.

\begin{acknowledgments}
The authors thank Gordon Berman, Kurt Wiesenfeld, Zeb Rocklin, Mike Chapman, Cristina Marchetti, Paul Umbanhowar, John Bush, Yves Couder, Paul Goldbart, and Andy Ruina, for insightful discussions; Kelimar Diaz, Nathan Hines, and Alex Hubbard  for help with data collection. This work was supported by the National Science Foundation (NSF) Physics of Living Systems (PoLS); the Army Research Lab Micro Autonomous Systems and Technology Collaborative Technology Alliance (ARL MAST CTA); Army Research Office (ARO);  National Defense Science and Engineering Graduate (NDSEG) Fellowship; and Defense Advanced Research Projects Agency (DARPA) Young Faculty Award (YFA). The authors declare no conflicts of interest. Data is available from the corresponding author upon request.
\end{acknowledgments}


\begin{thebibliography}{60}%
	\makeatletter
	\providecommand \@ifxundefined [1]{%
		\@ifx{#1\undefined}
	}%
	\providecommand \@ifnum [1]{%
		\ifnum #1\expandafter \@firstoftwo
		\else \expandafter \@secondoftwo
		\fi
	}%
	\providecommand \@ifx [1]{%
		\ifx #1\expandafter \@firstoftwo
		\else \expandafter \@secondoftwo
		\fi
	}%
	\providecommand \natexlab [1]{#1}%
	\providecommand \enquote  [1]{``#1''}%
	\providecommand \bibnamefont  [1]{#1}%
	\providecommand \bibfnamefont [1]{#1}%
	\providecommand \citenamefont [1]{#1}%
	\providecommand \href@noop [0]{\@secondoftwo}%
	\providecommand \href [0]{\begingroup \@sanitize@url \@href}%
	\providecommand \@href[1]{\@@startlink{#1}\@@href}%
	\providecommand \@@href[1]{\endgroup#1\@@endlink}%
	\providecommand \@sanitize@url [0]{\catcode `\\12\catcode `\$12\catcode
		`\&12\catcode `\#12\catcode `\^12\catcode `\_12\catcode `\%12\relax}%
	\providecommand \@@startlink[1]{}%
	\providecommand \@@endlink[0]{}%
	\providecommand \url  [0]{\begingroup\@sanitize@url \@url }%
	\providecommand \@url [1]{\endgroup\@href {#1}{\urlprefix }}%
	\providecommand \urlprefix  [0]{URL }%
	\providecommand \Eprint [0]{\href }%
	\providecommand \doibase [0]{http://dx.doi.org/}%
	\providecommand \selectlanguage [0]{\@gobble}%
	\providecommand \bibinfo  [0]{\@secondoftwo}%
	\providecommand \bibfield  [0]{\@secondoftwo}%
	\providecommand \translation [1]{[#1]}%
	\providecommand \BibitemOpen [0]{}%
	\providecommand \bibitemStop [0]{}%
	\providecommand \bibitemNoStop [0]{.\EOS\space}%
	\providecommand \EOS [0]{\spacefactor3000\relax}%
	\providecommand \BibitemShut  [1]{\csname bibitem#1\endcsname}%
	\let\auto@bib@innerbib\@empty
	\bibitem [{\citenamefont {Holmes}\ \emph {et~al.}(2006)\citenamefont {Holmes},
		\citenamefont {Full}, \citenamefont {Koditschek},\ and\ \citenamefont
		{Guckenheimer}}]{Holmes:2006ku}%
	\BibitemOpen
	\bibfield  {author} {\bibinfo {author} {\bibfnamefont {P.}~\bibnamefont
			{Holmes}}, \bibinfo {author} {\bibfnamefont {R.~J.}\ \bibnamefont {Full}},
		\bibinfo {author} {\bibfnamefont {D.}~\bibnamefont {Koditschek}}, \ and\
		\bibinfo {author} {\bibfnamefont {J.}~\bibnamefont {Guckenheimer}},\
	}\href@noop {} {\bibfield  {journal} {\bibinfo  {journal} {SIAM Review}\
		}\textbf {\bibinfo {volume} {48}},\ \bibinfo {pages} {207} (\bibinfo {year}
		{2006})}\BibitemShut {NoStop}%
	\bibitem [{\citenamefont {Hosoi}\ and\ \citenamefont
		{Goldman}(2015)}]{Hosoi:2015gy}%
	\BibitemOpen
	\bibfield  {author} {\bibinfo {author} {\bibfnamefont {A.~E.}\ \bibnamefont
			{Hosoi}}\ and\ \bibinfo {author} {\bibfnamefont {D.~I.}\ \bibnamefont
			{Goldman}},\ }\href@noop {} {\bibfield  {journal} {\bibinfo  {journal}
			{Annual Review of Fluid Mechanics}\ }\textbf {\bibinfo {volume} {47}},\
		\bibinfo {pages} {431} (\bibinfo {year} {2015})}\BibitemShut {NoStop}%
	\bibitem [{\citenamefont {Childress}\ \emph {et~al.}(2012)\citenamefont
		{Childress}, \citenamefont {Hosoi}, \citenamefont {Schultz},\ and\
		\citenamefont {Wang}}]{childress2012natural}%
	\BibitemOpen
	\bibfield  {author} {\bibinfo {author} {\bibfnamefont {S.}~\bibnamefont
			{Childress}}, \bibinfo {author} {\bibfnamefont {A.}~\bibnamefont {Hosoi}},
		\bibinfo {author} {\bibfnamefont {W.~W.}\ \bibnamefont {Schultz}}, \ and\
		\bibinfo {author} {\bibfnamefont {J.}~\bibnamefont {Wang}},\ }\href@noop {}
	{\emph {\bibinfo {title} {Natural locomotion in fluids and on surfaces:
				swimming, flying, and sliding}}},\ Vol.\ \bibinfo {volume} {155}\ (\bibinfo
	{publisher} {Springer},\ \bibinfo {year} {2012})\BibitemShut {NoStop}%
	\bibitem [{\citenamefont {Marchetti}\ \emph {et~al.}(2013)\citenamefont
		{Marchetti}, \citenamefont {Joanny}, \citenamefont {Ramaswamy}, \citenamefont
		{Liverpool}, \citenamefont {Prost}, \citenamefont {Rao},\ and\ \citenamefont
		{Simha}}]{marchetti2013}%
	\BibitemOpen
	\bibfield  {author} {\bibinfo {author} {\bibfnamefont {M.~C.}\ \bibnamefont
			{Marchetti}}, \bibinfo {author} {\bibfnamefont {J.~F.}\ \bibnamefont
			{Joanny}}, \bibinfo {author} {\bibfnamefont {S.}~\bibnamefont {Ramaswamy}},
		\bibinfo {author} {\bibfnamefont {T.~B.}\ \bibnamefont {Liverpool}}, \bibinfo
		{author} {\bibfnamefont {J.}~\bibnamefont {Prost}}, \bibinfo {author}
		{\bibfnamefont {M.}~\bibnamefont {Rao}}, \ and\ \bibinfo {author}
		{\bibfnamefont {R.~A.}\ \bibnamefont {Simha}},\ }\href@noop {} {\bibfield
		{journal} {\bibinfo  {journal} {Reviews of Modern Physics}\ }\textbf
		{\bibinfo {volume} {85}},\ \bibinfo {pages} {1143} (\bibinfo {year}
		{2013})}\BibitemShut {NoStop}%
	\bibitem [{\citenamefont {Drescher}\ \emph {et~al.}(2011)\citenamefont
		{Drescher}, \citenamefont {Dunkel}, \citenamefont {Cisneros}, \citenamefont
		{Ganguly},\ and\ \citenamefont {Goldstein}}]{drescher2011fluid}%
	\BibitemOpen
	\bibfield  {author} {\bibinfo {author} {\bibfnamefont {K.}~\bibnamefont
			{Drescher}}, \bibinfo {author} {\bibfnamefont {J.}~\bibnamefont {Dunkel}},
		\bibinfo {author} {\bibfnamefont {L.~H.}\ \bibnamefont {Cisneros}}, \bibinfo
		{author} {\bibfnamefont {S.}~\bibnamefont {Ganguly}}, \ and\ \bibinfo
		{author} {\bibfnamefont {R.~E.}\ \bibnamefont {Goldstein}},\ }\href@noop {}
	{\bibfield  {journal} {\bibinfo  {journal} {Proceedings of the National
				Academy of Sciences}\ }\textbf {\bibinfo {volume} {108}},\ \bibinfo {pages}
		{10940} (\bibinfo {year} {2011})}\BibitemShut {NoStop}%
	\bibitem [{\citenamefont {Palacci}\ \emph {et~al.}(2013)\citenamefont
		{Palacci}, \citenamefont {Sacanna}, \citenamefont {Steinberg}, \citenamefont
		{Pine},\ and\ \citenamefont {Chaikin}}]{palacci2013living}%
	\BibitemOpen
	\bibfield  {author} {\bibinfo {author} {\bibfnamefont {J.}~\bibnamefont
			{Palacci}}, \bibinfo {author} {\bibfnamefont {S.}~\bibnamefont {Sacanna}},
		\bibinfo {author} {\bibfnamefont {A.~P.}\ \bibnamefont {Steinberg}}, \bibinfo
		{author} {\bibfnamefont {D.~J.}\ \bibnamefont {Pine}}, \ and\ \bibinfo
		{author} {\bibfnamefont {P.~M.}\ \bibnamefont {Chaikin}},\ }\href@noop {}
	{\bibfield  {journal} {\bibinfo  {journal} {Science}\ ,\ \bibinfo {pages}
			{1230020}} (\bibinfo {year} {2013})}\BibitemShut {NoStop}%
	\bibitem [{\citenamefont {Kantsler}\ \emph {et~al.}(2013)\citenamefont
		{Kantsler}, \citenamefont {Dunkel},\ and\ \citenamefont
		{Polin}}]{Kantsler:2013ge}%
	\BibitemOpen
	\bibfield  {author} {\bibinfo {author} {\bibfnamefont {V.}~\bibnamefont
			{Kantsler}}, \bibinfo {author} {\bibfnamefont {J.}~\bibnamefont {Dunkel}}, \
		and\ \bibinfo {author} {\bibfnamefont {M.}~\bibnamefont {Polin}},\ }in\
	\href@noop {} {\emph {\bibinfo {booktitle} {Proceedings of the National
				Academy of Sciences}}}\ (\bibinfo {year} {2013})\ pp.\ \bibinfo {pages}
	{1--9}\BibitemShut {NoStop}%
	\bibitem [{\citenamefont {Frymier}\ \emph {et~al.}(1995)\citenamefont
		{Frymier}, \citenamefont {Ford}, \citenamefont {Berg},\ and\ \citenamefont
		{Cummings}}]{frymier1995three}%
	\BibitemOpen
	\bibfield  {author} {\bibinfo {author} {\bibfnamefont {P.~D.}\ \bibnamefont
			{Frymier}}, \bibinfo {author} {\bibfnamefont {R.~M.}\ \bibnamefont {Ford}},
		\bibinfo {author} {\bibfnamefont {H.~C.}\ \bibnamefont {Berg}}, \ and\
		\bibinfo {author} {\bibfnamefont {P.~T.}\ \bibnamefont {Cummings}},\
	}\href@noop {} {\bibfield  {journal} {\bibinfo  {journal} {Proceedings of the
				National Academy of Sciences}\ }\textbf {\bibinfo {volume} {92}},\ \bibinfo
		{pages} {6195} (\bibinfo {year} {1995})}\BibitemShut {NoStop}%
	\bibitem [{\citenamefont {Takagi}\ \emph {et~al.}(2014)\citenamefont {Takagi},
		\citenamefont {Palacci}, \citenamefont {Braunschweig}, \citenamefont
		{Shelley},\ and\ \citenamefont {Zhang}}]{takagi2014hydrodynamic}%
	\BibitemOpen
	\bibfield  {author} {\bibinfo {author} {\bibfnamefont {D.}~\bibnamefont
			{Takagi}}, \bibinfo {author} {\bibfnamefont {J.}~\bibnamefont {Palacci}},
		\bibinfo {author} {\bibfnamefont {A.~B.}\ \bibnamefont {Braunschweig}},
		\bibinfo {author} {\bibfnamefont {M.~J.}\ \bibnamefont {Shelley}}, \ and\
		\bibinfo {author} {\bibfnamefont {J.}~\bibnamefont {Zhang}},\ }\href@noop {}
	{\bibfield  {journal} {\bibinfo  {journal} {Soft Matter}\ }\textbf {\bibinfo
			{volume} {10}},\ \bibinfo {pages} {1784} (\bibinfo {year}
		{2014})}\BibitemShut {NoStop}%
	\bibitem [{\citenamefont {Reichhardt}\ and\ \citenamefont
		{Reichhardt}(2014)}]{reichhardt2014active}%
	\BibitemOpen
	\bibfield  {author} {\bibinfo {author} {\bibfnamefont {C.}~\bibnamefont
			{Reichhardt}}\ and\ \bibinfo {author} {\bibfnamefont {C.~O.}\ \bibnamefont
			{Reichhardt}},\ }\href@noop {} {\bibfield  {journal} {\bibinfo  {journal}
			{Physical Review E}\ }\textbf {\bibinfo {volume} {90}},\ \bibinfo {pages}
		{012701} (\bibinfo {year} {2014})}\BibitemShut {NoStop}%
	\bibitem [{\citenamefont {Li}\ \emph {et~al.}(2015)\citenamefont {Li},
		\citenamefont {Pullin}, \citenamefont {Haldane}, \citenamefont {Lam},
		\citenamefont {Fearing},\ and\ \citenamefont {Full}}]{Li:2015jn}%
	\BibitemOpen
	\bibfield  {author} {\bibinfo {author} {\bibfnamefont {C.}~\bibnamefont
			{Li}}, \bibinfo {author} {\bibfnamefont {A.~O.}\ \bibnamefont {Pullin}},
		\bibinfo {author} {\bibfnamefont {D.~W.}\ \bibnamefont {Haldane}}, \bibinfo
		{author} {\bibfnamefont {H.~K.}\ \bibnamefont {Lam}}, \bibinfo {author}
		{\bibfnamefont {R.~S.}\ \bibnamefont {Fearing}}, \ and\ \bibinfo {author}
		{\bibfnamefont {R.~J.}\ \bibnamefont {Full}},\ }\href@noop {} {\bibfield
		{journal} {\bibinfo  {journal} {Bioinspiration {\&} Biomimetics}\ }\textbf
		{\bibinfo {volume} {10}},\ \bibinfo {pages} {1} (\bibinfo {year}
		{2015})}\BibitemShut {NoStop}%
	\bibitem [{\citenamefont {Spagna}\ \emph {et~al.}(2007)\citenamefont {Spagna},
		\citenamefont {Goldman}, \citenamefont {Lin}, \citenamefont {Koditschek},\
		and\ \citenamefont {Full}}]{Spagna:2007kf}%
	\BibitemOpen
	\bibfield  {author} {\bibinfo {author} {\bibfnamefont {J.~C.}\ \bibnamefont
			{Spagna}}, \bibinfo {author} {\bibfnamefont {D.~I.}\ \bibnamefont {Goldman}},
		\bibinfo {author} {\bibfnamefont {P.-C.}\ \bibnamefont {Lin}}, \bibinfo
		{author} {\bibfnamefont {D.~E.}\ \bibnamefont {Koditschek}}, \ and\ \bibinfo
		{author} {\bibfnamefont {R.~J.}\ \bibnamefont {Full}},\ }\href@noop {}
	{\bibfield  {journal} {\bibinfo  {journal} {Bioinspiration {\&} Biomimetics}\
		}\textbf {\bibinfo {volume} {2}},\ \bibinfo {pages} {9} (\bibinfo {year}
		{2007})}\BibitemShut {NoStop}%
	\bibitem [{\citenamefont {Kelley}\ \emph {et~al.}(1997)\citenamefont {Kelley},
		\citenamefont {Arnold},\ and\ \citenamefont {Gladstone}}]{kelley1997effects}%
	\BibitemOpen
	\bibfield  {author} {\bibinfo {author} {\bibfnamefont {K.}~\bibnamefont
			{Kelley}}, \bibinfo {author} {\bibfnamefont {S.}~\bibnamefont {Arnold}}, \
		and\ \bibinfo {author} {\bibfnamefont {J.}~\bibnamefont {Gladstone}},\
	}\href@noop {} {\bibfield  {journal} {\bibinfo  {journal} {Functional
				Ecology}\ }\textbf {\bibinfo {volume} {11}},\ \bibinfo {pages} {189}
		(\bibinfo {year} {1997})}\BibitemShut {NoStop}%
	\bibitem [{\citenamefont {Park}\ \emph {et~al.}(2008)\citenamefont {Park},
		\citenamefont {Hwang}, \citenamefont {Nam}, \citenamefont {Martinez},
		\citenamefont {Austin},\ and\ \citenamefont {Ryu}}]{park2008enhanced}%
	\BibitemOpen
	\bibfield  {author} {\bibinfo {author} {\bibfnamefont {S.}~\bibnamefont
			{Park}}, \bibinfo {author} {\bibfnamefont {H.}~\bibnamefont {Hwang}},
		\bibinfo {author} {\bibfnamefont {S.-W.}\ \bibnamefont {Nam}}, \bibinfo
		{author} {\bibfnamefont {F.}~\bibnamefont {Martinez}}, \bibinfo {author}
		{\bibfnamefont {R.~H.}\ \bibnamefont {Austin}}, \ and\ \bibinfo {author}
		{\bibfnamefont {W.~S.}\ \bibnamefont {Ryu}},\ }\href@noop {} {\bibfield
		{journal} {\bibinfo  {journal} {PloS one}\ }\textbf {\bibinfo {volume} {3}},\
		\bibinfo {pages} {e2550} (\bibinfo {year} {2008})}\BibitemShut {NoStop}%
	\bibitem [{\citenamefont {McGeer}(1990)}]{McGeer:1990uk}%
	\BibitemOpen
	\bibfield  {author} {\bibinfo {author} {\bibfnamefont {T.}~\bibnamefont
			{McGeer}},\ }\href@noop {} {\bibfield  {journal} {\bibinfo  {journal}
			{International Journal of Robotics Research}\ }\textbf {\bibinfo {volume}
			{9}},\ \bibinfo {pages} {62} (\bibinfo {year} {1990})}\BibitemShut {NoStop}%
	\bibitem [{\citenamefont {Coleman}\ and\ \citenamefont
		{Ruina}(1998)}]{Coleman:1998uk}%
	\BibitemOpen
	\bibfield  {author} {\bibinfo {author} {\bibfnamefont {M.~J.}\ \bibnamefont
			{Coleman}}\ and\ \bibinfo {author} {\bibfnamefont {A.}~\bibnamefont
			{Ruina}},\ }\href@noop {} {\bibfield  {journal} {\bibinfo  {journal}
			{Physical Review Letters}\ }\textbf {\bibinfo {volume} {80}},\ \bibinfo
		{pages} {3658} (\bibinfo {year} {1998})}\BibitemShut {NoStop}%
	\bibitem [{\citenamefont {Saranli}\ \emph {et~al.}(2001)\citenamefont
		{Saranli}, \citenamefont {Buehler},\ and\ \citenamefont
		{Koditschek}}]{Saranli:2001vk}%
	\BibitemOpen
	\bibfield  {author} {\bibinfo {author} {\bibfnamefont {U.}~\bibnamefont
			{Saranli}}, \bibinfo {author} {\bibfnamefont {M.}~\bibnamefont {Buehler}}, \
		and\ \bibinfo {author} {\bibfnamefont {D.~E.}\ \bibnamefont {Koditschek}},\
	}\href@noop {} {\bibfield  {journal} {\bibinfo  {journal} {International
				Journal of Robotics Research}\ }\textbf {\bibinfo {volume} {20}},\ \bibinfo
		{pages} {616} (\bibinfo {year} {2001})}\BibitemShut {NoStop}%
	\bibitem [{\citenamefont {McInroe}\ \emph {et~al.}(2016)\citenamefont
		{McInroe}, \citenamefont {Astley}, \citenamefont {Gong}, \citenamefont
		{Kawano}, \citenamefont {Schiebel}, \citenamefont {Rieser}, \citenamefont
		{Choset}, \citenamefont {Blob},\ and\ \citenamefont
		{Goldman}}]{mcinroe2016tail}%
	\BibitemOpen
	\bibfield  {author} {\bibinfo {author} {\bibfnamefont {B.}~\bibnamefont
			{McInroe}}, \bibinfo {author} {\bibfnamefont {H.~C.}\ \bibnamefont {Astley}},
		\bibinfo {author} {\bibfnamefont {C.}~\bibnamefont {Gong}}, \bibinfo {author}
		{\bibfnamefont {S.~M.}\ \bibnamefont {Kawano}}, \bibinfo {author}
		{\bibfnamefont {P.~E.}\ \bibnamefont {Schiebel}}, \bibinfo {author}
		{\bibfnamefont {J.~M.}\ \bibnamefont {Rieser}}, \bibinfo {author}
		{\bibfnamefont {H.}~\bibnamefont {Choset}}, \bibinfo {author} {\bibfnamefont
			{R.~W.}\ \bibnamefont {Blob}}, \ and\ \bibinfo {author} {\bibfnamefont
			{D.~I.}\ \bibnamefont {Goldman}},\ }\href@noop {} {\bibfield  {journal}
		{\bibinfo  {journal} {Science}\ }\textbf {\bibinfo {volume} {353}},\ \bibinfo
		{pages} {154} (\bibinfo {year} {2016})}\BibitemShut {NoStop}%
	\bibitem [{\citenamefont {Qian}\ and\ \citenamefont
		{Goldman}(2015{\natexlab{a}})}]{Qian:2015gz}%
	\BibitemOpen
	\bibfield  {author} {\bibinfo {author} {\bibfnamefont {F.}~\bibnamefont
			{Qian}}\ and\ \bibinfo {author} {\bibfnamefont {D.~I.}\ \bibnamefont
			{Goldman}},\ }in\ \href@noop {} {\emph {\bibinfo {booktitle} {SPIE Defense +
				Security}}},\ \bibinfo {editor} {edited by\ \bibinfo {editor} {\bibfnamefont
			{T.}~\bibnamefont {George}}, \bibinfo {editor} {\bibfnamefont {A.~K.}\
			\bibnamefont {Dutta}}, \ and\ \bibinfo {editor} {\bibfnamefont {M.~S.}\
			\bibnamefont {Islam}}}\ (\bibinfo  {publisher} {SPIE},\ \bibinfo {year}
	{2015})\ p.\ \bibinfo {pages} {94671U}\BibitemShut {NoStop}%
	\bibitem [{\citenamefont {Bechinger}\ \emph {et~al.}(2016)\citenamefont
		{Bechinger}, \citenamefont {Di~Leonardo}, \citenamefont {L{\"o}wen},
		\citenamefont {Reichhardt}, \citenamefont {Volpe},\ and\ \citenamefont
		{Volpe}}]{bechinger2016active}%
	\BibitemOpen
	\bibfield  {author} {\bibinfo {author} {\bibfnamefont {C.}~\bibnamefont
			{Bechinger}}, \bibinfo {author} {\bibfnamefont {R.}~\bibnamefont
			{Di~Leonardo}}, \bibinfo {author} {\bibfnamefont {H.}~\bibnamefont
			{L{\"o}wen}}, \bibinfo {author} {\bibfnamefont {C.}~\bibnamefont
			{Reichhardt}}, \bibinfo {author} {\bibfnamefont {G.}~\bibnamefont {Volpe}}, \
		and\ \bibinfo {author} {\bibfnamefont {G.}~\bibnamefont {Volpe}},\
	}\href@noop {} {\bibfield  {journal} {\bibinfo  {journal} {Reviews of Modern
				Physics}\ }\textbf {\bibinfo {volume} {88}},\ \bibinfo {pages} {045006}
		(\bibinfo {year} {2016})}\BibitemShut {NoStop}%
	\bibitem [{\citenamefont {Thoms}\ \emph {et~al.}(2017)\citenamefont {Thoms},
		\citenamefont {Yu}, \citenamefont {Kang},\ and\ \citenamefont
		{Li}}]{Li2017mm}%
	\BibitemOpen
	\bibfield  {author} {\bibinfo {author} {\bibfnamefont {G.}~\bibnamefont
			{Thoms}}, \bibinfo {author} {\bibfnamefont {S.}~\bibnamefont {Yu}}, \bibinfo
		{author} {\bibfnamefont {Y.}~\bibnamefont {Kang}}, \ and\ \bibinfo {author}
		{\bibfnamefont {C.}~\bibnamefont {Li}},\ }\href@noop {} {\enquote {\bibinfo
			{title} {Induced vibrations facilitate traversal of cluttered obstacles},}\
	}\bibinfo {howpublished} {APS March Meeting} (\bibinfo {year} {2017}),\
	\bibinfo {note}
	{http://meetings.aps.org/Meeting/MAR17/Session/Y12.9}\BibitemShut {NoStop}%
	\bibitem [{\citenamefont {Gray}(1953)}]{gray1953undulatory}%
	\BibitemOpen
	\bibfield  {author} {\bibinfo {author} {\bibfnamefont {J.}~\bibnamefont
			{Gray}},\ }\href@noop {} {\bibfield  {journal} {\bibinfo  {journal} {Journal
				of Cell Science}\ }\textbf {\bibinfo {volume} {3}},\ \bibinfo {pages} {551}
		(\bibinfo {year} {1953})}\BibitemShut {NoStop}%
	\bibitem [{\citenamefont {Gray}\ and\ \citenamefont
		{Hancock}(1955)}]{gray1955propulsion}%
	\BibitemOpen
	\bibfield  {author} {\bibinfo {author} {\bibfnamefont {J.}~\bibnamefont
			{Gray}}\ and\ \bibinfo {author} {\bibfnamefont {G.}~\bibnamefont {Hancock}},\
	}\href@noop {} {\bibfield  {journal} {\bibinfo  {journal} {Journal of
				Experimental Biology}\ }\textbf {\bibinfo {volume} {32}},\ \bibinfo {pages}
		{802} (\bibinfo {year} {1955})}\BibitemShut {NoStop}%
	\bibitem [{\citenamefont {Gray}\ and\ \citenamefont
		{Lissmann}(1964)}]{gray1964locomotion}%
	\BibitemOpen
	\bibfield  {author} {\bibinfo {author} {\bibfnamefont {J.}~\bibnamefont
			{Gray}}\ and\ \bibinfo {author} {\bibfnamefont {H.~W.}\ \bibnamefont
			{Lissmann}},\ }\href@noop {} {\bibfield  {journal} {\bibinfo  {journal}
			{Journal of Experimental Biology}\ }\textbf {\bibinfo {volume} {41}},\
		\bibinfo {pages} {135} (\bibinfo {year} {1964})}\BibitemShut {NoStop}%
	\bibitem [{\citenamefont {Sfakiotakis}\ \emph {et~al.}(1999)\citenamefont
		{Sfakiotakis}, \citenamefont {Lane},\ and\ \citenamefont
		{Davies}}]{sfakiotakis1999review}%
	\BibitemOpen
	\bibfield  {author} {\bibinfo {author} {\bibfnamefont {M.}~\bibnamefont
			{Sfakiotakis}}, \bibinfo {author} {\bibfnamefont {D.~M.}\ \bibnamefont
			{Lane}}, \ and\ \bibinfo {author} {\bibfnamefont {J.~B.~C.}\ \bibnamefont
			{Davies}},\ }\href@noop {} {\bibfield  {journal} {\bibinfo  {journal} {IEEE
				Journal of oceanic engineering}\ }\textbf {\bibinfo {volume} {24}},\ \bibinfo
		{pages} {237} (\bibinfo {year} {1999})}\BibitemShut {NoStop}%
	\bibitem [{\citenamefont {Gillis}(1996)}]{gillis1996undulatory}%
	\BibitemOpen
	\bibfield  {author} {\bibinfo {author} {\bibfnamefont {G.~B.}\ \bibnamefont
			{Gillis}},\ }\href@noop {} {\bibfield  {journal} {\bibinfo  {journal}
			{American Zoologist}\ }\textbf {\bibinfo {volume} {36}},\ \bibinfo {pages}
		{656} (\bibinfo {year} {1996})}\BibitemShut {NoStop}%
	\bibitem [{\citenamefont {Juarez}\ \emph {et~al.}(2010)\citenamefont {Juarez},
		\citenamefont {Lu}, \citenamefont {Sznitman},\ and\ \citenamefont
		{Arratia}}]{juarez2010motility}%
	\BibitemOpen
	\bibfield  {author} {\bibinfo {author} {\bibfnamefont {G.}~\bibnamefont
			{Juarez}}, \bibinfo {author} {\bibfnamefont {K.}~\bibnamefont {Lu}}, \bibinfo
		{author} {\bibfnamefont {J.}~\bibnamefont {Sznitman}}, \ and\ \bibinfo
		{author} {\bibfnamefont {P.~E.}\ \bibnamefont {Arratia}},\ }\href@noop {}
	{\bibfield  {journal} {\bibinfo  {journal} {EPL (Europhysics Letters)}\
		}\textbf {\bibinfo {volume} {92}},\ \bibinfo {pages} {44002} (\bibinfo {year}
		{2010})}\BibitemShut {NoStop}%
	\bibitem [{\citenamefont {Maladen}\ \emph {et~al.}(2009)\citenamefont
		{Maladen}, \citenamefont {Ding}, \citenamefont {Li},\ and\ \citenamefont
		{Goldman}}]{maladen2009undulatory}%
	\BibitemOpen
	\bibfield  {author} {\bibinfo {author} {\bibfnamefont {R.~D.}\ \bibnamefont
			{Maladen}}, \bibinfo {author} {\bibfnamefont {Y.}~\bibnamefont {Ding}},
		\bibinfo {author} {\bibfnamefont {C.}~\bibnamefont {Li}}, \ and\ \bibinfo
		{author} {\bibfnamefont {D.~I.}\ \bibnamefont {Goldman}},\ }\href@noop {}
	{\bibfield  {journal} {\bibinfo  {journal} {science}\ }\textbf {\bibinfo
			{volume} {325}},\ \bibinfo {pages} {314} (\bibinfo {year}
		{2009})}\BibitemShut {NoStop}%
	\bibitem [{\citenamefont {Sharpe}\ \emph {et~al.}(2015)\citenamefont {Sharpe},
		\citenamefont {Koehler}, \citenamefont {Kuckuk}, \citenamefont {Serrano},
		\citenamefont {Vela}, \citenamefont {Mendelson},\ and\ \citenamefont
		{Goldman}}]{sharpe2015locomotor}%
	\BibitemOpen
	\bibfield  {author} {\bibinfo {author} {\bibfnamefont {S.~S.}\ \bibnamefont
			{Sharpe}}, \bibinfo {author} {\bibfnamefont {S.~A.}\ \bibnamefont {Koehler}},
		\bibinfo {author} {\bibfnamefont {R.~M.}\ \bibnamefont {Kuckuk}}, \bibinfo
		{author} {\bibfnamefont {M.}~\bibnamefont {Serrano}}, \bibinfo {author}
		{\bibfnamefont {P.~A.}\ \bibnamefont {Vela}}, \bibinfo {author}
		{\bibfnamefont {J.}~\bibnamefont {Mendelson}}, \ and\ \bibinfo {author}
		{\bibfnamefont {D.~I.}\ \bibnamefont {Goldman}},\ }\href@noop {} {\bibfield
		{journal} {\bibinfo  {journal} {Journal of Experimental Biology}\ }\textbf
		{\bibinfo {volume} {218}},\ \bibinfo {pages} {440} (\bibinfo {year}
		{2015})}\BibitemShut {NoStop}%
	\bibitem [{\citenamefont {Lockery}\ \emph {et~al.}(2008)\citenamefont
		{Lockery}, \citenamefont {Lawton}, \citenamefont {Doll}, \citenamefont
		{Faumont}, \citenamefont {Coulthard}, \citenamefont {Thiele}, \citenamefont
		{Chronis}, \citenamefont {McCormick}, \citenamefont {Goodman},\ and\
		\citenamefont {Pruitt}}]{lockery2008artificial}%
	\BibitemOpen
	\bibfield  {author} {\bibinfo {author} {\bibfnamefont {S.~R.}\ \bibnamefont
			{Lockery}}, \bibinfo {author} {\bibfnamefont {K.~J.}\ \bibnamefont {Lawton}},
		\bibinfo {author} {\bibfnamefont {J.~C.}\ \bibnamefont {Doll}}, \bibinfo
		{author} {\bibfnamefont {S.}~\bibnamefont {Faumont}}, \bibinfo {author}
		{\bibfnamefont {S.~M.}\ \bibnamefont {Coulthard}}, \bibinfo {author}
		{\bibfnamefont {T.~R.}\ \bibnamefont {Thiele}}, \bibinfo {author}
		{\bibfnamefont {N.}~\bibnamefont {Chronis}}, \bibinfo {author} {\bibfnamefont
			{K.~E.}\ \bibnamefont {McCormick}}, \bibinfo {author} {\bibfnamefont {M.~B.}\
			\bibnamefont {Goodman}}, \ and\ \bibinfo {author} {\bibfnamefont {B.~L.}\
			\bibnamefont {Pruitt}},\ }\href@noop {} {\bibfield  {journal} {\bibinfo
			{journal} {Journal of neurophysiology}\ }\textbf {\bibinfo {volume} {99}},\
		\bibinfo {pages} {3136} (\bibinfo {year} {2008})}\BibitemShut {NoStop}%
	\bibitem [{\citenamefont {Majmudar}\ \emph {et~al.}(2012)\citenamefont
		{Majmudar}, \citenamefont {Keaveny}, \citenamefont {Zhang},\ and\
		\citenamefont {Shelley}}]{majmudar2012experiments}%
	\BibitemOpen
	\bibfield  {author} {\bibinfo {author} {\bibfnamefont {T.}~\bibnamefont
			{Majmudar}}, \bibinfo {author} {\bibfnamefont {E.~E.}\ \bibnamefont
			{Keaveny}}, \bibinfo {author} {\bibfnamefont {J.}~\bibnamefont {Zhang}}, \
		and\ \bibinfo {author} {\bibfnamefont {M.~J.}\ \bibnamefont {Shelley}},\
	}\href@noop {} {\bibfield  {journal} {\bibinfo  {journal} {Journal of The
				Royal Society Interface}\ ,\ \bibinfo {pages} {rsif20110856}} (\bibinfo
		{year} {2012})}\BibitemShut {NoStop}%
	\bibitem [{\citenamefont {Gray}\ and\ \citenamefont
		{Lissmann}(1950)}]{gray1950kinetics}%
	\BibitemOpen
	\bibfield  {author} {\bibinfo {author} {\bibfnamefont {J.}~\bibnamefont
			{Gray}}\ and\ \bibinfo {author} {\bibfnamefont {H.~W.}\ \bibnamefont
			{Lissmann}},\ }\href@noop {} {\bibfield  {journal} {\bibinfo  {journal}
			{Journal of Experimental Biology}\ }\textbf {\bibinfo {volume} {26}},\
		\bibinfo {pages} {354} (\bibinfo {year} {1950})}\BibitemShut {NoStop}%
	\bibitem [{\citenamefont {Gans}(1975)}]{gans1975tetrapod}%
	\BibitemOpen
	\bibfield  {author} {\bibinfo {author} {\bibfnamefont {C.}~\bibnamefont
			{Gans}},\ }\href@noop {} {\bibfield  {journal} {\bibinfo  {journal} {American
				Zoologist}\ }\textbf {\bibinfo {volume} {15}},\ \bibinfo {pages} {455}
		(\bibinfo {year} {1975})}\BibitemShut {NoStop}%
	\bibitem [{\citenamefont {Murphy}\ \emph {et~al.}(2008)\citenamefont {Murphy},
		\citenamefont {Tadokoro}, \citenamefont {Nardi}, \citenamefont {Jacoff},
		\citenamefont {Fiorini}, \citenamefont {Choset},\ and\ \citenamefont
		{Erkmen}}]{Murphy2008}%
	\BibitemOpen
	\bibfield  {author} {\bibinfo {author} {\bibfnamefont {R.~R.}\ \bibnamefont
			{Murphy}}, \bibinfo {author} {\bibfnamefont {S.}~\bibnamefont {Tadokoro}},
		\bibinfo {author} {\bibfnamefont {D.}~\bibnamefont {Nardi}}, \bibinfo
		{author} {\bibfnamefont {A.}~\bibnamefont {Jacoff}}, \bibinfo {author}
		{\bibfnamefont {P.}~\bibnamefont {Fiorini}}, \bibinfo {author} {\bibfnamefont
			{H.}~\bibnamefont {Choset}}, \ and\ \bibinfo {author} {\bibfnamefont {A.~M.}\
			\bibnamefont {Erkmen}},\ }\enquote {\bibinfo {title} {Search and rescue
			robotics},}\ in\ \href {\doibase 10.1007/978-3-540-30301-5_51} {\emph
		{\bibinfo {booktitle} {Springer Handbook of Robotics}}},\ \bibinfo {editor}
	{edited by\ \bibinfo {editor} {\bibfnamefont {B.}~\bibnamefont {Siciliano}}\
		and\ \bibinfo {editor} {\bibfnamefont {O.}~\bibnamefont {Khatib}}}\ (\bibinfo
	{publisher} {Springer Berlin Heidelberg},\ \bibinfo {address} {Berlin,
		Heidelberg},\ \bibinfo {year} {2008})\ pp.\ \bibinfo {pages}
	{1151--1173}\BibitemShut {NoStop}%
	\bibitem [{\citenamefont {Transeth}\ \emph {et~al.}(2008)\citenamefont
		{Transeth}, \citenamefont {Leine}, \citenamefont {Glocker}, \citenamefont
		{Pettersen},\ and\ \citenamefont {Liljeb{\"a}ck}}]{transeth2008snake}%
	\BibitemOpen
	\bibfield  {author} {\bibinfo {author} {\bibfnamefont {A.~A.}\ \bibnamefont
			{Transeth}}, \bibinfo {author} {\bibfnamefont {R.~I.}\ \bibnamefont {Leine}},
		\bibinfo {author} {\bibfnamefont {C.}~\bibnamefont {Glocker}}, \bibinfo
		{author} {\bibfnamefont {K.~Y.}\ \bibnamefont {Pettersen}}, \ and\ \bibinfo
		{author} {\bibfnamefont {P.}~\bibnamefont {Liljeb{\"a}ck}},\ }\href@noop {}
	{\bibfield  {journal} {\bibinfo  {journal} {IEEE Transactions on Robotics}\
		}\textbf {\bibinfo {volume} {24}},\ \bibinfo {pages} {88} (\bibinfo {year}
		{2008})}\BibitemShut {NoStop}%
	\bibitem [{\citenamefont {Liljeb{\"a}ck}\ \emph
		{et~al.}(2010{\natexlab{a}})\citenamefont {Liljeb{\"a}ck}, \citenamefont
		{Pettersen}, \citenamefont {Stavdahl},\ and\ \citenamefont
		{Gravdahl}}]{Liljeback:2010eb}%
	\BibitemOpen
	\bibfield  {author} {\bibinfo {author} {\bibfnamefont {P.}~\bibnamefont
			{Liljeb{\"a}ck}}, \bibinfo {author} {\bibfnamefont {K.~Y.}\ \bibnamefont
			{Pettersen}}, \bibinfo {author} {\bibfnamefont {{\O}.}~\bibnamefont
			{Stavdahl}}, \ and\ \bibinfo {author} {\bibfnamefont {J.~T.}\ \bibnamefont
			{Gravdahl}},\ }\href@noop {} {\bibfield  {journal} {\bibinfo  {journal} {IEEE
				Transactions on Robotics}\ }\textbf {\bibinfo {volume} {26}},\ \bibinfo
		{pages} {781} (\bibinfo {year} {2010}{\natexlab{a}})}\BibitemShut {NoStop}%
	\bibitem [{\citenamefont {Liljeb{\"a}ck}\ \emph
		{et~al.}(2010{\natexlab{b}})\citenamefont {Liljeb{\"a}ck}, \citenamefont
		{Pettersen},\ and\ \citenamefont {Stavdahl}}]{Liljeback:2010tg}%
	\BibitemOpen
	\bibfield  {author} {\bibinfo {author} {\bibfnamefont {P.}~\bibnamefont
			{Liljeb{\"a}ck}}, \bibinfo {author} {\bibfnamefont {K.~Y.}\ \bibnamefont
			{Pettersen}}, \ and\ \bibinfo {author} {\bibfnamefont {{\O}.}~\bibnamefont
			{Stavdahl}},\ }\href@noop {} {\bibfield  {journal} {\bibinfo  {journal}
			{ICRA}\ } (\bibinfo {year} {2010}{\natexlab{b}})}\BibitemShut {NoStop}%
	\bibitem [{\citenamefont {Travers}\ \emph {et~al.}(2016)\citenamefont
		{Travers}, \citenamefont {Whitman}, \citenamefont {Schiebel}, \citenamefont
		{Goldman},\ and\ \citenamefont {Choset}}]{travers2016shape}%
	\BibitemOpen
	\bibfield  {author} {\bibinfo {author} {\bibfnamefont {M.~J.}\ \bibnamefont
			{Travers}}, \bibinfo {author} {\bibfnamefont {J.}~\bibnamefont {Whitman}},
		\bibinfo {author} {\bibfnamefont {P.}~\bibnamefont {Schiebel}}, \bibinfo
		{author} {\bibfnamefont {D.}~\bibnamefont {Goldman}}, \ and\ \bibinfo
		{author} {\bibfnamefont {H.}~\bibnamefont {Choset}},\ }in\ \href@noop {}
	{\emph {\bibinfo {booktitle} {Robotics: Science and Systems}}}\ (\bibinfo
	{year} {2016})\BibitemShut {NoStop}%
	\bibitem [{\citenamefont {Aguilar}\ \emph {et~al.}(2016)\citenamefont
		{Aguilar}, \citenamefont {Zhang}, \citenamefont {Qian}, \citenamefont
		{Kingsbury}, \citenamefont {McInroe}, \citenamefont {Mazouchova},
		\citenamefont {Li}, \citenamefont {Maladen}, \citenamefont {Gong},
		\citenamefont {Travers}, \citenamefont {Hatton}, \citenamefont {Choset},
		\citenamefont {Umbanhowar},\ and\ \citenamefont {Goldman}}]{Aguilar:2016bq}%
	\BibitemOpen
	\bibfield  {author} {\bibinfo {author} {\bibfnamefont {J.}~\bibnamefont
			{Aguilar}}, \bibinfo {author} {\bibfnamefont {T.}~\bibnamefont {Zhang}},
		\bibinfo {author} {\bibfnamefont {F.}~\bibnamefont {Qian}}, \bibinfo {author}
		{\bibfnamefont {M.}~\bibnamefont {Kingsbury}}, \bibinfo {author}
		{\bibfnamefont {B.}~\bibnamefont {McInroe}}, \bibinfo {author} {\bibfnamefont
			{N.}~\bibnamefont {Mazouchova}}, \bibinfo {author} {\bibfnamefont
			{C.}~\bibnamefont {Li}}, \bibinfo {author} {\bibfnamefont {R.}~\bibnamefont
			{Maladen}}, \bibinfo {author} {\bibfnamefont {C.}~\bibnamefont {Gong}},
		\bibinfo {author} {\bibfnamefont {M.}~\bibnamefont {Travers}}, \bibinfo
		{author} {\bibfnamefont {R.~L.}\ \bibnamefont {Hatton}}, \bibinfo {author}
		{\bibfnamefont {H.}~\bibnamefont {Choset}}, \bibinfo {author} {\bibfnamefont
			{P.~B.}\ \bibnamefont {Umbanhowar}}, \ and\ \bibinfo {author} {\bibfnamefont
			{D.~I.}\ \bibnamefont {Goldman}},\ }\href@noop {} {\bibfield  {journal}
		{\bibinfo  {journal} {Reports on Progress in Physics}\ ,\ \bibinfo {pages}
			{1}} (\bibinfo {year} {2016})}\BibitemShut {NoStop}%
	\bibitem [{\citenamefont {Qian}\ and\ \citenamefont
		{Goldman}(2015{\natexlab{b}})}]{Qian:2015ua}%
	\BibitemOpen
	\bibfield  {author} {\bibinfo {author} {\bibfnamefont {F.}~\bibnamefont
			{Qian}}\ and\ \bibinfo {author} {\bibfnamefont {D.~I.}\ \bibnamefont
			{Goldman}},\ }\href@noop {} {\bibfield  {journal} {\bibinfo  {journal}
			{Robotics: Science and Systems}\ ,\ \bibinfo {pages} {1}} (\bibinfo {year}
		{2015}{\natexlab{b}})}\BibitemShut {NoStop}%
	\bibitem [{\citenamefont {Pazouki}\ \emph {et~al.}(2017)\citenamefont
		{Pazouki}, \citenamefont {Kwarta}, \citenamefont {Williams}, \citenamefont
		{Likos}, \citenamefont {Serban}, \citenamefont {Jayakumar},\ and\
		\citenamefont {Negrut}}]{pazouki2017compliant}%
	\BibitemOpen
	\bibfield  {author} {\bibinfo {author} {\bibfnamefont {A.}~\bibnamefont
			{Pazouki}}, \bibinfo {author} {\bibfnamefont {M.}~\bibnamefont {Kwarta}},
		\bibinfo {author} {\bibfnamefont {K.}~\bibnamefont {Williams}}, \bibinfo
		{author} {\bibfnamefont {W.}~\bibnamefont {Likos}}, \bibinfo {author}
		{\bibfnamefont {R.}~\bibnamefont {Serban}}, \bibinfo {author} {\bibfnamefont
			{P.}~\bibnamefont {Jayakumar}}, \ and\ \bibinfo {author} {\bibfnamefont
			{D.}~\bibnamefont {Negrut}},\ }\href@noop {} {\bibfield  {journal} {\bibinfo
			{journal} {Physical Review E}\ }\textbf {\bibinfo {volume} {96}},\ \bibinfo
		{pages} {042905} (\bibinfo {year} {2017})}\BibitemShut {NoStop}%
	\bibitem [{sup()}]{supplemental}%
	\BibitemOpen
	\href@noop {} {}\bibinfo {note} {See Supplemental Material at [URL will be
		inserted by publisher] for [further experiment and simulation details;
		simulation validation details; number and duration of head collisions as a
		function of $d$; comparisons between simulation and single-post model;
		distances between original and remapped states; and distances between nearest
		single and multi-peg states].}\BibitemShut {Stop}%
	\bibitem [{\citenamefont {Hirose}(1993)}]{Hirose:1993}%
	\BibitemOpen
	\bibfield  {author} {\bibinfo {author} {\bibfnamefont {S.}~\bibnamefont
			{Hirose}},\ }\href@noop {} {\emph {\bibinfo {title} {Biologically Inspired
				Robots: Serpentile Locomotors and Manipulators}}}\ (\bibinfo  {publisher}
	{Oxford University Press},\ \bibinfo {year} {1993})\BibitemShut {NoStop}%
	\bibitem [{\citenamefont {Hirose}\ and\ \citenamefont
		{Yamada}(2009)}]{hirose2009snake}%
	\BibitemOpen
	\bibfield  {author} {\bibinfo {author} {\bibfnamefont {S.}~\bibnamefont
			{Hirose}}\ and\ \bibinfo {author} {\bibfnamefont {H.}~\bibnamefont
			{Yamada}},\ }\href@noop {} {\bibfield  {journal} {\bibinfo  {journal} {IEEE
				Robotics \& Automation Magazine}\ }\textbf {\bibinfo {volume} {16}},\
		\bibinfo {pages} {88} (\bibinfo {year} {2009})}\BibitemShut {NoStop}%
	\bibitem [{\citenamefont {Tasora}\ \emph {et~al.}(2016)\citenamefont {Tasora},
		\citenamefont {Serban}, \citenamefont {Mazhar}, \citenamefont {Pazouki},
		\citenamefont {Melanz}, \citenamefont {Fleischmann}, \citenamefont {Taylor},
		\citenamefont {Sugiyama},\ and\ \citenamefont {Negrut}}]{Chrono2016}%
	\BibitemOpen
	\bibfield  {author} {\bibinfo {author} {\bibfnamefont {A.}~\bibnamefont
			{Tasora}}, \bibinfo {author} {\bibfnamefont {R.}~\bibnamefont {Serban}},
		\bibinfo {author} {\bibfnamefont {H.}~\bibnamefont {Mazhar}}, \bibinfo
		{author} {\bibfnamefont {A.}~\bibnamefont {Pazouki}}, \bibinfo {author}
		{\bibfnamefont {D.}~\bibnamefont {Melanz}}, \bibinfo {author} {\bibfnamefont
			{J.}~\bibnamefont {Fleischmann}}, \bibinfo {author} {\bibfnamefont
			{M.}~\bibnamefont {Taylor}}, \bibinfo {author} {\bibfnamefont
			{H.}~\bibnamefont {Sugiyama}}, \ and\ \bibinfo {author} {\bibfnamefont
			{D.}~\bibnamefont {Negrut}},\ }in\ \href@noop {} {\emph {\bibinfo {booktitle}
			{High Performance Computing in Science and Engineering – Lecture Notes in
				Computer Science}}},\ \bibinfo {editor} {edited by\ \bibinfo {editor}
		{\bibfnamefont {T.}~\bibnamefont {Kozubek}}}\ (\bibinfo  {publisher}
	{Springer},\ \bibinfo {year} {2016})\ pp.\ \bibinfo {pages}
	{19--49}\BibitemShut {NoStop}%
	\bibitem [{whe()}]{wheelnote}%
	\BibitemOpen
	\href@noop {} {}\bibinfo {note} {We used a slightly modified perpendicular
		force relation in the simulation to achieve better agreement between
		experimental and simulation trajectories, $F_{\perp,\mathrm{sim}} = 1.2
		F_{\perp,\mathrm{exp}}$.}\BibitemShut {Stop}%
	\bibitem [{\citenamefont {Johnson}(1987)}]{johnson1987contact}%
	\BibitemOpen
	\bibfield  {author} {\bibinfo {author} {\bibfnamefont {K.~L.}\ \bibnamefont
			{Johnson}},\ }\href@noop {} {\emph {\bibinfo {title} {Contact mechanics}}}\
	(\bibinfo  {publisher} {Cambridge University Press},\ \bibinfo {year}
	{1987})\BibitemShut {NoStop}%
	\bibitem [{fit()}]{fitnote}%
	\BibitemOpen
	\href@noop {} {}\bibinfo {note} {We note the small angle approximation ($\sin
		\theta \approx \theta$) is valid for angles we measure, so a fit to the
		function expected for far-field wave diffraction, $\theta_{q} = 180/\pi
		\sin^{-1}(D/d)$, is indistinguishable from the fit we have
		chosen.}\BibitemShut {Stop}%
	\bibitem [{\citenamefont {Couder}\ and\ \citenamefont
		{Fort}(2006)}]{Couder:2006ix}%
	\BibitemOpen
	\bibfield  {author} {\bibinfo {author} {\bibfnamefont {Y.}~\bibnamefont
			{Couder}}\ and\ \bibinfo {author} {\bibfnamefont {E.}~\bibnamefont {Fort}},\
	}\href@noop {} {\bibfield  {journal} {\bibinfo  {journal} {Physical Review
				Letters}\ }\textbf {\bibinfo {volume} {97}},\ \bibinfo {pages} {114}
		(\bibinfo {year} {2006})}\BibitemShut {NoStop}%
	\bibitem [{\citenamefont {Bush}(2015)}]{Bush:2015ev}%
	\BibitemOpen
	\bibfield  {author} {\bibinfo {author} {\bibfnamefont {J.~W.~M.}\
			\bibnamefont {Bush}},\ }\href@noop {} {\bibfield  {journal} {\bibinfo
			{journal} {Annual Review of Fluid Mechanics}\ }\textbf {\bibinfo {volume}
			{47}},\ \bibinfo {pages} {269} (\bibinfo {year} {2015})}\BibitemShut
	{NoStop}%
	\bibitem [{\citenamefont {Schiebel}\ \emph {et~al.}(2018)\citenamefont
		{Schiebel}, \citenamefont {Rieser}, \citenamefont {Hubbard}, \citenamefont
		{Chen}, \citenamefont {Rocklin},\ and\ \citenamefont
		{Goldman}}]{schiebel2018}%
	\BibitemOpen
	\bibfield  {author} {\bibinfo {author} {\bibfnamefont {P.~E.}\ \bibnamefont
			{Schiebel}}, \bibinfo {author} {\bibfnamefont {J.~M.}\ \bibnamefont
			{Rieser}}, \bibinfo {author} {\bibfnamefont {A.~M.}\ \bibnamefont {Hubbard}},
		\bibinfo {author} {\bibfnamefont {L.}~\bibnamefont {Chen}}, \bibinfo {author}
		{\bibfnamefont {Z.}~\bibnamefont {Rocklin}}, \ and\ \bibinfo {author}
		{\bibfnamefont {D.~I.}\ \bibnamefont {Goldman}},\ }\href@noop {} {\bibfield
		{journal} {\bibinfo  {journal} {submitted}\ } (\bibinfo {year}
		{2018})}\BibitemShut {NoStop}%
	\bibitem [{\citenamefont {Li}\ \emph {et~al.}(2013)\citenamefont {Li},
		\citenamefont {Zhang},\ and\ \citenamefont {Goldman}}]{li2013}%
	\BibitemOpen
	\bibfield  {author} {\bibinfo {author} {\bibfnamefont {C.}~\bibnamefont
			{Li}}, \bibinfo {author} {\bibfnamefont {T.}~\bibnamefont {Zhang}}, \ and\
		\bibinfo {author} {\bibfnamefont {D.~I.}\ \bibnamefont {Goldman}},\
	}\href@noop {} {\bibfield  {journal} {\bibinfo  {journal} {Science}\ }\textbf
		{\bibinfo {volume} {339}},\ \bibinfo {pages} {1408} (\bibinfo {year}
		{2013})}\BibitemShut {NoStop}%
	\bibitem [{\citenamefont {Jayne}(1986)}]{jayne1986kinematics}%
	\BibitemOpen
	\bibfield  {author} {\bibinfo {author} {\bibfnamefont {B.~C.}\ \bibnamefont
			{Jayne}},\ }\href@noop {} {\bibfield  {journal} {\bibinfo  {journal}
			{Copeia}\ ,\ \bibinfo {pages} {915}} (\bibinfo {year} {1986})}\BibitemShut
	{NoStop}%
	\bibitem [{\citenamefont {Marvi}\ \emph {et~al.}(2014)\citenamefont {Marvi},
		\citenamefont {Gong}, \citenamefont {Gravish}, \citenamefont {Astley},
		\citenamefont {Travers}, \citenamefont {Hatton}, \citenamefont {Mendelson},
		\citenamefont {Choset}, \citenamefont {Hu},\ and\ \citenamefont
		{Goldman}}]{marvi2014sidewinding}%
	\BibitemOpen
	\bibfield  {author} {\bibinfo {author} {\bibfnamefont {H.}~\bibnamefont
			{Marvi}}, \bibinfo {author} {\bibfnamefont {C.}~\bibnamefont {Gong}},
		\bibinfo {author} {\bibfnamefont {N.}~\bibnamefont {Gravish}}, \bibinfo
		{author} {\bibfnamefont {H.}~\bibnamefont {Astley}}, \bibinfo {author}
		{\bibfnamefont {M.}~\bibnamefont {Travers}}, \bibinfo {author} {\bibfnamefont
			{R.~L.}\ \bibnamefont {Hatton}}, \bibinfo {author} {\bibfnamefont {J.~R.}\
			\bibnamefont {Mendelson}}, \bibinfo {author} {\bibfnamefont {H.}~\bibnamefont
			{Choset}}, \bibinfo {author} {\bibfnamefont {D.~L.}\ \bibnamefont {Hu}}, \
		and\ \bibinfo {author} {\bibfnamefont {D.~I.}\ \bibnamefont {Goldman}},\
	}\href@noop {} {\bibfield  {journal} {\bibinfo  {journal} {Science}\ }\textbf
		{\bibinfo {volume} {346}},\ \bibinfo {pages} {224} (\bibinfo {year}
		{2014})}\BibitemShut {NoStop}%
	\bibitem [{\citenamefont {Iagnemma}\ \emph {et~al.}(2003)\citenamefont
		{Iagnemma}, \citenamefont {Golda}, \citenamefont {Spenko},\ and\
		\citenamefont {Dubowsky}}]{iagnemma2003experimental}%
	\BibitemOpen
	\bibfield  {author} {\bibinfo {author} {\bibfnamefont {K.}~\bibnamefont
			{Iagnemma}}, \bibinfo {author} {\bibfnamefont {D.}~\bibnamefont {Golda}},
		\bibinfo {author} {\bibfnamefont {M.}~\bibnamefont {Spenko}}, \ and\ \bibinfo
		{author} {\bibfnamefont {S.}~\bibnamefont {Dubowsky}},\ }\href@noop {}
	{\bibfield  {journal} {\bibinfo  {journal} {Experimental Robotics VIII}\ ,\
			\bibinfo {pages} {654}} (\bibinfo {year} {2003})}\BibitemShut {NoStop}%
	\bibitem [{\citenamefont {Wong}(2009)}]{wong2009terramechanics}%
	\BibitemOpen
	\bibfield  {author} {\bibinfo {author} {\bibfnamefont {J.~Y.}\ \bibnamefont
			{Wong}},\ }\href@noop {} {\emph {\bibinfo {title} {Terramechanics and
				off-road vehicle engineering: terrain behaviour, off-road vehicle performance
				and design}}}\ (\bibinfo  {publisher} {Butterworth-Heinemann},\ \bibinfo
	{year} {2009})\BibitemShut {NoStop}%
	\bibitem [{\citenamefont {Floreano}\ \emph {et~al.}(2010)\citenamefont
		{Floreano}, \citenamefont {Zufferey}, \citenamefont {Srinivasan},\ and\
		\citenamefont {Ellington}}]{floreano2010flying}%
	\BibitemOpen
	\bibfield  {author} {\bibinfo {author} {\bibfnamefont {D.}~\bibnamefont
			{Floreano}}, \bibinfo {author} {\bibfnamefont {J.-C.}\ \bibnamefont
			{Zufferey}}, \bibinfo {author} {\bibfnamefont {M.~V.}\ \bibnamefont
			{Srinivasan}}, \ and\ \bibinfo {author} {\bibfnamefont {C.}~\bibnamefont
			{Ellington}},\ }\href@noop {} {\emph {\bibinfo {title} {Flying insects and
				robots}}}\ (\bibinfo  {publisher} {Springer},\ \bibinfo {year}
	{2010})\BibitemShut {NoStop}%
	\bibitem [{\citenamefont {Turpin}\ \emph {et~al.}(2012)\citenamefont {Turpin},
		\citenamefont {Michael},\ and\ \citenamefont {Kumar}}]{turpin2012trajectory}%
	\BibitemOpen
	\bibfield  {author} {\bibinfo {author} {\bibfnamefont {M.}~\bibnamefont
			{Turpin}}, \bibinfo {author} {\bibfnamefont {N.}~\bibnamefont {Michael}}, \
		and\ \bibinfo {author} {\bibfnamefont {V.}~\bibnamefont {Kumar}},\
	}\href@noop {} {\bibfield  {journal} {\bibinfo  {journal} {Autonomous
				Robots}\ }\textbf {\bibinfo {volume} {33}},\ \bibinfo {pages} {143} (\bibinfo
		{year} {2012})}\BibitemShut {NoStop}%
	\bibitem [{\citenamefont {Batchelor}(1970)}]{batchelor1970slender}%
	\BibitemOpen
	\bibfield  {author} {\bibinfo {author} {\bibfnamefont {G.}~\bibnamefont
			{Batchelor}},\ }\href@noop {} {\bibfield  {journal} {\bibinfo  {journal}
			{Journal of Fluid Mechanics}\ }\textbf {\bibinfo {volume} {44}},\ \bibinfo
		{pages} {419} (\bibinfo {year} {1970})}\BibitemShut {NoStop}%
	\bibitem [{\citenamefont {Machado}\ \emph {et~al.}(2012)\citenamefont
		{Machado}, \citenamefont {Moreira}, \citenamefont {Flores},\ and\
		\citenamefont {Lankarani}}]{machado2012compliant}%
	\BibitemOpen
	\bibfield  {author} {\bibinfo {author} {\bibfnamefont {M.}~\bibnamefont
			{Machado}}, \bibinfo {author} {\bibfnamefont {P.}~\bibnamefont {Moreira}},
		\bibinfo {author} {\bibfnamefont {P.}~\bibnamefont {Flores}}, \ and\ \bibinfo
		{author} {\bibfnamefont {H.~M.}\ \bibnamefont {Lankarani}},\ }\href@noop {}
	{\bibfield  {journal} {\bibinfo  {journal} {Mechanism and Machine Theory}\
		}\textbf {\bibinfo {volume} {53}},\ \bibinfo {pages} {99} (\bibinfo {year}
		{2012})}\BibitemShut {NoStop}%
\end{thebibliography}

%

\clearpage
\begin{figure}[ht!]
	\includegraphics[width=\columnwidth]{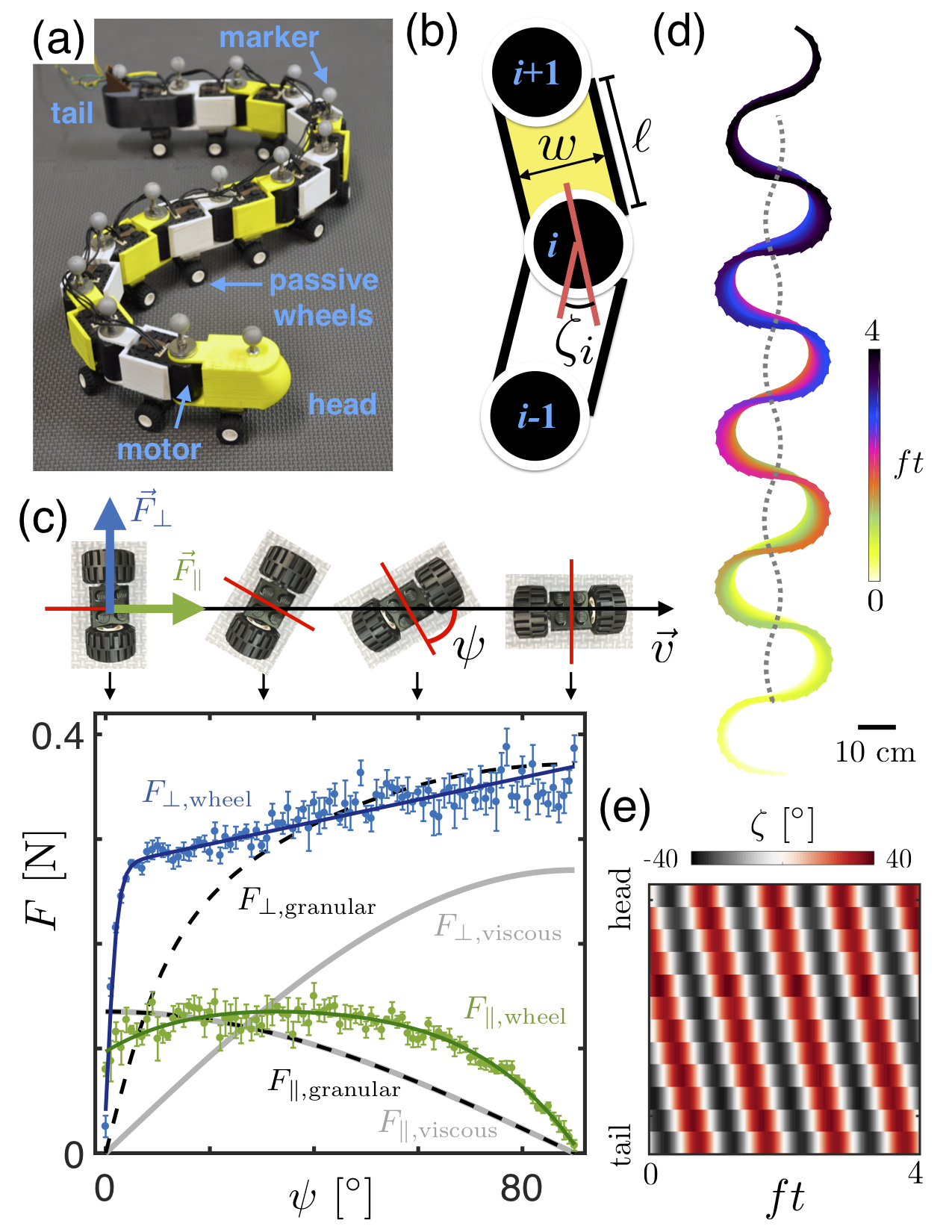}
	\caption{\textbf{Robophysical snake movement.} (a) A photo of the robophysical snake. $12$ servo motors were linked together with $3$D-printed plastic brackets to create a robotic snake. Passive wheels were affixed to the base of each segment to create an anisotropic friction with the ground. (b) A schematic of three adjacent servo motors (black circles). The angular position of each motor, $\zeta_i$, was driven as a function of time. (c) Experimentally-measured wheel friction relations. Two passive wheels were connected to a single axle and translated at $v = 10$~mm/s over a rubber substrate under constant load. $\psi$, the orientation of the wheels relative to the direction of motion, was varied in $1^\circ$ increments. Points show average steady-state forces along (green) and perpendicular to (blue) rolling direction, error bars represent variation over five experiments, and curves show fits to data. For comparison, drag forces on a submerged rod moving through $300$-$\mu$m glass beads (dashed black lines, adapted from~\cite{sharpe2015locomotor}) and moving through a viscous fluid  (light gray lines, see, e.g.,~\cite{batchelor1970slender}) are shown. For granular and viscous curves, forces are scaled so that each $F_{\parallel}$ has the same maximum value as $F_{\parallel,\mbox{wheel}}$. (d) Snapshots of the robot position and configurations while moving in a post-free environment are colored by elapsed time. The dashed gray curve shows the corresponding center-of-geometry trajectory. (e) A space-time plot of experimentally measured $\zeta_i$ from head is shown over four undulations in a single experiment ($d=5.7$~cm).}
	\label{fig:robot}
\end{figure}

\begin{figure}[ht!]
	\includegraphics[width=\columnwidth]{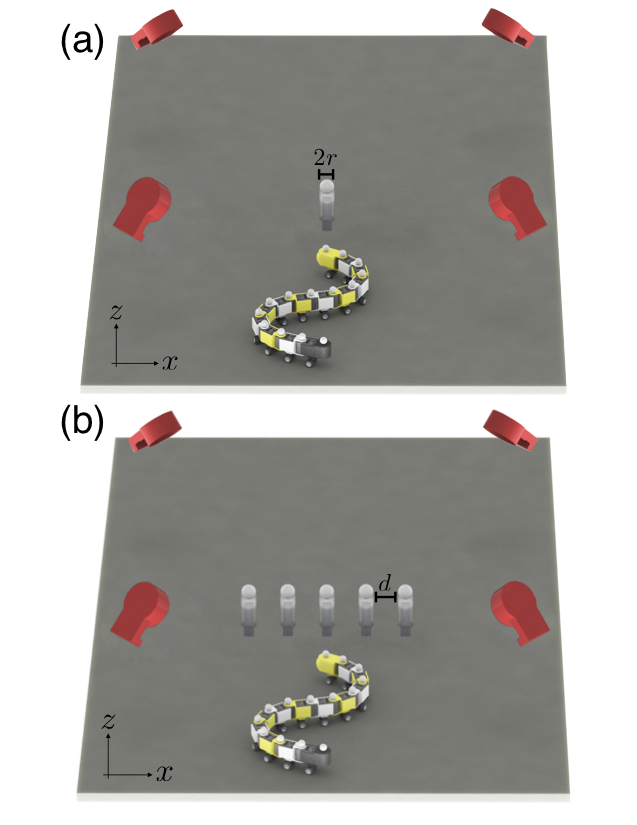}
	\caption{\textbf{Single- and multi-post experimental configurations.} (a) Schematic of the single-post setup.  A single vertical post is rigidly affixed to an otherwise homogeneous substrate. (b) Snapshots of robot configurations and locations (colored by time) throughout an interaction with the single post. (b) Schematic of the multi-post setup. A single row of five evenly-spaced vertical posts are rigidly affixed to the substrate. }
	\label{fig:setups}
\end{figure}

\begin{figure}[ht!]
	\includegraphics[width=\columnwidth]{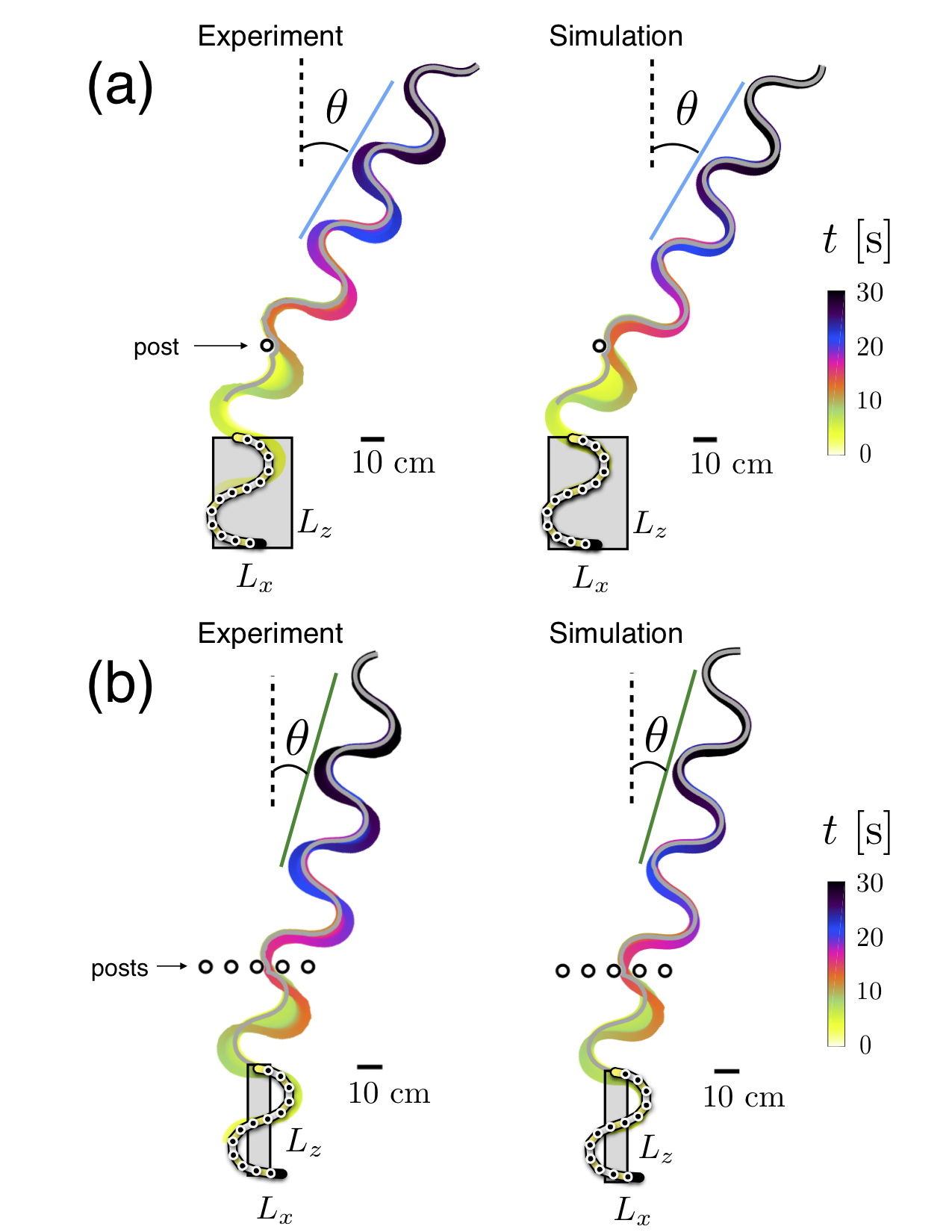}
	\caption{\textbf{Scattering interactions with single- and multi-post arrays.} (a) Snapshots of robot configurations and locations (colored by time) throughout an interaction with the single post. Experiment and simulation are shown for a nearly-identical collision, and the resulting rotations are similar (experiment: $\theta = 30.4^\circ$; simulation: $\theta = 29.2^\circ$). (b) Snapshots of the robot positions and configurations throughout an interaction with the posts.  The collisions and resulting reorientations are similar (experiment: $\theta = 15.9^\circ$; simulation: $\theta = 15.3^\circ$).}
	\label{fig:scatter}
\end{figure}

\begin{figure*}[ht!]
	\includegraphics[width=1.7\columnwidth]{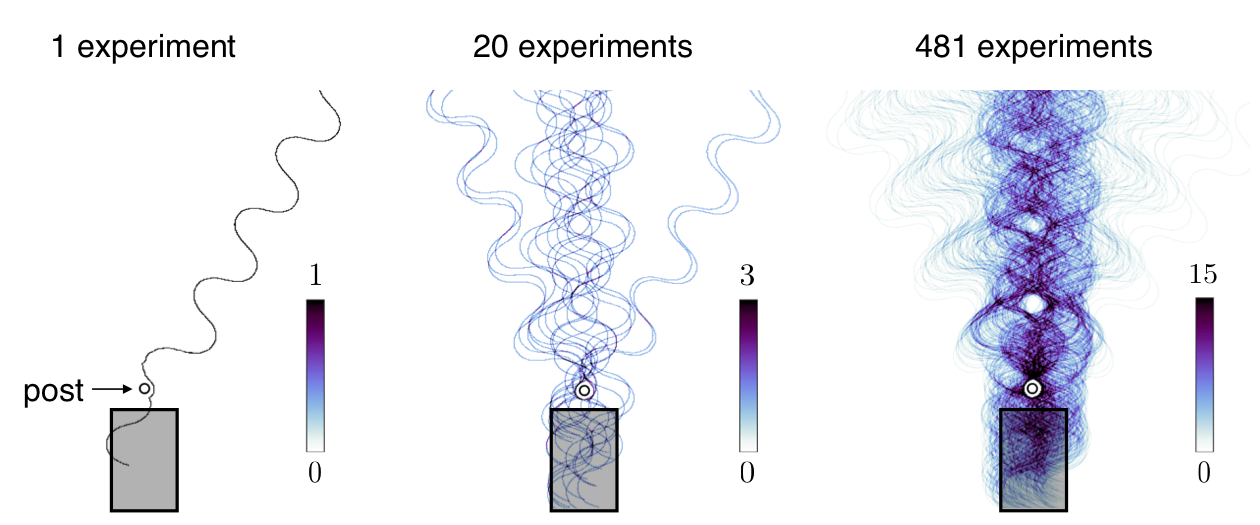}
	\caption{\textbf{Emergence of a single-post scattering pattern.} Left: binary image created from head trajectory in Fig.~\ref{fig:scatter}b. Middle: Summation of binary images from $20$ arbitrarily-chosen initial conditions. Right: Summation of binary images over $481$ initial conditions, evenly sampled within the gray box shown in Fig.~\ref{fig:scatter}b. In each panel, the color of each pixel indicates the number of experiments that traveled through the corresponding point in space.}
	\label{fig:singlepeg}
\end{figure*}

\begin{figure}[ht!]
	\includegraphics[width=\columnwidth]{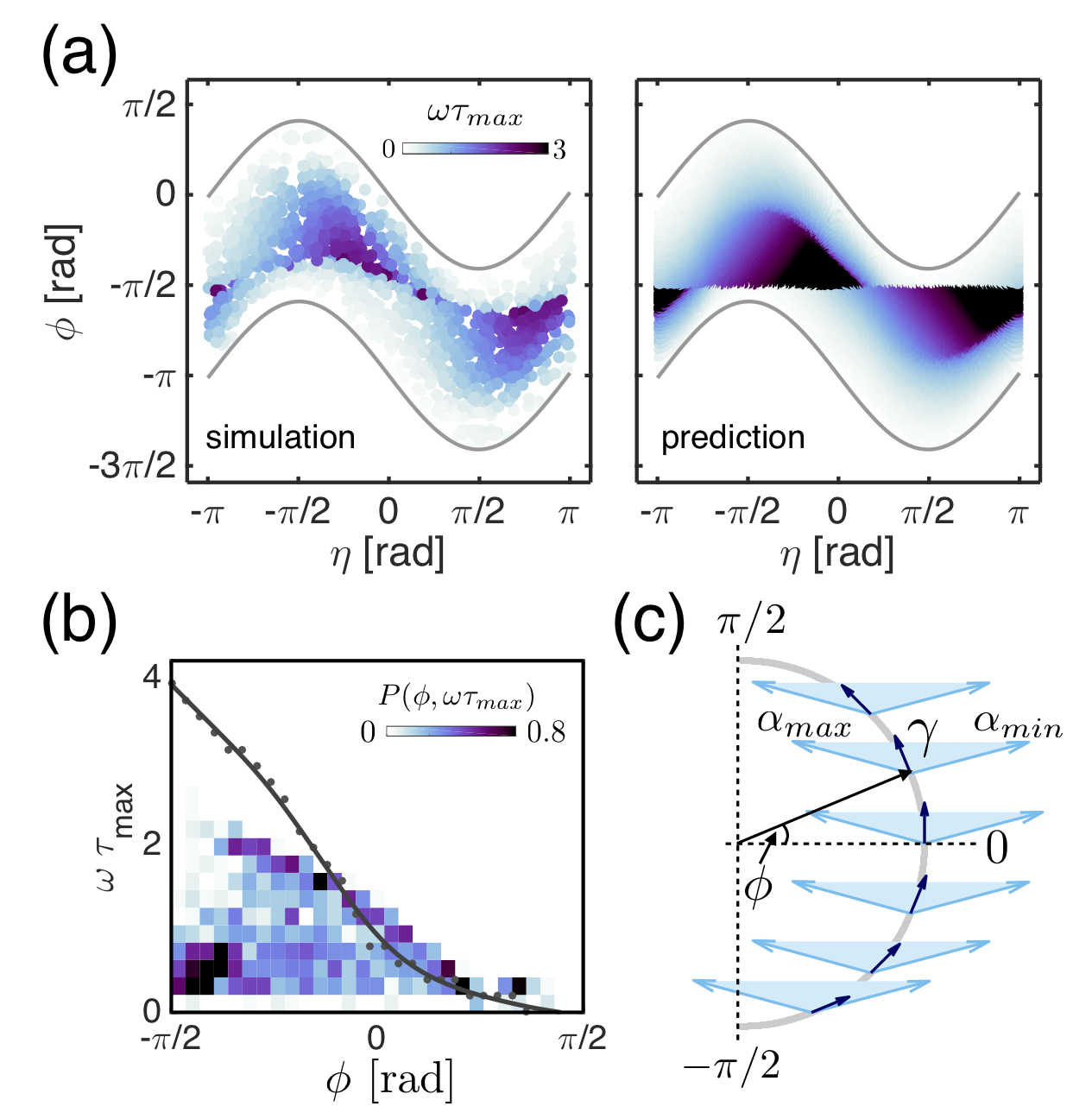}
	\caption{\textbf{Single post collision states and durations.} (a) Impact location, $\phi$, and wave phase, $\eta$, of  colored by the duration of the resulting head-post collision for $1,200$ simulations (left)  and predicted for $85,000$ points (right). (b) Two-dimensional probability distribution of contact durations as a function of impact location of the post (right side only). The gray points and splined fit show model-predicted behavior. (c) Sketch comparing the range of velocity vectors swept out in a single cycle (blue triangles) to the local post tangent,$\gamma$ for several impact locations, $\phi$. There are more states that can be pinned near the leading surface of the post, and these additional states can be pinned for longer $\tau_{max}$.}
	\label{fig:singlepoststates}
\end{figure}

\begin{figure*}[ht!]
	\includegraphics[width=1.7\columnwidth]{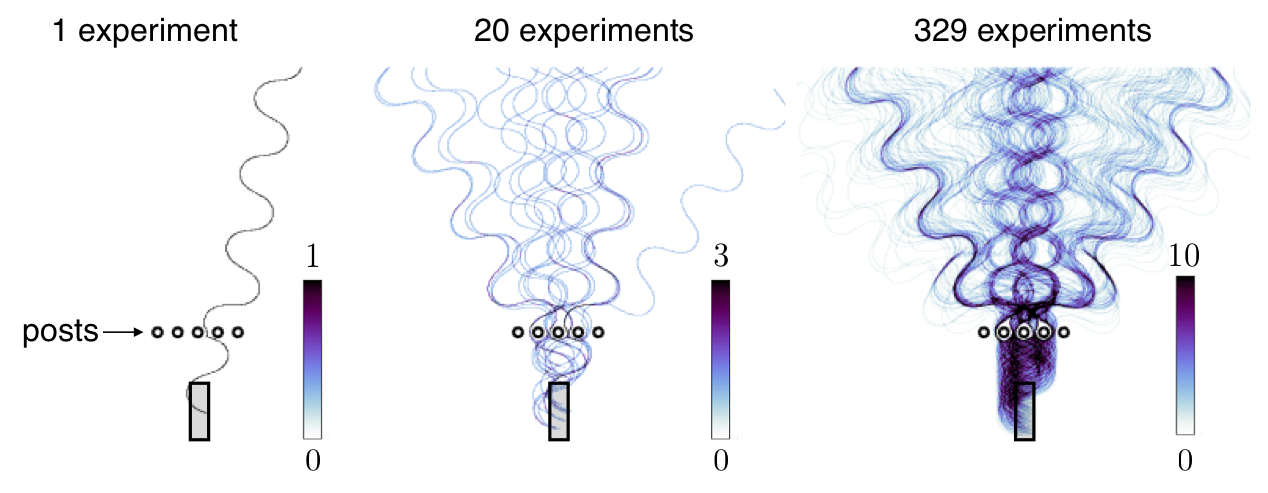}
	\caption{\textbf{Emergence of a multi-post scattering pattern.}  Left: binary image created from head trajectory in Fig.~\ref{fig:scatter}b. Middle: Summation of binary images from $20$ arbitrarily-chosen initial conditions. Right: Summation of binary images over $329$ initial conditions, evenly sampled within the gray box. In each panel, the color of each pixel indicates the number of experiments that traveled through the corresponding point in space.}
	\label{fig:multiBuildup}
\end{figure*}

\begin{figure}[h]
	\includegraphics[width=\columnwidth]{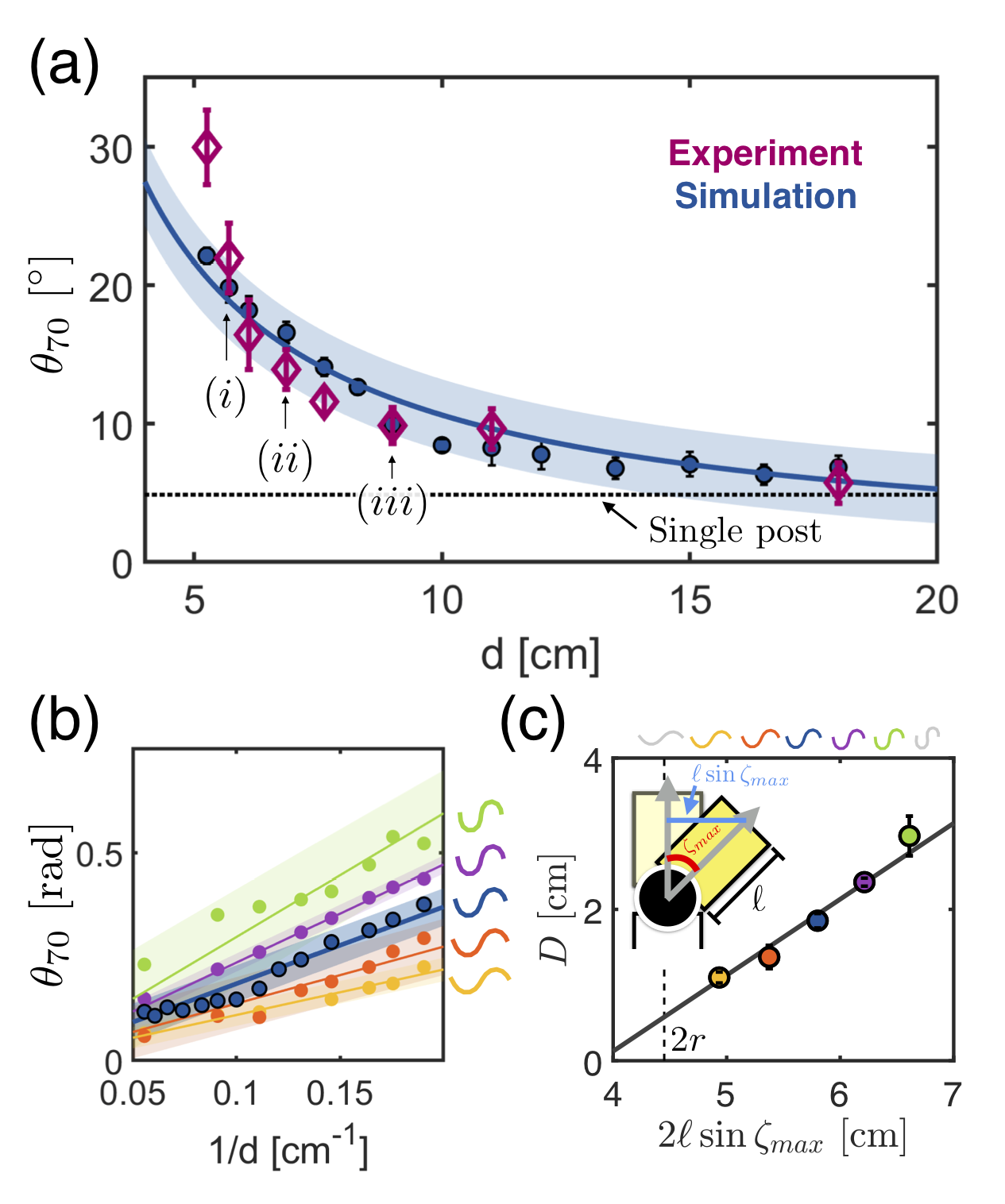}
	\caption{\textbf{Distribution spread dependence on post spacing and segment angular amplitude.} (a) $\theta_{70}$ vs $d$ for experiment (maroon diamonds) and simulation (blue circles). Numerically labeled points result from corresponding distribution in Fig.~\ref{fig:preferredPaths}. Bootstrapping is used to determine $95\%$ confidence intervals associated with each quantile measurement, and bounds of this confidence interval are indicated by error bars. The curve shows the fit of $\theta_{70} = 180/\pi (D/d)$ to the simulation, with the shaded region indicating the $95\%$ prediction bounds for the fit. The single-post value is shown by the horizontal line. Inset: $\theta_{70}$ vs $1/d$ for different $\zeta_{max}$ (varied in simulation).  Data points are measured from distributions at specified spacing, lines show fits to the data, and shaded regions indicate $95\%$ prediction bounds for each fit. Corresponding wave shapes are shown to the right. (b) The fit parameter, $D$, for different $\zeta_{max}$. $D$ is linearly related to the full perpendicular distance each segment sweeps out in one period. (c) Schematic of single motor and two adjacent segments. The perpendicular distance swept out by a single segment during a full cycle is given by $2\ell \sin \zeta_{max}$.}
	\label{fig:diffraction}
\end{figure}

\begin{figure}[h!]
	\includegraphics[width=\columnwidth]{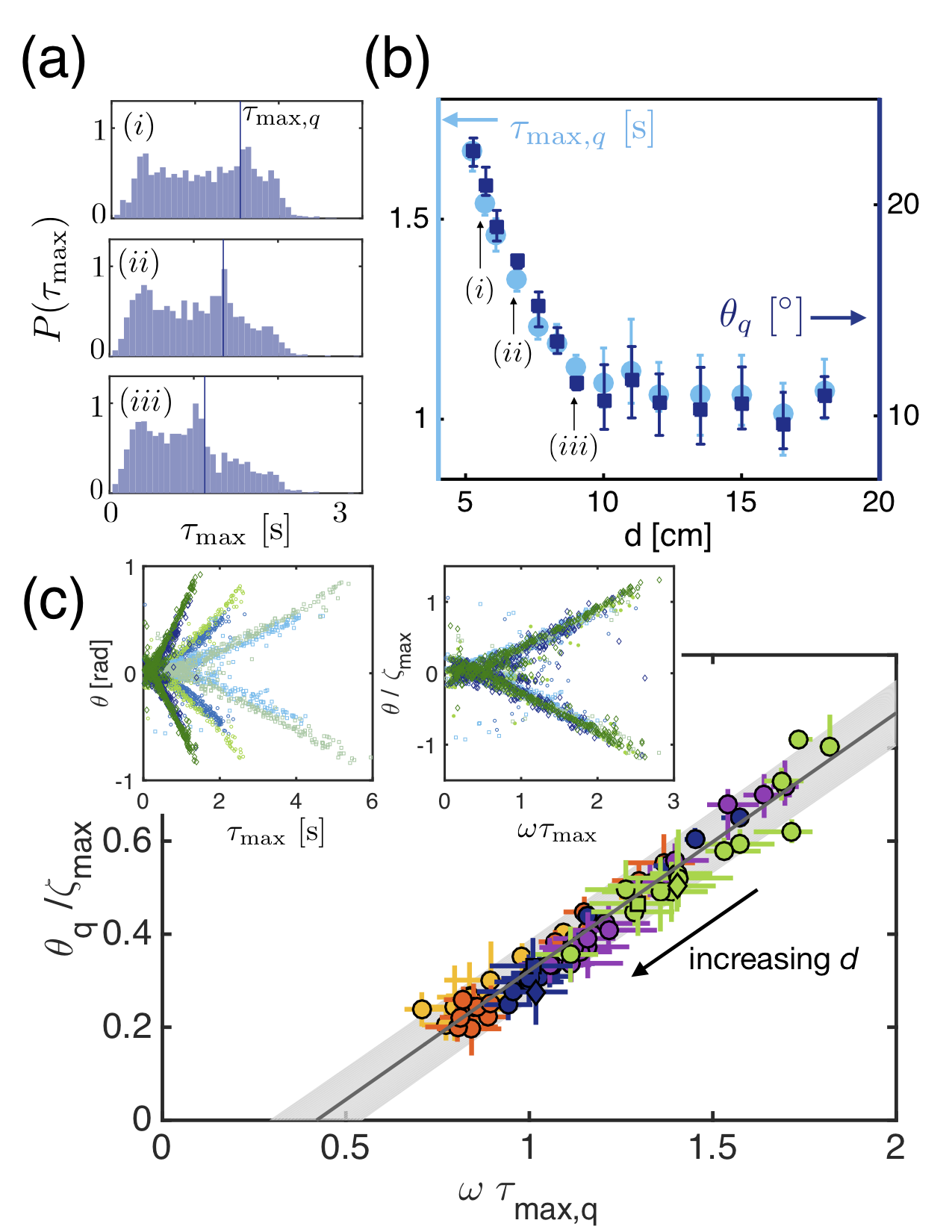}
	\caption{\textbf{Collision duration sets reorientation.} (a) Probability distributions of the maximal contact duration, $\tau_{max}$, for three post configurations (with spacing increasing down the column). Vertical lines show $\tau_{max,70}$, the $70^{\mathrm{th}}$-quantile of each $\tau_{max}$ distribution, which decreases as $d$ increases ($(i) \to (iii)$). (b) The spacing-dependence of $\tau_{max,70}$ (light blue, left axis) and $\theta_{70}$ (dark blue, right axis). Here, distributions only include simulations for which there was a head-post collision. Errorbars show the bootstrapping-estimated $95\%$ confidence interval for each value. (c) $\theta_{70}$ and $\tau_{max,70}$ (non-dimensionalized by $\zeta_{max}$ and $\omega$, respectively) plotted against each other, shows the dependence can be described by a single line (of slope $0.55 \pm 0.03$ and y-intercept $-0.23 \pm 0.04$) over a range of $\zeta_{max}$ (indicated by color and consistent with Fig.~\ref{fig:diffraction}), $f$ ($\square: 0.075$~Hz; $\circ: 0.15$~Hz; $\diamond: 0.3$~Hz), and $d$. Inset: $\theta$ vs $\tau$ for $\zeta_{max} = 0.605$~rad (blue) and $\zeta_{max} = 0.705$~rad (green) for raw (left) and non-dimensionalized (right) data. Symbols indicate frequency and are consistent with main plot. 
	} 
	\label{fig:collapse}
\end{figure}

\clearpage
\onecolumngrid
\appendix*
\setcounter{figure}{0}

\makeatletter 
\renewcommand{\thefigure}{S\@arabic\c@figure}
\makeatother

\setcounter{table}{0}

\makeatletter 
\renewcommand{\thetable}{S\@arabic\c@table}
\makeatother

\setcounter{page}{1}

\makeatletter
\renewcommand{\thepage}{S\arabic{page}} 
\makeatother

\section*{Supplementary information}
\begin{center}
	\textbf{The dynamics of scattering in undulatory active collisions}
	
	Jennifer M. Rieser, Perrin E. Schiebel, Arman Pazouki, Feifei Qian, Zachary Goddard,\\ Andrew Zangwill,
	Dan Negrut, Daniel I. Goldman
\end{center}

\subsection{Experiment}

Our $13$-segment-robotic snake, shown in Fig.~\ref{fig:robot}a, had $12$~Dynamixel AX-12A servo motors connected together with custom-designed 3D-printed plastic brackets, and a Robotis CM-700 controller was programmed to command the angular position of each motor to vary sinusoidally with time and position along the body. Robot segments were  $3.7$~cm wide, and $3$-cm tall all interior segments were $5.1$~cm long. The head, $6.0$~cm long, added a nearly-spherical nose cap to the interior segment design, and the tail, $7.5$~cm long, was adapted to have a cylindrical cap. The robot mass was $1.13$~kg and the fully-extended length was around $80$~cm. 

The snake moved in a model heterogeneous terrain, created from a level wooden platform (dimensions $2.4$~m wide x $3.6$~m long) covered by a firm rubber mat. Obstacles consisted of a single row of vertical polycarbonate posts (radius, $r=0.023$~m) anchored to the platform (see Fig.~\ref{fig:setups}). Before each experiment, the robot motor configuration was reset and the robot was manually positioned and oriented so the initial heading was transverse to the post row.  Positions of infrared-reflective markers atop each robot segment were identified and recorded at $120$~Hz by four Optitrack Flex13 infrared cameras (positions were accurate to within $0.1$~mm). Using the tracking data, we quantified the final heading of the robot, $\theta$ (see Fig.~\ref{fig:scatter}b), for each trajectory by identifying and fitting lines to the extrema of segment trajectories (for at least three undulations) after the tail had moved beyond the post row.

\subsection{Wheel friction}
\label{suppmat:wheels}
To characterize the robot-substrate interaction forces during movement, we designed a custom, 3D printed bracket to attach a single pair of Lego wheels to a 6-axis force-torque transducer (Nano~43, ATI Industrial Automation, Apex, NC, USA) and mounted the force sensor to a 6-axis industrial robot arm (Denso VS087A2-AV6-NNN-NNN). The robot arm was programmed to repeatedly execute the following automated procedure: (1) rotate the wheels by some angle, $\psi$, relative to the dragging direction and begin recording forces at $1$~kH; (2) lower the wheels to the a predetermined height, $H$, at which point wheel contact with the substrate (ethylene-vinyl acetate (EVA) Soft Linking Mats) had been established and the normal load on the wheels was comparable to the weight of a robot segment; (3) horizontally drag the wheels $40$~cm across the substrate at a constant speed, $v = 10$~mm/s; (4) raise the wheels, stop recording forces, and return to the initial position. 

Five trials were performed per $\psi$, which was varied from $0^{\circ}$ to $90^{\circ}$ (parallel to perpendicular to the wheel axle) in increments of one degree. For each trial, forces were decomposed into components along the wheel axle, $F_{\perp}$, and along the preferred rolling direction, $F_{\parallel}$. Force components quickly reached and subsequently maintained a near-constant value for most of the dragging distance, therefore, we estimated the steady-state values by averaging each component over the five trials within this near-constant window. Functions were fit to $F_{\perp}$ and $F_{\parallel}$ (shown as the curves in Fig.~\ref{fig:robot}d) these forces could be incorporated into the Chrono simulation. Numerical values of fit parameters along with corresponding $95\%$ confidence intervals are given in Table~\ref{tab:fitParams}.
\begin{equation*}
F_\perp(\psi)=a\psi+\frac{b}{1+e^(-\psi/c)}+d
\end{equation*}
\begin{equation*}
F_{\parallel} (\psi)=p_1 \psi^4+p_2 \psi^3+p_3 \psi^2+p_4 \psi+p_5
\end{equation*}
\begin{center}
	\begin{table}
		\centering
		\begin{tabular}{ | c | c | c | }
			\hline
			parameter & value & $95\%$ confidence interval  \\ \hline
			$a$ & $0.0011$ & $(0.0010,0.0012)$ \\ \hline
			$b$ & $0.47$ & $(0.42,0.51)$ \\ \hline
			$c$ & $1.1$ & $(0.98,1.21)$ \\ \hline
			$d$ & $-0.19$ & $(-0.24,-0.15)$ \\ \hline
			$p_1$ & $-7.8 \times 10^{-9}$ & $(-1.2\times 10^{-8},-3.1 \times 10^{-9})$ \\ \hline
			$p_2$ & $1.2 \times 10^{-6}$ & $(2.9 \times 10^{-7},2.1 \times 10^{-6})$ \\ \hline
			$p_3$ & $-8.7 \times 10^{-5}$ & $(-1.4 \times 10^{-4},-3.2 \times 10^{-5})$ \\ \hline
			$p_4$ & $0.0030$ & $(0.0018,0.0043)$ \\ \hline
			$p_5$ & $0.097$ & $(0.089,0.110)$ \\ \hline
		\end{tabular} 
		\caption{fit parameter values and $95\%$ confidence intervals for $F_\parallel$ and $F_\perp$.}
		\label{tab:fitParams}
	\end{table}
\end{center}

\subsection{Simulation}
\label{suppmat:simValidation}

The simulation-based studies conducted relied on an open-source simulation framework called Chrono~\cite{Chrono2016}. For a constrained multibody dynamics problem, Chrono formulates a set of index three differential-algebraic equations whose solution captures the time evolution of the dynamic system. All simulation results reported here were obtained using a half implicit, first order, symplectic Euler time integration method and a successive over-relaxation iteration scheme. Geometric overlaps between contacting objects was used to approximate local deformations at contact points. The contact force between mating surfaces was calculated via a Hertzian contact force model~\cite{johnson1987contact}, 

\begin{equation*}
F_n=k_n \delta_n-g_n v_n^r
\end{equation*}
\begin{equation*}
F_t=k_t \delta_t,
\end{equation*}
where the subscripts $n$ and $t$ denote the contact force components, $F_n$ and $F_t$, in the normal and tangential directions, respectively; $\delta_n$ is the overlap of two interacting bodies; and $v_n^r$ is the relative velocity of the bodies at the contact point. For the contact of parallel cylinders,  $k_n= \pi/4 Y^* l$ is the contact stiffness modulus and $k_t = 2k_n/7$. Here $l$ is the cylinder length, i.e. the height of a segment, and $Y^*$ is the effective Young's modulus, defined based on Young's modulus, $Y$, and Poisson's ratio, $\nu$, of the mating surfaces as
\begin{equation*}
1/Y^* =(1-\nu_1^2)/Y_1 +(1-\nu_2^2)/Y_2.
\end{equation*}

Contact forces between a post and a segment with a flat surface were calculated in a similar fashion. To allow for larger integration time-steps and thus reduce simulation time, the value of Young's modulus was chosen to be smaller than the actual one. Drawing on a sensitivity analysis that quantified the impact of relaxing $Y$ on the accuracy of the simulation results, we used $Y=2.5 \times 10^6$ and $\nu=0.4$. The damping coefficient, $g_n$, depends on the material coefficient of restitution and collision scenario~\cite{machado2012compliant}. We used a larger value, $g_n$ ($\sim 10^3$), to enforce a plastic contact.

The geometry of the snake model was modeled through a set of shape primitives such as box and cylinders. The body components were connected by revolute joints, which removed five out of six relative degrees of freedom. Additional light-weight cylinders were positioned on the joints to facilitate, from a geometric perspective, a smooth interaction of the segments with the cylindrical posts. Table~\ref{tab:simDetails} shows parameter values used.

\begin{center}
	\begin{table}
		\centering
		\begin{tabular}{ | c | c | c | }
			\hline
			\multirow{8}{*}{Snake geometry} 
			& Segment length & $5.1$~cm \\
			& Segment height & $3.5$~cm \\
			& Segment width & $3.2$~cm \\
			& Head radius & $1.92$~cm \\ 
			& Tail radius & $1.8$~cm \\ 
			& Tail height & $3.5$~cm \\ 
			& Joint radius & $1.85$~cm \\ 
			& Density & $1.2$~g/cm$^3$ \\ \hline
			\multirow{2}{*}{Snake motion} 
			& Wave amplitude ($\zeta_{max}$) & $0.605$~rad \\
			& Wave frequency ($f$) & $0.15$~Hz \\ \hline
			\multirow{3}{*}{post} 
			& Radius & $2.25$~cm\\
			& Height & $20$~cm\\
			& Density & $1.2$~g/cm$^3$\\ 
			\hline
		\end{tabular}
		\caption{Attributes of the snake and posts in the simulation.}
		\label{tab:simDetails}
	\end{table}
\end{center}

\begin{figure}[h!]
	\includegraphics[width=0.5\columnwidth]{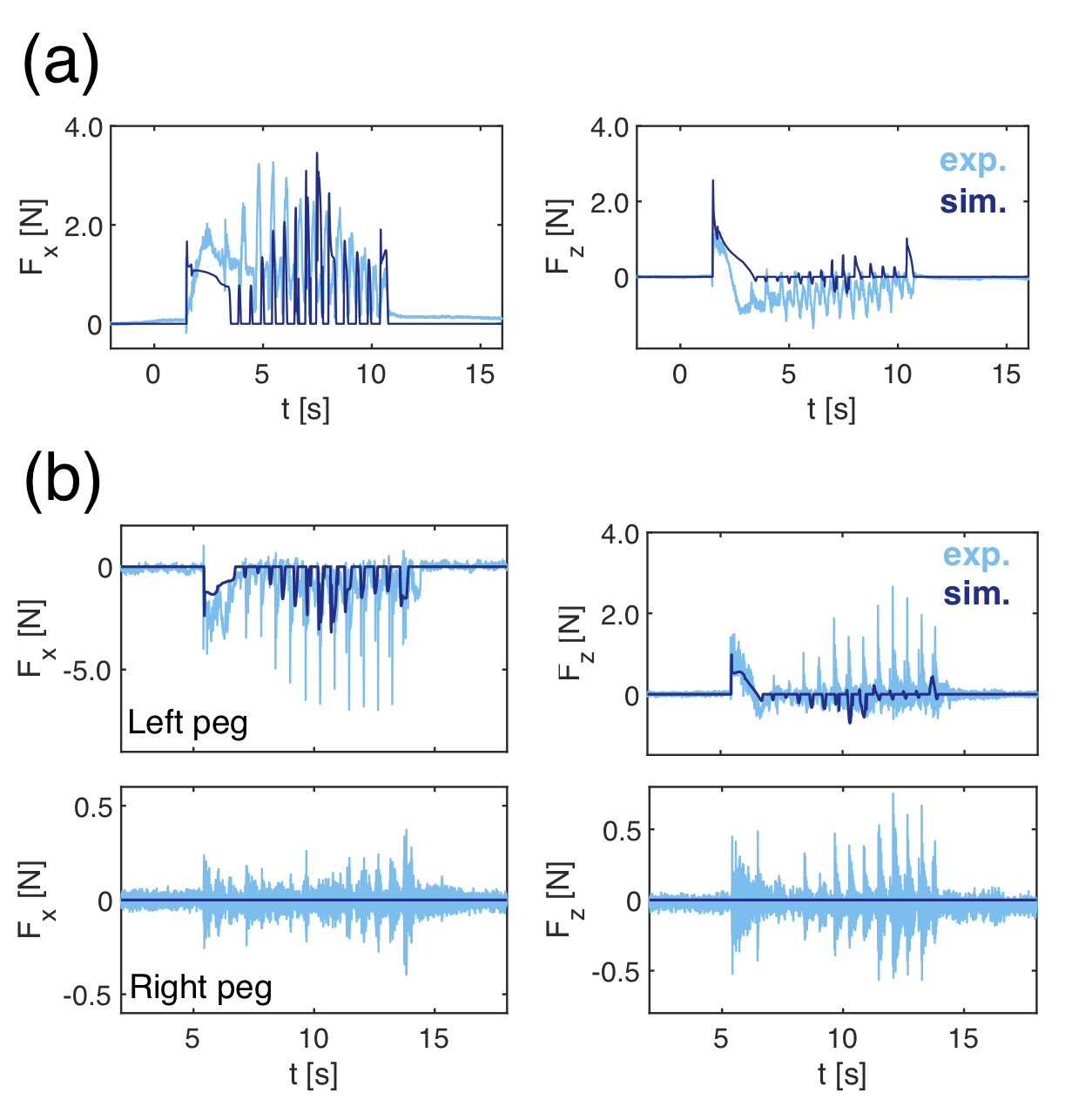}
	\caption{\textbf{Experimental and simulation forces on posts.} (a) Forces exerted onto the post by the robot during the experimental and simulation trajectory shown in Fig.~\ref{fig:scatter}a. (b) Forces exerted onto the posts by the robot during the experimental and simulation trajectory shown in Fig.~\ref{fig:scatter}b.}
	\label{fig:forces}
\end{figure}

Simulations were then validated by comparing experimental and simulated trajectories and forces for snake interacting with a single post. In experiments, forces exerted by the robot during collisions with the post were recorded by mounting the post to an ATI Nano 43 6-axis force-torque transducer. Forces exerted by the robot onto the posts for the trials shown in Fig.~\ref{fig:scatter} are shown in Fig.~\ref{fig:forces}a (single post) and Fig.~\ref{fig:forces}b (multi-post). This comparison is representative of agreement between simulation and experiment: trajectories for similar collsion states produced nearly-identical trajectories and forces were often comparable and exhibited similar structure in both simulation and experiment. While there were some quantitative differences between simulation and experimental forces, these discrepancies did not seem to affect the kinematic agreement. A time step convergence analysis revealed that forces and resulting trajectories were insensitive to the time step, $\Delta t$, for $\Delta t<6\times10^{-4}$~s. $\Delta t = 10^{-4}$~s was selected for all the simulations presented here.

In the multi-post simulations, the accuracy of the results improved significantly when, to mirror the presence of the revolute joints in the physical prototype, the snake model was augmented with spheres connecting the boxes used for the snake segments. The diameter of the connecting spheres was identical to the width of the robotic snake. The width of the cubic segments in the simulation was slightly reduced from that of the robotic snake to bury the edges inside the spherical joints and prevent the edge contact, particularly at large time step. Table~\ref{tab:simDetails} summarizes the attributes of the snake and the posts for simulations presented here. 

\begin{figure}[h!]
	\includegraphics[width=0.5\columnwidth]{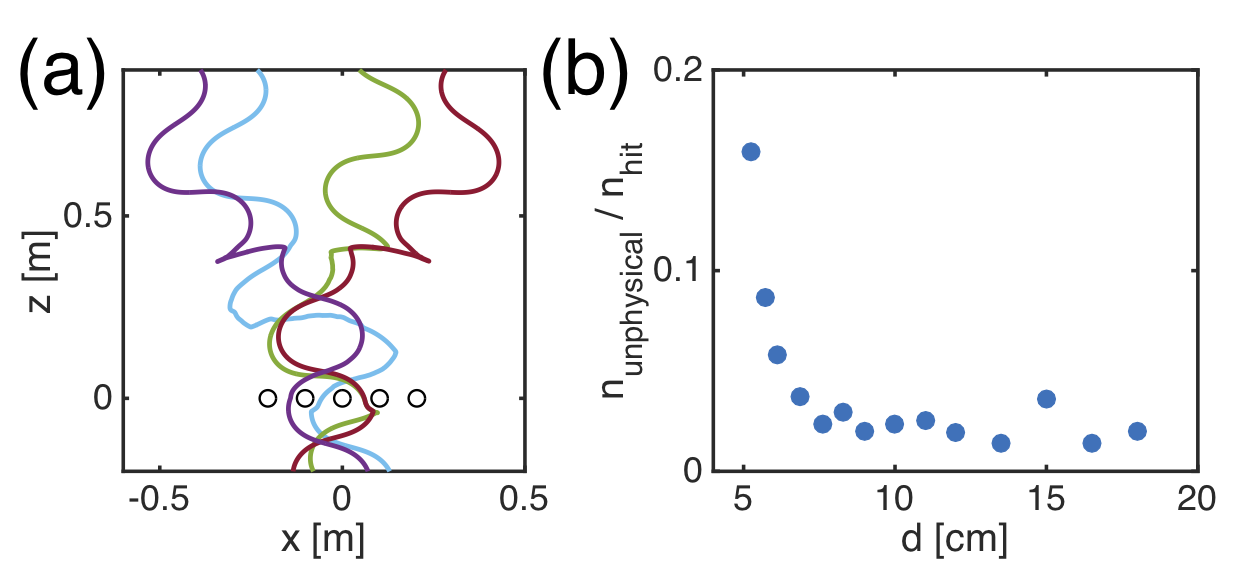}
	\caption{\textbf{Unphysical simulation trajectories.} (a) Four examples of unphysical trajectories for $d = 5.7$~cm.  The robot becomes pinned after the head has cleared the posts, and as a result, the body is rapidly reoriented.  These situations do not occur in the experiment. (b) Fraction of simulations for each spacing that are unphysical.}
	\label{fig:unhealthy}
\end{figure}

At least $1,000$ simulations were run for each post configuration, and for each configuration, there were a few trajectories which were not physical. These typically occurred when the tail of the snake became stuck on the post, causing the entire snake to rapidly change direction. Four representative examples are shown in Fig.~\ref{fig:unhealthy}a. These were identified and removed from further analysis using the following criteria: if, at any point after the head has moved beyond the post row, (1) velocity of head is at least twice as large as maximum head velocity for the freely-moving snake, $v_{head} \ge 2 v_{max,free}$ and (2) force on the head does not exceed a nominal value, chosen here to be $F_{head} \le 0.01$~N. Fig.~\ref{fig:unhealthy}b shows (for $\zeta_{max} = 0.605$~rad), as a function of post spacing, how many unphysical trajectories occurred relative to the number of simulations that had collisions with the posts.

\subsection{Single post contact times}
\label{suppmat:duration}
\begin{figure}[h!]
	\includegraphics[width=0.7\columnwidth]{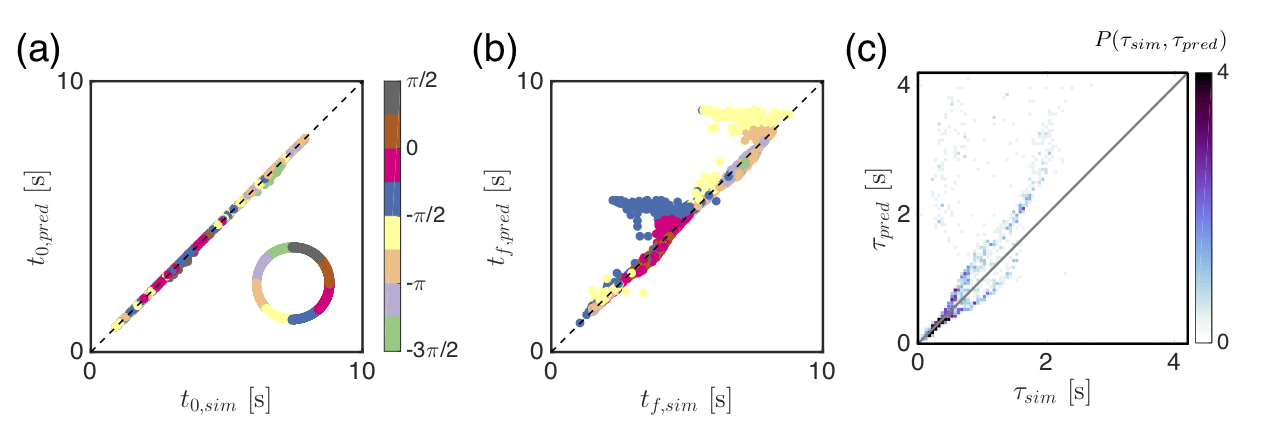}
	\caption{\textbf{Predicted and actual initial and final contact times.} Particle-post contact was broken when there was a component of the velocity that pointed away from the post.  (a) initial contact time, predicted from geometry, agrees well with actual initial contact time (for all impact locations). (b) Predicted vs actual final contact time.  Final times agree reasonably well for most impact locations, with exceptions near the leading surface of the post. (c) Two-dimensional probability map of predicted and actual durations.  Most predictions fall along the line $y = x$, but collisions near the leading surface of the post tend to be over-predicted by the simple contact rules. }
	\label{fig:SuppTimes}
\end{figure}

Fig.~\ref{fig:SuppTimes}a shows predicted vs actual initial contact times ($t_{0,pred}$ vs $t_{0,sim}$) starting from the same initial condition. The predicted initial contact time, $t_{0,pred}$, was determined from geometry.  We required that the head and the post cannot overlap, and assumed that a circle with diameter equal to the snake width was traveling along an unobstructed head trajectory originating at the specified initial condition. We defined $t_0$ as the first time for which there was any overlap between the post and the circle. The color of each point indicates where on the post, in one of eight segments, the initial contact occurred. Regardless of initial contact location, geometry was an excellent predictor of initial contact time as all points fall along the line $t_{0,pred} = t_{0,sim}$.  

Fig.~\ref{fig:SuppTimes}b shows predicted vs actual final contact times ($t_{f,pred}$ vs $t_{f,sim}$) starting from the same initial condition. To find the final contact time, $t_{f,pred}$, we assumed that the circular particle with diameter equal to the width of the snake is trying to achieve the freely-moving velocity but can only move forward with the component of this velocity that is along the local post tangent (i.e., the particle cannot penetrate the post surface). The final time is predicted by identifying the first time for which there is no component of the freely-moving velocity driving into the post. The final contact times are reasonably well predicted for most initial contact locations (most final times fall along or close to the line $t_{f,pred} = t_{f,sim}$), with the exception of near the leading surface of the post.  This discrepancy arises from the stringent requirement that the particle cannot move backward.  Near the leading surface of the post, a small move backward can occur in the simulations and often results in a grazing collision of small duration.  With no backward motion allowed, these collisions can last significant fractions of an undulation cycle.

Fig.~\ref{fig:SuppTimes}c shows a two-dimensional probability density predicted vs actual contact durations, $\tau = t_f - t_0$. Many of the durations lie along the line $\tau_{pred} = \tau_{sim}$. There are two distinct branches for larger $\tau$, one which over-predicts and one which under-predicts.  Large over-predictions originate from grazing collisions near the leading surface of the post. Smaller over- and under-predictions are possible for many impact locations.

\begin{figure}[h!]
	\includegraphics[width=0.8\columnwidth]{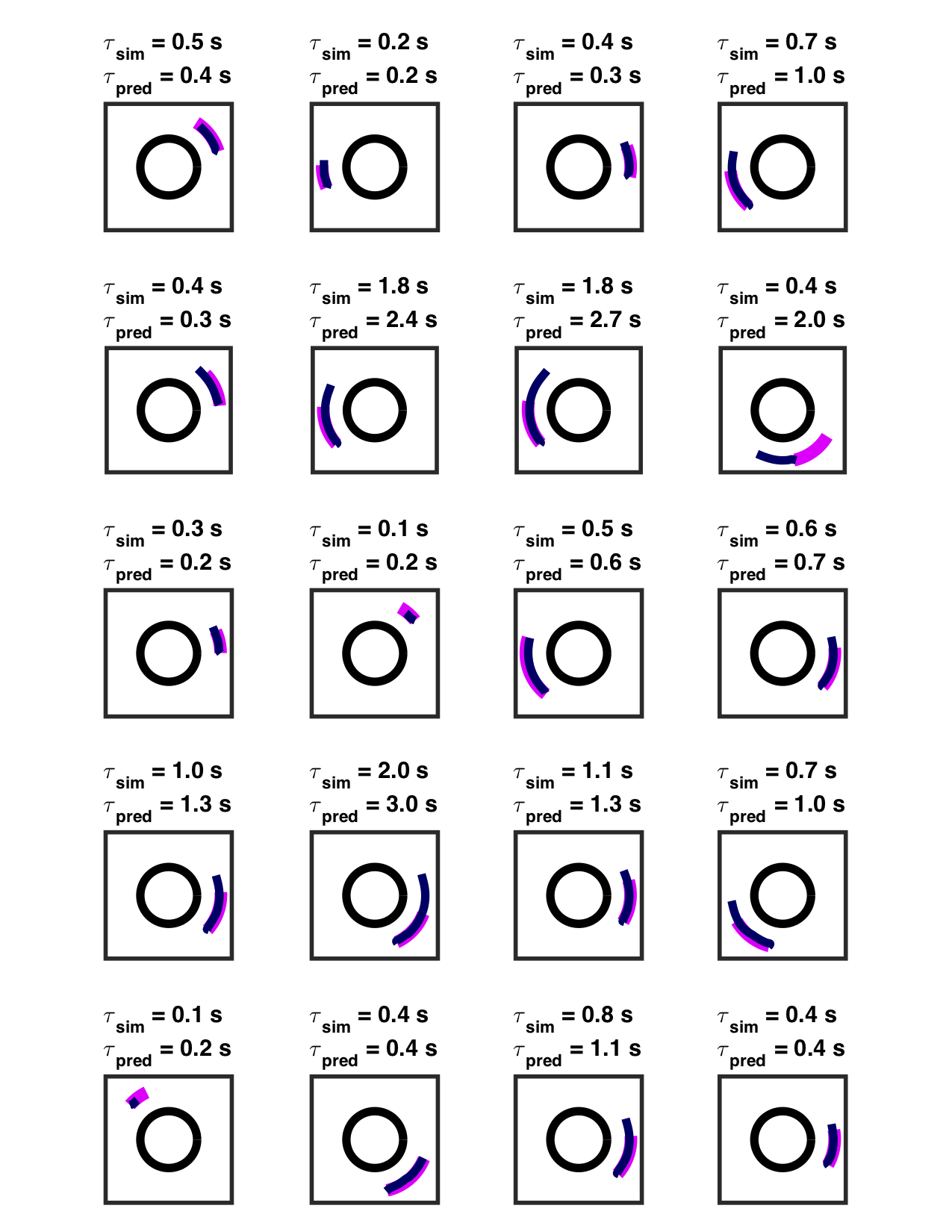}
	\caption{\textbf{Single-post contact trajectories.} $20$ randomly-selected  head trajectories from simulation (dark blue) and  model (magenta) during head-post contact. In each plot, the black circle represents the post. Durations associated with each trajectory are shown above each plot.}
	\label{fig:comparisons}
\end{figure}

Fig.~\ref{fig:comparisons} shows the path traced by the robot head while in contact with the post for $20$ randomly-selected single post simulations, along with the model prediction for each simulation. Actual and predicted durations for each set of trajectories is shown above each plot.  

The largest discrepancies between prediction and simulation result from collisions close to the leading surface of the post ($\phi \sim -\pi/2$), arising from the requirement that the $\vec{v}_{tan} \cdot \hat{z} \geq 0$. In the simulation, these collisions can be grazing, with the robot head sliding off the other side of the post shortly after initial contact (e.g., second row, fourth column of Fig.~\ref{fig:comparisons}). For the prediction, however, the particle is always pinned until the driving velocity is reoriented to align with the local post tangent. The range of durations possible as well as relative likelihood for a contact duration at a given impact location is shown in Fig.~\ref{fig:singlepoststates}c. Near the leading surface of the post, the robot can slide easily and lose contact quickly. 

\subsection{Small and large $\zeta_{max}$: Qualitatively different spacing dependence}
\begin{figure}[h!]
	\includegraphics[width=0.5\columnwidth]{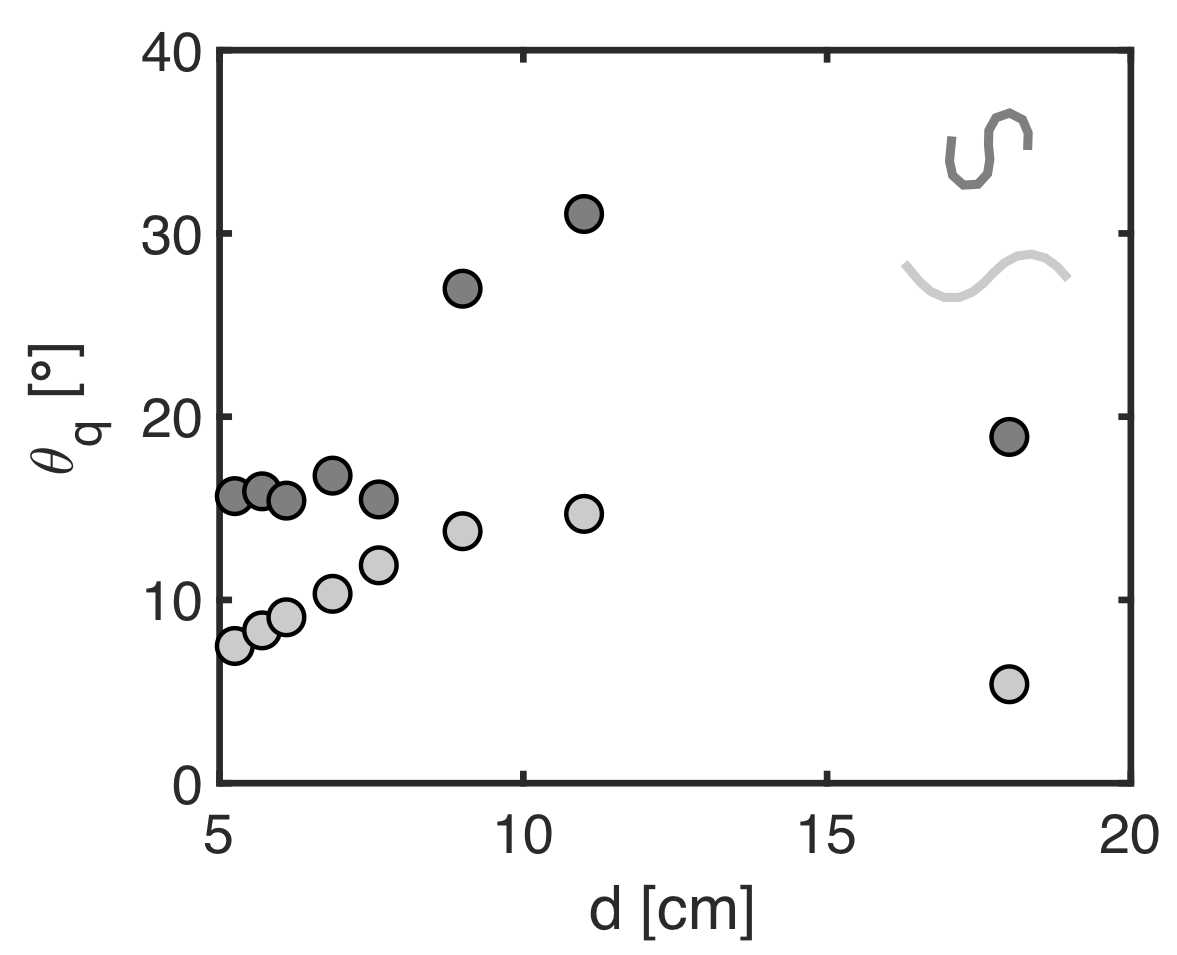}
	\caption{\textbf{Scattering angle distribution dependence on spacing for large and small angular amplitude.} When $\zeta_{max}$ is outside of the range presented in the main text, the qualitative dependence of $\theta_{70}$ on $d$ changes. The light gray points show this dependence for small $\zeta_{max}$, and the dark gray points show the dependence for large $\zeta_{max}$}
	\label{fig:spacing}
\end{figure}

We find that the distribution dependence on spacing presented in Fig.~\ref{fig:diffraction} does not hold for all $\zeta_{max}$ angular amplitudes of oscillation.  If $\zeta_{max}$ is sufficiently small, the distance swept out in a single cycle, $2\ell \zeta_{max}$ does not exceed the post diameter, $2r$. The qualitative behavior change we observe for small $\zeta_{max}$ is consistent with this observation, falling to the left of the dashed line in Fig.~\ref{fig:diffraction}c. For large $\zeta_{max}$, the body becomes very curved and points along the body in the direction of travel are no longer monotonically increasing from tail to head. We suspect that this may set a qualitative change in behavior as well.  Fig.~\ref{fig:spacing} shows the dependence of the spread of the distributions, $\theta_{70}$, on the inter-post spacing, $d$, for two amplitudes with qualitatively different behavior.

\subsection{Multi-post configuration: One dominant head collision}
\label{suppmat:head}
\begin{figure}[h!]
	\includegraphics[width=0.5\columnwidth]{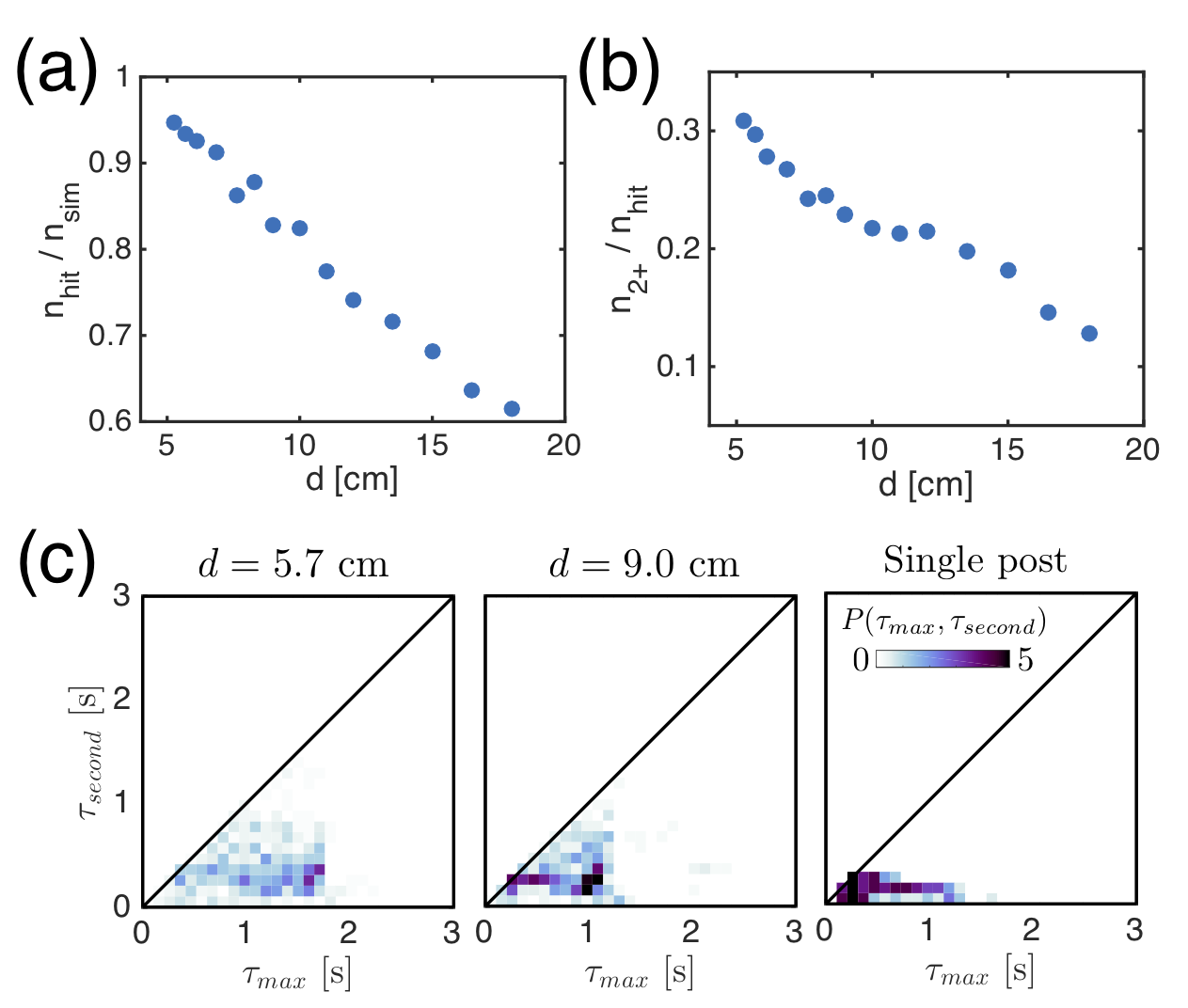}
	\caption{\textbf{Fraction of simulations with head collisions.} (a) Fraction of simulations for which at least one head-post collision occurred as a function of spacing. (b) Fraction of simulations in (a) for which two or more collisions occurred. (c) Probability maps for second-longest vs longest durations for $d = 5.7$~cm  (left) to $d = 9.0$~cm (middle) to single post (right). Only  simulations for which there were at least two collisions are shown here.}
	\label{fig:hit}
\end{figure}

To demonstrate that there is one dominant collision in the multi-post configuration, we first show that most simulations, even for small spacings, had one head-post collision. Fig.~\ref{fig:hit}a shows how many of the simulations, $n_{hit}$, had at least one collision between the head of the snake and the post row relative to the total number of simulations, $n_{sim}$. The number of simulations in which two or more collisions occurred, $n_{2+}$, compared to $n_{hit}$ is shown in Fig.~\ref{fig:hit}b.    

Of the simulations in which multiple collisions occur, we next show that the second-longest collision is typically not of comparable duration. In Fig.~\ref{fig:hit}c, two-dimensional probability densities of second-longest vs longest duration are shown for four post configurations.  If collisions were of comparable durations, the density of points would lie along the black lines in each plot. However, in each case, most of the points are concentrated below the line. This, along with the decreasing number of simulations for which multiple collisions occur, confirms that there is typically one dominant collision.

\subsection{Single- and multi-post collision states}
\label{suppmat:stateDistances}

\begin{figure}[h!]
	\includegraphics[width=0.8\columnwidth]{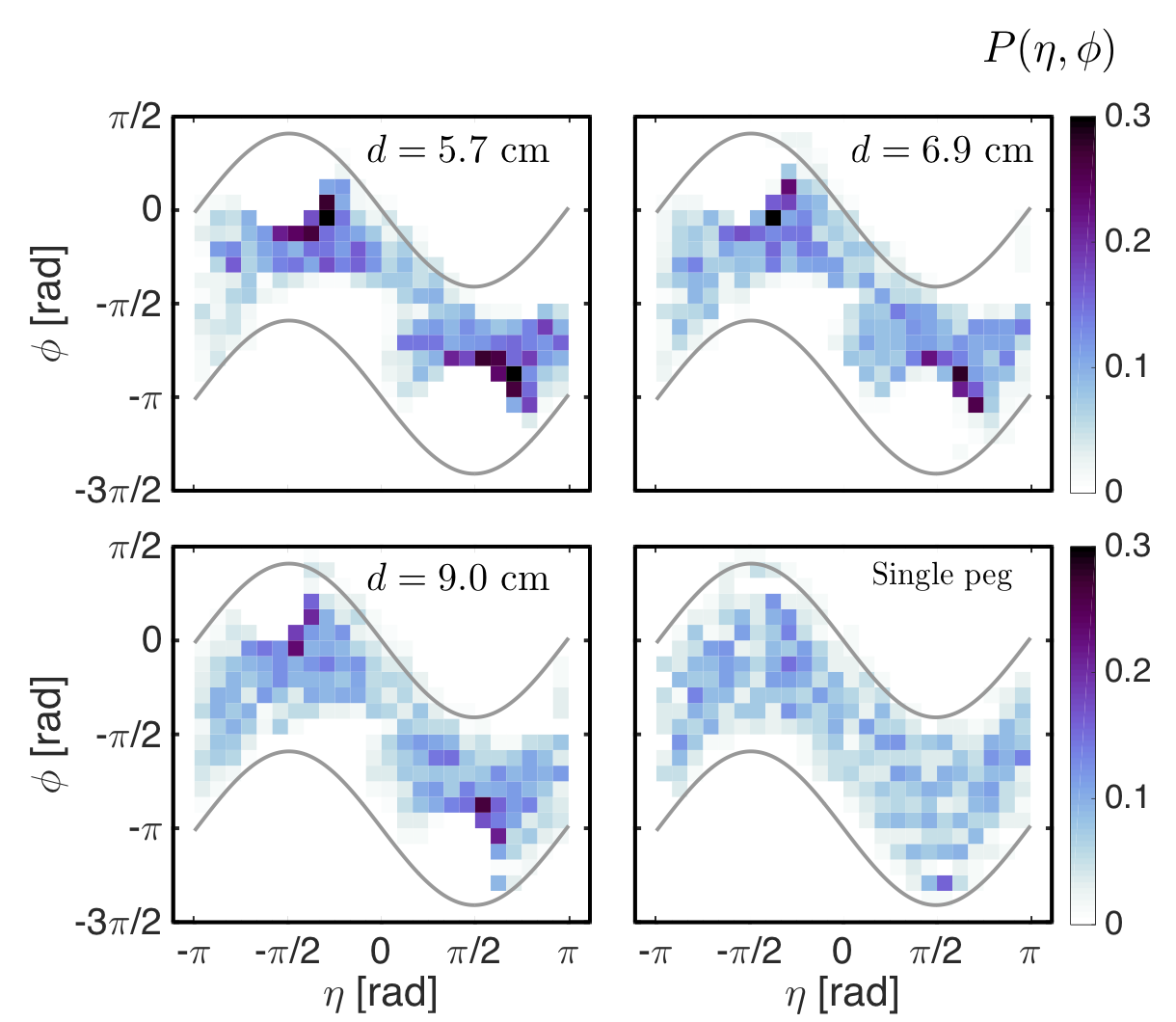}
	\caption{\textbf{Densities of collision states.} Densities of states shift around as spacing is changed. In the single-post case, all allowed states are evenly sampled.  As spacing is decreased when multiple posts are present, some regions become inaccessible and others more favored.}
	\label{fig:densities}
\end{figure}

Fig.~\ref{fig:densities} shows how the density collision states depends on the inter-post spacing.  For the single-post simulations (bottom right), all allowed states are evenly-sampled.  As spacing is decreased, certain undulation-phase and impact location collision states become inaccessible, and others become more likely to occur.  These excluded regions become larger as spacing becomes smaller, and the non-uniformity of the densities of remaining states becomes more pronounced.

For each post configuration, initial conditions within the relevant region were randomly generated. Therefore, we did not necessarily have information about precisely the same collision for single- and multi-post configurations. Therefore, to determine how collision states were influenced by the presence and location of additional posts, we identified, in $(\eta,\phi)$ space, the single-post state closest to each multi-post state by minimizing $\delta_{\eta\phi} = \sqrt{(\eta_s-\eta_m)^2+(\phi_s-\phi_m)^2}$. 

\begin{figure}[h!]
	\includegraphics[width=0.5\columnwidth]{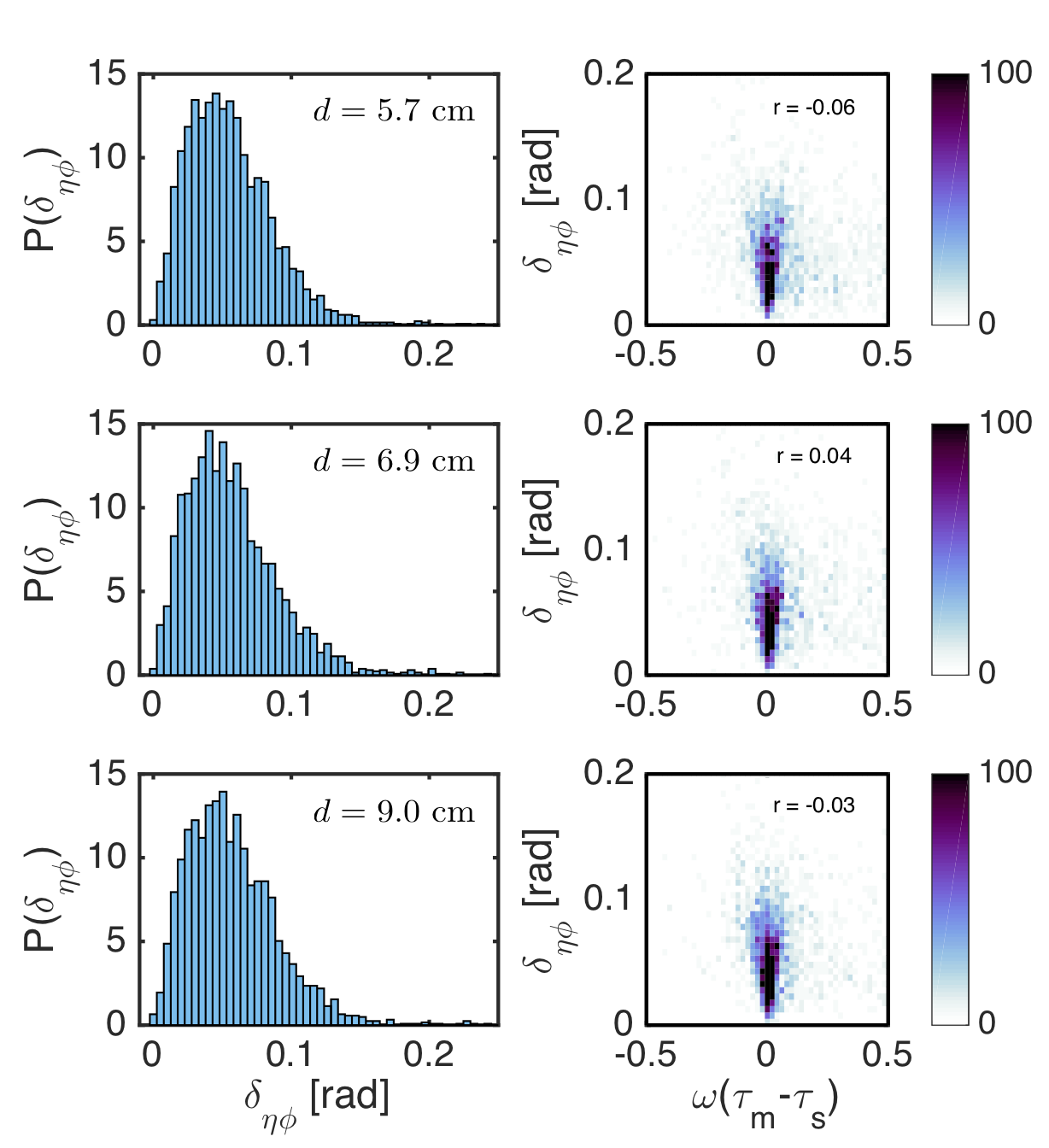}
	\caption{\textbf{Distances between single- and multi-post states.} Each row shows data for a spacing specified in left plot. Left column: Probability distributions of distances between nearest single and multi-peg states (nearest is defined by the smallest euclidean distance between states in $(\eta,\phi)$-space). Right column: Two dimensional PDFs showing that there is no significant correlation between deviation from $\omega \tau_m = \omega \tau_s$ line and distance between nearest single- and multi-peg collision states. }
	\label{fig:distances}
\end{figure}

The distributions of distances between the single- and multi-post states are shown in the left column of Fig.~\ref{fig:distances}. These distributions do not depend on post spacing, and in all cases, distances are typically small, so single-post points assigned to multi-post states are nearby in $(\eta,\phi)$ space. As a final check, we show in the right column of Fig.~\ref{fig:distances} that there is no significant correlation between $\delta_{\eta\phi}$ and deviation from the $\omega \tau_m = \omega \tau_s$ trend. Two-dimensional PDFs for two post configurations are shown, and the corresponding correlation coefficient for each spacing is given in each plot.

\clearpage
\subsection{Supplementary movies}
\vspace{5mm}
Movies available upon request. 
\vspace{5mm}

Movie 1. \textbf{Robotic snake in single-post environment.} A view of the robotic snake moving toward and interacting with a single post.  The sliding/pushing head-post interaction is visible here.
\label{mov:singlePostRobot}
\vspace{5mm}

Movie 2. \textbf{Robotic snake in multi-post environment.} An overhead view of several experiments in which the robotic snake moving toward, interacting with, and subsequently exiting the multi-post array (here, $d = 5.7$~cm). The final heading depends on the initial placement of robot, which is varied along the fore-aft direction here. 
\label{mov:multiPostRobot}
\vspace{5mm}

Movie 3. \textbf{Simulated snake in multi-post environment.} Three examples of the simulated snake interacting with a multi-post array ($d = 5.7$~cm).
\label{mov:multiPostSim}
\vspace{5mm}

Movie 4. \textbf{Emergence of preferred directions.} Summation of binary images created from the head trajectory of the robot in each of $329$ experiments for $d = 5.7$~cm.  Trajectories from different initial positions are added in a randomized order. As more experiments are included, a more complete picture of possible interactions and outcomes appears and preferred scattering directions emerge.
\label{mov:preferredDirections}

\end{document}